\documentclass[a4paper,11pt]{article}
\usepackage{jinstpub} 
\usepackage{lineno}
\usepackage{hyperref}
\usepackage{booktabs}
\usepackage{subcaption}
\usepackage{multirow}
\usepackage{float}
\usepackage{upgreek,bm}
\title{First Production of Skipper-CCD Modules for the DAMIC-M Experiment}

\author[a,b]{H.~Lin,}
\author[a,c,*]{M.~Traina,\note[*]{Corresponding author.}}
\author[d]{S.~Paul,}

\author[a]{K.~Aggarwal,}
\author[e]{I.~Arnquist,}
\author[c]{N.~Castell\'{o}-Mor,}
\author[a]{A.\,E.~Chavarria,}
\author[a]{M.~Conde,}
\author[f]{C.~De Dominicis,}
\author[a]{M.~Huehn,}
\author[b]{S.~Hope,}
\author[e]{T.~Hossbach,}
\author[f]{L.~Iddir,}
\author[g]{I.~Lawson,}
\author[d]{R.~Lou,}
\author[d]{S.~Munagavalasa,}
\author[b]{D.~Norcini,}
\author[d,f,h]{P.~Privitera,}
\author[d]{B.~Roach,}
\author[a]{R.~Roehnelt,}
\author[e]{N.~Rocco,}
\author[e]{R.~Saldanha,}
\author[e]{T.~Schleider,}
\author[d]{R.~Smida,}
\author[d]{B.~Stillwell,}
\author[a]{A.~Vellozzi,}
\author[b]{D.~Venegas-Vargas,}
\author[d]{R.~Yajur,}

\author[i]{N.~Avalos,}
\author[j]{X.~Bertou,}
\author[j]{C.~Bourgeois,}
\author[d]{J.~Brandt,}
\author[c]{C.~Centeno~Lorca,}
\author[d]{J.~Cuevas-Zepeda,} 
\author[k]{A.~Dastgheibi-Fard,}
\author[j]{O.~Deligny,}
\author[d]{R.~Desani,}
\author[f]{M.~Dhellot,}
\author[c]{J.~Duarte-Campderros,}
\author[i]{E.~Estrada,}
\author[l]{D.~Florin,}
\author[l]{N.~Gadola,}
\author[f]{R.~Ga\"{i}or,}
\author[l]{E.-L.~Gkougkousis,}
\author[c]{J.~Gonz\'{a}lez S\'{a}nchez,}
\author[c]{B.~Kavanagh,}
\author[l]{B.~Kilminster,}
\author[f]{A.~Letessier Selvon,}
\author[j]{P.~Loaiza,}
\author[a]{D.~Loeb,}
\author[c]{A.~Lopez-Virto,}
\author[c]{D.~Moya Martin,}
\author[d]{J.~Noonan,}
\author[a]{X.~Ni,}
\author[c]{M.~Perez~Martinez,}
\author[a]{D.~Peterson,}
\author[j]{O.~Pochon,}
\author[j]{D.~Reynet,}
\author[l]{P.~Robmann,}
\author[m]{M.~Settimo,}
\author[b]{S.~Smee,}
\author[a]{T.~Van~Wechel,}
\author[c]{R.~Vilar,}
\author[j]{P.~Vallerand,}
\author[l]{A.~Vollhardt,}
\author[l]{D.~Wolf,}
\author[b]{C.~Zhu,}
\author[f]{Y.~Zhu.}

\affiliation[a]{Center for Experimental Nuclear Physics and Astrophysics, University of Washington, Seattle, WA, United States}

\affiliation[b]{Department of Physics and Astronomy, Johns Hopkins University, Baltimore, MD , USA}

\affiliation[c]{Instituto de F\'{i}sica de Cantabria (IFCA), CSIC - Universidad de Cantabria, Santander, Spain}

\affiliation[d]{Kavli Institute for Cosmological Physics and The Enrico Fermi Institute, The University of Chicago, Chicago, IL, United States}

\affiliation[e]{Pacific Northwest National Laboratory (PNNL), Richland, WA, United States}

\affiliation[f]{Laboratoire de physique nucl\'{e}aire et des hautes \'{e}nergies (LPNHE), Sorbonne Universit\'{e}, Universit\'{e} Paris Cit\'{e}, CNRS/IN2P3, Paris, France}

\affiliation[g]{SNOLAB, Lively, ON, Canada}

\affiliation[h]{Laboratoire de Physique, École Normale Supérieure, Sorbonne Université, Université Paris Cité, CNRS/IN2P3, Paris, France}

\affiliation[i]{Centro At\'{o}mico Bariloche and Instituto Balseiro, Comisi\'{o}n Nacional de Energ\'{i}a At\'{o}mica (CNEA), Consejo Nacional de Investigaciones Cient\'{i}ficas y T\'{e}cnicas (CONICET), Universidad Nacional de Cuyo (UNCUYO), San Carlos de Bariloche, Argentina}

\affiliation[j]{IJCLab, CNRS/IN2P3, Universit\'{e} Paris-Saclay, Orsay, France}

\affiliation[k]{LPSC LSM, CNRS/IN2P3, Universit\'{e} Grenoble-Alpes, Grenoble, France}

\affiliation[l]{Universit\"{a}t Z\"{u}rich Physik Institut, Z\"{u}rich, Switzerland}

\affiliation[m]{SUBATECH, Nantes Universit\'{e}, IMT Atlantique, CNRS-IN2P3, Nantes, France}

\emailAdd{mtraina@uw.edu}

\abstract{The DAMIC-M experiment will search for sub-GeV dark matter particles with a large array of silicon skipper charge-coupled devices (CCDs) at the Modane Underground Laboratory (LSM) in France. After five years of development, we recently completed the production of 28 CCD modules at the University of Washington, each consisting of four 9-megapixel skipper CCDs. Material screening and background controls were implemented to meet stringent radio-purity targets, while extensive testing was employed to select science-grade CCDs for the modules and confirm their excellent performance after fabrication. Further testing at LSM will select 26 of these modules (${\sim}$350 g active mass) to be installed and operated in the DAMIC-M detector in early 2026.}


\begin{document}
\maketitle
\flushbottom

\section{Introduction}
\label{sec:intro}

Dark matter is a non-baryonic, non-luminous form of matter whose existence is strongly supported by astrophysical and cosmological observations \cite{Bertone:2016nfn}. A leading hypothesis suggests that dark matter consists of Weakly Interacting Massive Particles (WIMPs) with mass exceeding that of a proton (${\sim}1$~GeV/c$^2$)~\cite{Drukier:1984vhf,Goodman:1984dc,Billard:2021uyg}. Searches for WIMPs in the past decades have consistently returned null results. This has shifted the attention of the community toward lighter sub-GeV dark matter particles, e.g., from a ``hidden sector,'' in recent years~\cite{Boehm:2003ha,Hooper:2008im,Pospelov:2007mp,Knapen:2017xzo}. The DAMIC-M (DArk Matter In CCDs at Modane) detector is designed to search for ionization signals of only a few electrons produced by sub-GeV dark matter in  silicon skipper charge-coupled devices (CCDs) \cite{Privitera:2024tpq}. The first deployment of DAMIC-M will consist of 26 CCD modules (${\sim}350$~g total active mass), each with four 9-megapixel CCDs, in a shielded cryostat in the Modane Underground Laboratory (LSM) in France.
The detector was designed, and components were carefully selected, to achieve a background rate ${\lesssim}1~\text{event}\cdot \text{keV}^{-1} \text{kg}^{-1} \text{day}^{-1}$ in the sub-keV region of interest (ROI) for the dark matter search.
The CCD modules were developed over five years at the Center for Experimental Nuclear Physics and Astrophysics (CENPA) at the University of Washington (UW).
The design was validated by operating two prototype modules in the Low Background Chamber (LBC) test stand at LSM~\cite{DAMIC-M:2024ooa}, which has already provided world-leading results in the search for sub-GeV dark matter~\cite{DAMIC-M:2025luv}.

The DAMIC-M collaboration fabricated 28 CCD modules at UW between July and December 2024. A total of 188 single CCDs were tested in a temporary package in cryogenic test chambers, which resulted in 113 science-grade CCDs selected to fabricate CCD modules. The CCD modules were then tested to confirm science-grade performance. Strict protocols to mitigate radioactive contamination of detector materials were implemented from the onset to meet DAMIC-M radio-purity requirements.
Radioactive contaminants such as $^{3}$H (from cosmogenic activation of the CCD silicon), $^{210}$Pb (from surface deposition of radon progeny) and ubiquitous primordial U/Th/K dominate the radioactive background spectrum \cite{DAMIC:2021crr, DAMIC:2021esz}.
Cosmogenic activation is mitigated with shielded transport and storage to decrease the cosmogenic neutron flux.
Radon plate-out is suppressed using low-radon facilities and nitrogen-flushed environments.
U/Th/K control requires sourcing of high-purity materials, chemical cleaning, and clean room operations.
At the same time, all processes must include electrostatic discharge (ESD) precautions to prevent damage to the CCDs.

This paper details the packaging and testing activities conducted at UW for the production of the DAMIC-M CCD modules. It is organized as follows. In Section \ref{sec:moduleprod}, we provide an overview of the activities, illustrating the main components of CCD modules, as well as infrastructure and logistics for packaging, testing, and background mitigation. In Section \ref{sec:diepackaging}, we report the details of single-die (CCD) and module packaging: the temporary die-test package, the fabrication of the CCD pitch adapters, and packaging procedures. Section \ref{sec:mitigation} delves into the low-background protocols. In Section \ref{sec:ccdtesting}, we describe the setups and procedures used for single-die and module testing, as well as CCD selection criteria. 
In Section \ref{sec:results}, we present the fabrication results, highlighting the yield of science-grade devices, CCD module performance, and radioactive contamination. Finally, in Section \ref{sec:conclusion}, we summarize our findings and discuss their relevance to the objectives of the DAMIC-M experiment.
 
\section{CCD Module Production}
\label{sec:moduleprod}

\begin{figure}[t]
\includegraphics[width=1.00\textwidth]{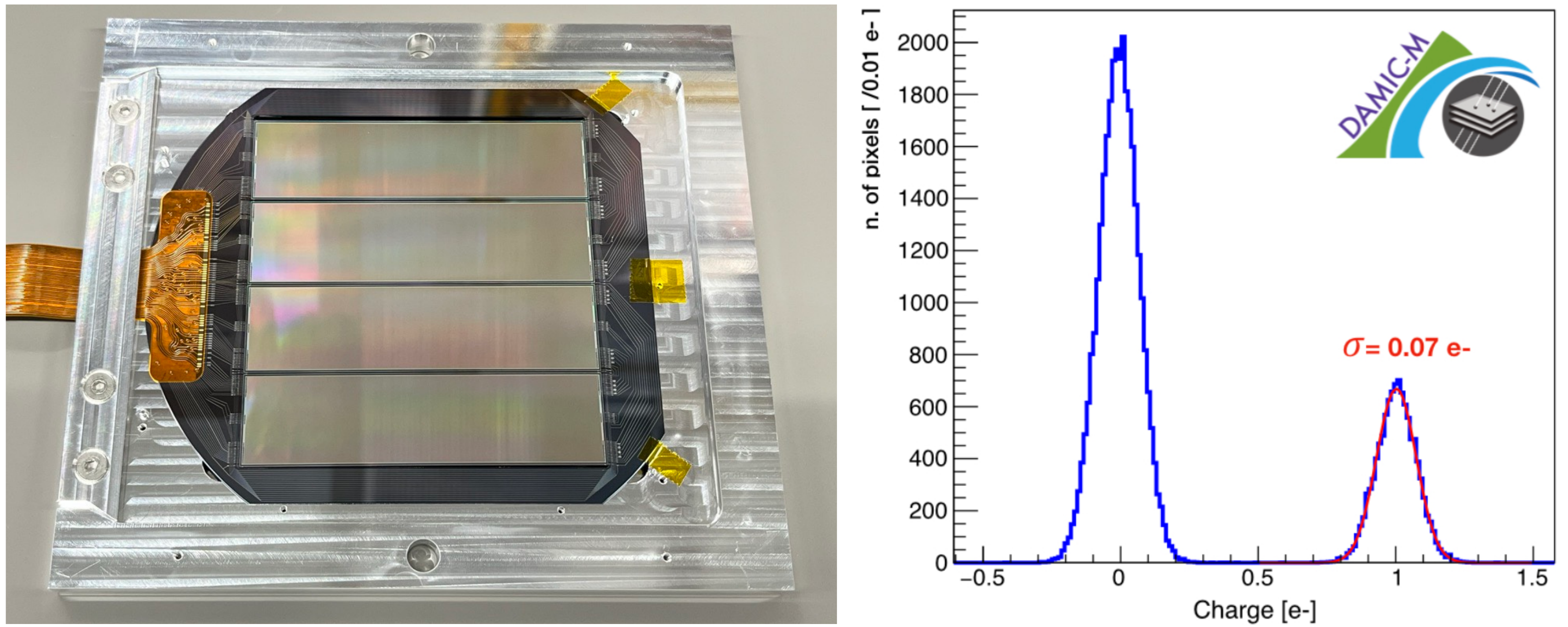}
\caption{Left: DAMIC-M CCD module sitting in its aluminum box. The four CCDs (silver rectangles) and the flex cable (orange) are glued under and over the pitch adapter (dark grey), respectively. Right: pixel charge distribution with single-electron resolution ($0.07~e^-$, $3000$~NDCM) from a DAMIC-M skipper CCD.}
\label{fig:modser}  %
\end{figure}

The DAMIC-M CCD module consists of three key components: four 9-Megapixel CCDs, one silicon pitch adapter, and one flex cable. Each DAMIC-M CCD has 6144$\times$1536 pixels of size 15\,$\upmu$m\,$\times$15\,$\upmu$m, and is 675\,$\upmu$m thick. Charge readout is performed with skipper amplifiers, one at each corner of the rectangular pixel array. Skipper amplifiers enable non-destructive charge measurements (NDCM), thereby suppressing electronic noise on the pixel charge measurement by a factor $\sqrt{N_\text{NDCM}}$. Resolution of individual charge carriers is possible with as few as 100 NDCMs.
The pitch adapter consists of aluminum traces patterned on a silicon frame, onto which four CCDs are glued.
It acts as an interposer between the flex cable, which carries the electrical signals from/to the electronics, and the CCDs.
The use of the pitch adapter maximizes the distance from the CCDs to the flex cable, one of the most radioactive components of the module.
Figure~\ref{fig:modser} shows a DAMIC-M CCD module (left) and a pixel charge distribution with single-electron resolution from a DAMIC-M skipper CCD (right).
Both CCDs and pitch adapters were produced starting with n-doped, high-resistivity (${>}10~\text{k}\Omega\cdot\text{cm}$) silicon wafers from an ingot grown by Topsil in Denmark \cite{topsil}, and sliced by Shin-Etsu Handotai in the UK \cite{sehe}. DAMIC-M CCDs were fabricated by Teledyne-DALSA in Quebec, Canada \cite{dalsa}. Various precautions were adopted from ingot growth until delivery of the diced CCD wafers to mitigate radioactive contamination, including transportation in a 16-ton steel shield in a container across the North Atlantic and North America, storage in a nitrogen-flushed environment under ${\sim}2000~\text{m}$ of granite in the SNOLAB underground laboratory in Ontario, Canada, and use of a 1.5-ton stainless steel shield  for storage during production at Teledyne-DALSA. 

\subsection{Infrastructure}
\label{sec:uwinfrastructure}

The fabrication of DAMIC-M modules was performed at UW, as key resources were readily available on site. CCD packaging and testing were conducted in the DAMIC-M laboratory in the Physics and Astronomy Building (PAB). Storage space with sufficient overburden and nitrogen-flushed enclosures was made available at the North Physics Laboratory (NPL). Additionally, the Washington Nanofabrication Facility (WNF) \cite{wnf} on campus enabled the fabrication of the silicon pitch adapters.

\begin{figure}[h!]
\centering
\includegraphics[width=0.63\textwidth]{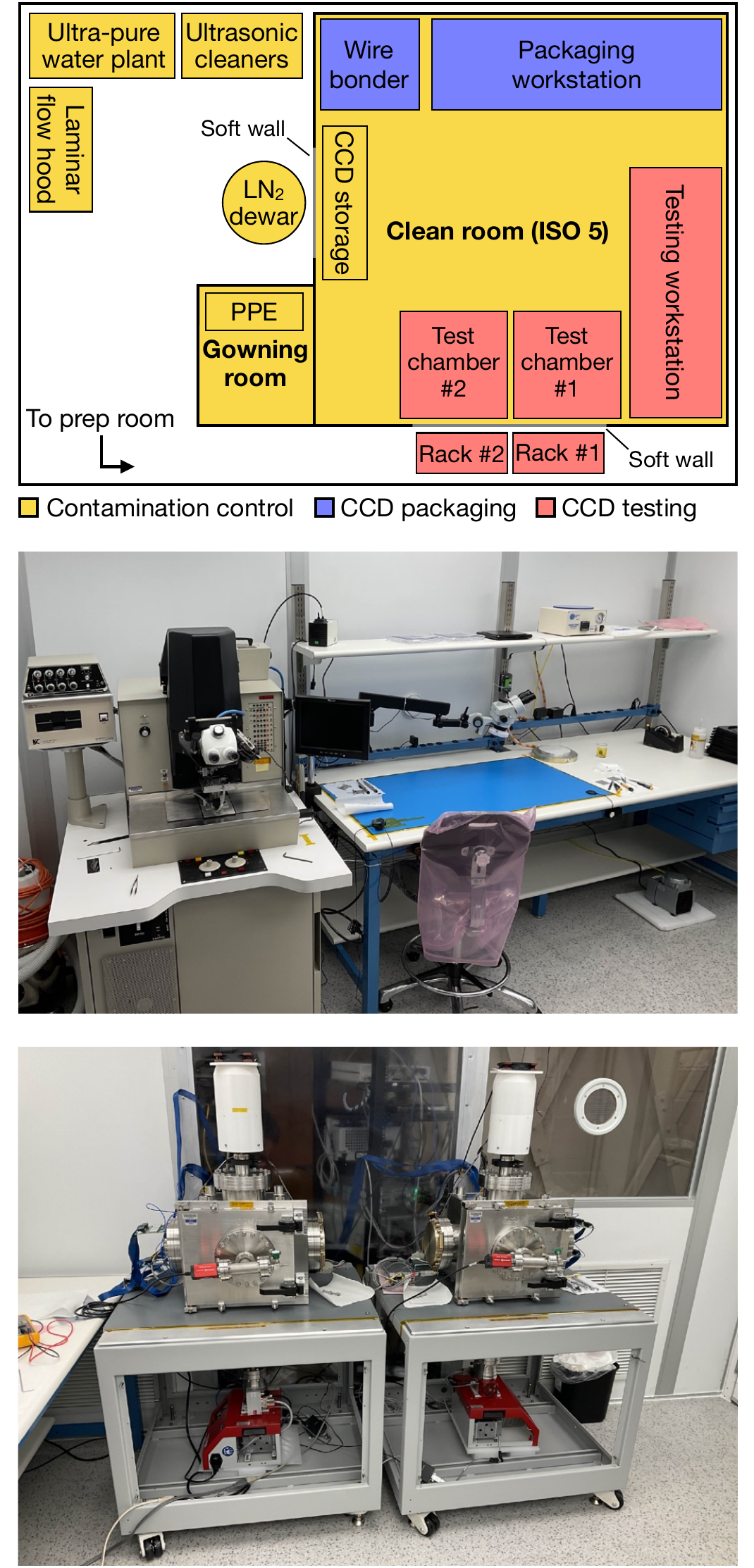}

\caption{Top: diagram of the DAMIC-M PAB lab at UW (not to scale). PPE stands for personal protective equipment. Middle, Bottom: photos of the packaging workstation and CCD test chambers inside the PAB lab clean room, respectively.}
\label{fig:pablab}  %
\end{figure}

The DAMIC-M group at UW has exclusive access to 80 m$^2$ of laboratory space in the PAB. The laboratory includes device packaging and testing equipment within an 18-m$^2$ ISO class 5 clean room, which is accessible through a 3.5-m$^2$ gowning room. The clean room features antistatic flooring and work surfaces (tables, shelving, etc.). It is equipped with tools for handling and packaging large-area semiconductor devices, including an automatic K\&S 1470 wire bonder with 10~cm~$\times$~10~cm table travel, an ESD-safe packaging workstation (ESD-safe mat, ground monitors, ionizing air gun, etc.), and antistatic storage cabinets flushed with liquid nitrogen (LN$_2$) boil-off. Two new, fully-equipped cryogenic test setups for CCD testing were provided by the DAMIC-M group at Johns Hopkins University (JHU) and commissioned in the PAB clean room. Racks with CCD read-out electronics and cryogenic controllers were installed outside the clean room to mitigate contamination. The PAB laboratory additionally features a laminar flow hood, an ultra-pure (${>}10$\,M$\Omega\cdot\text{cm}$) water plant, two ultrasonic cleaners, additional storage cabinets, a preparation room with a fume hood, a 1-ton crane, and a mezzanine. The clean room, preparation room, and mezzanine are equipped with humidity, temperature, and radon sensors, which are kept operational and consulted regularly during activities. A custom humidification system using ultra-pure water was designed to ensure relative humidity (RH) ${>}35\%$ whenever handling CCDs, while preserving laminar airflow inside the clean room. Figure \ref{fig:pablab} shows a diagram of the DAMIC-M lab at UW (top), a photo of the packaging workstation (middle), and a photo of the two CCD test chambers (bottom).

\begin{figure}[t]
\centering
\includegraphics[width=0.99\textwidth]{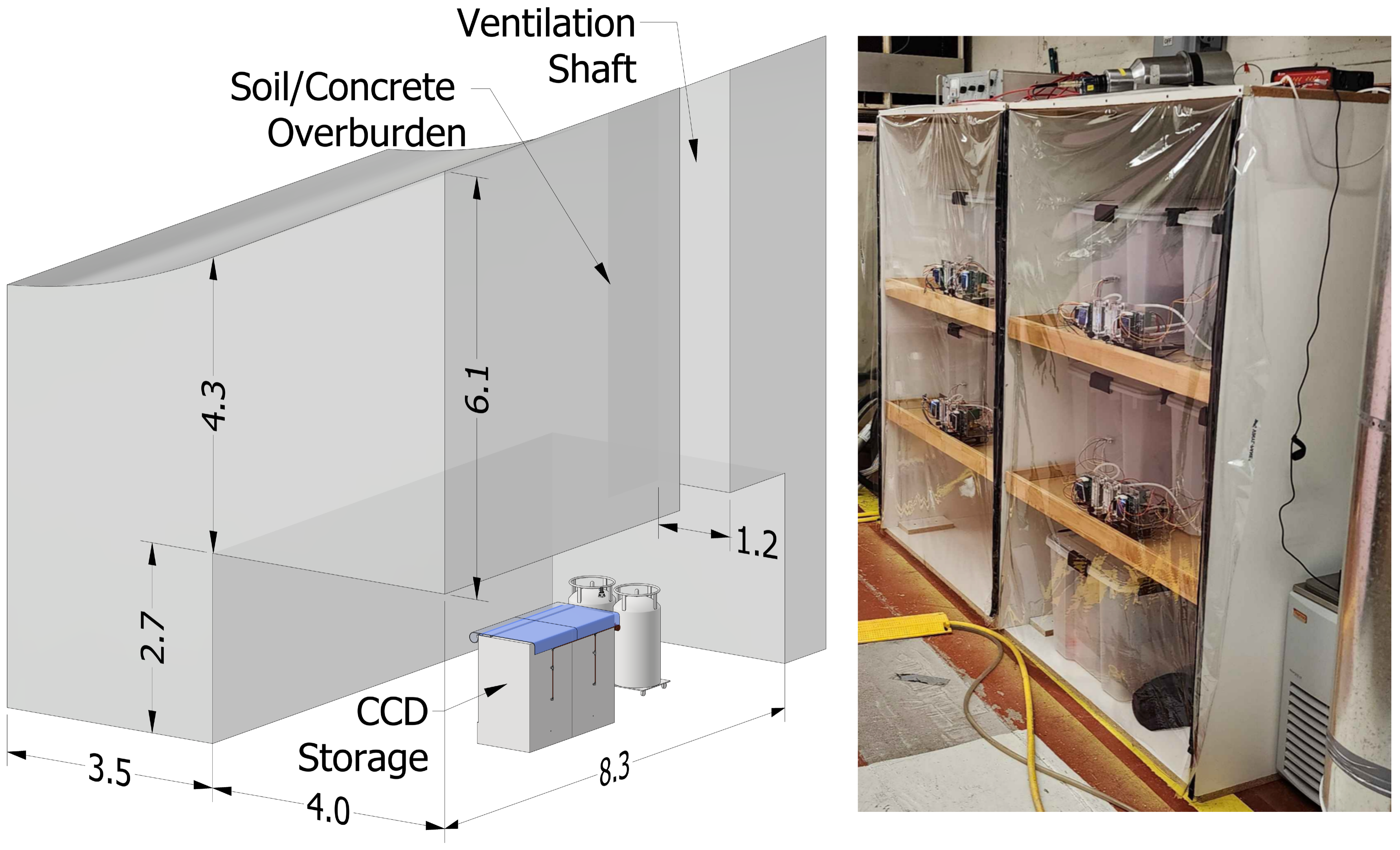}
\caption{Left: rendering of the Gravity Garage at NPL (dimensions in meters). Horizontal distance between shaft and CCD storage unit is 1.2 m. Right: CCD storage units in Gravity Garage. Apparatus for measurement of high-energy neutrons (see Sec. \ref{sec:mitigation}) sits on top of the shelving unit.}
\label{fig:gravitygarage}  
\end{figure}

To mitigate cosmogenic activation during storage, we used the Gravity Garage in NPL (${\sim}35~\text{m}^2$), which offers between $4.3~$m and $6.1~$m of soil/concrete overburden, thereby suppressing tritium production in silicon by a factor of ${\sim}$50. We set up antistatic enclosures with LN$_2$ boil-off supply lines to store diced CCD wafers (pre-test), tested dice, CCD modules (post-test), silicon wafers for pitch adapters, and other low-background items for DAMIC-M. Figure \ref{fig:gravitygarage} shows a rendering of the Gravity Garage and overburden with dimensions and location of the CCD storage space (left), and a photo of the CCD storage units (right), including the apparatus used to measure the high-energy neutron flux (see Sec. \ref{sec:mitigation}) sitting on top of the shelves. 

The Washington Nanofabrication Facility (WNF) is an open-access 1400-m$^2$ user facility with capabilities for semiconductor processing and packaging. The DAMIC-M group at UW has access to tools and technical support at WNF, including suites for lithography, etching, evaporation, deposition, annealing, and dicing. The pitch adapter fabrication was conducted mainly at WNF, with only laser cutting (to produce a window in the processed wafer) performed at NPL. After laser cutting, wafers would head back to WNF for dicing and final cleaning. The fabrication of pitch adapters is detailed in Sec. \ref{sec:diepackaging}.

\subsection{Logistics}
\label{sec:logistics}

\begin{figure}[t!]
\centering
\includegraphics[width=0.7\textwidth]{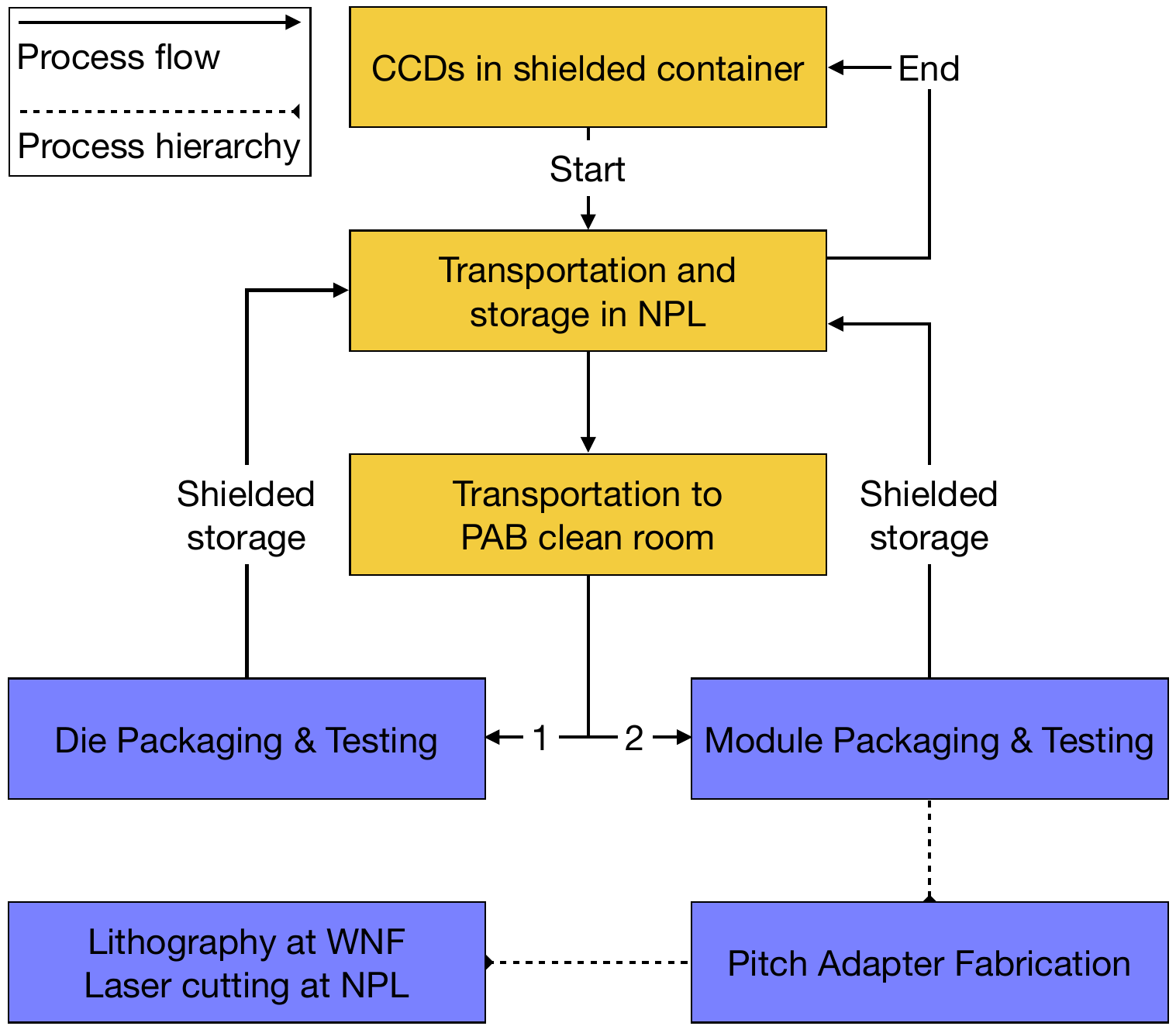}

\caption{Process flow for CCD module production at UW. Yellow boxes indicate logistics and background mitigation stages. Blue boxes correspond to packaging and testing operations. Retrieval of CCD wafers from shielded container sets the start of activities. Deposit of CCD modules in shielded container concludes the activities.}
\label{fig:flowchart}  %
\end{figure}

Activities began in July 2024, with the shipment of the shielded container from SNOLAB to Vancouver, Canada. In Vancouver, two cassettes with 23 and 24 diced wafers, respectively, were retrieved and transported in their sealed boxes by car to Seattle (${\sim}3$~h). The arrival of the wafers at UW sets the start of the logistics and background mitigation workflow, following the steps outlined in the flowchart shown in Fig. \ref{fig:flowchart}. Wafers spent about a month in the Gravity Garage while the PAB lab was being prepared, including test chamber commissioning and thorough decontamination of the clean room and other working spaces.

Similar process flows were implemented for die and module packaging and testing. In both cases, the process involved the controlled storage and transfer of dice on a wafer basis. A daily transfer routine was established to streamline operations, balancing wafer deliveries with the return of tested dice for storage. All handling of the CCD wafers in NPL took place in a dedicated clean laminar flow hood equipped with ESD protections. Wafers were carefully prepared and transported from NPL to PAB (${\sim}10$-min trip) inside a two-layer antistatic, nitrogen-flushed container. 
Wafers were brought into the PAB cleanroom inside the inner antistatic container, where the wire bonder operator and their assistant performed the packaging. 
Systematic tracking of the individual dice had to be implemented since there are no identifiable labels on the dice.
Packaged dice were temporarily stored in a nitrogen-flushed, antistatic cabinet prior to installation in test chambers.
Die testing, from installation to retrieval of the CCDs, spanned 24 hours.
Afterward, tested dice were placed in wafer carriers following a predefined arrangement, with dice from the same wafer in one carrier, and stored in the cabinet.
Storage was organized with diced wafers and packaged/tested dice assigned to predefined compartments.
The entire process was thoroughly logged on a wafer basis, noting any possible contamination and ESD-safety events. We packaged and tested 188 dice in 35 days, between September and October 2024.
The die packaging and testing routine follows:
\begin{itemize}  
    \item Wafer transfer: 2 (1) diced wafers were transported from NPL to PAB.  
    \item Die packaging: 6 dice were packaged in PAB, within 4 hours.
    \item Die testing: 6 dice were tested in PAB, including thermal cycle, within 24 hours.  
    \item Die return: 1 (2) wafers were transported back to NPL, in the return trip after new wafers were delivered.  
\end{itemize}

The assembly of one CCD module required up to four wafers, with substantially different packaging and testing procedures compared to individual dice (see Sec. \ref{sec:diepackaging} and Sec. \ref{sec:ccdtesting}). Moreover, pitch adapter fabrication was conducted in parallel, which required efficient allocation of personnel and additional logistics. We packaged and tested 28 CCD modules in 31 days, between November and December 2024.
The module packaging and testing routine follows:
\begin{itemize}  
    \item CCD selection: 4 science-grade CCDs were selected to package one CCD module.
    \item Wafer transfer: up to 6 wafers were transported from NPL to PAB.
    \item Module packaging: 3 CCD modules were packaged in PAB, within 6 hours.
    \item Module testing: 3 CCD modules were tested, including thermal cycle, within 48 hours.  
    \item Shipment prep: 3 CCD modules were prepared for shipment in PAB.
    \item Module return: 3 tested CCD modules were transported back to NPL.  
\end{itemize}

\section{Die-test and CCD Module Packaging}
\label{sec:diepackaging}

This section details the single-die packaging procedures, including the design of a temporary CCD package for testing (Sec. \ref{sec:diepack}), pitch adapter fabrication at the WNF (Sec. \ref{sec:pitchadapt}) and CCD module packaging with selected dice (Sec. \ref{sec:modulepack}).
All components of the CCD module were selected for low radioactivity following an exhaustive assay program (Sec.~\ref{sec:radioresults}).

\subsection{Packaging for Die Test}
\label{sec:diepack}

\begin{figure}[tbp]
\includegraphics[height=4.74cm]{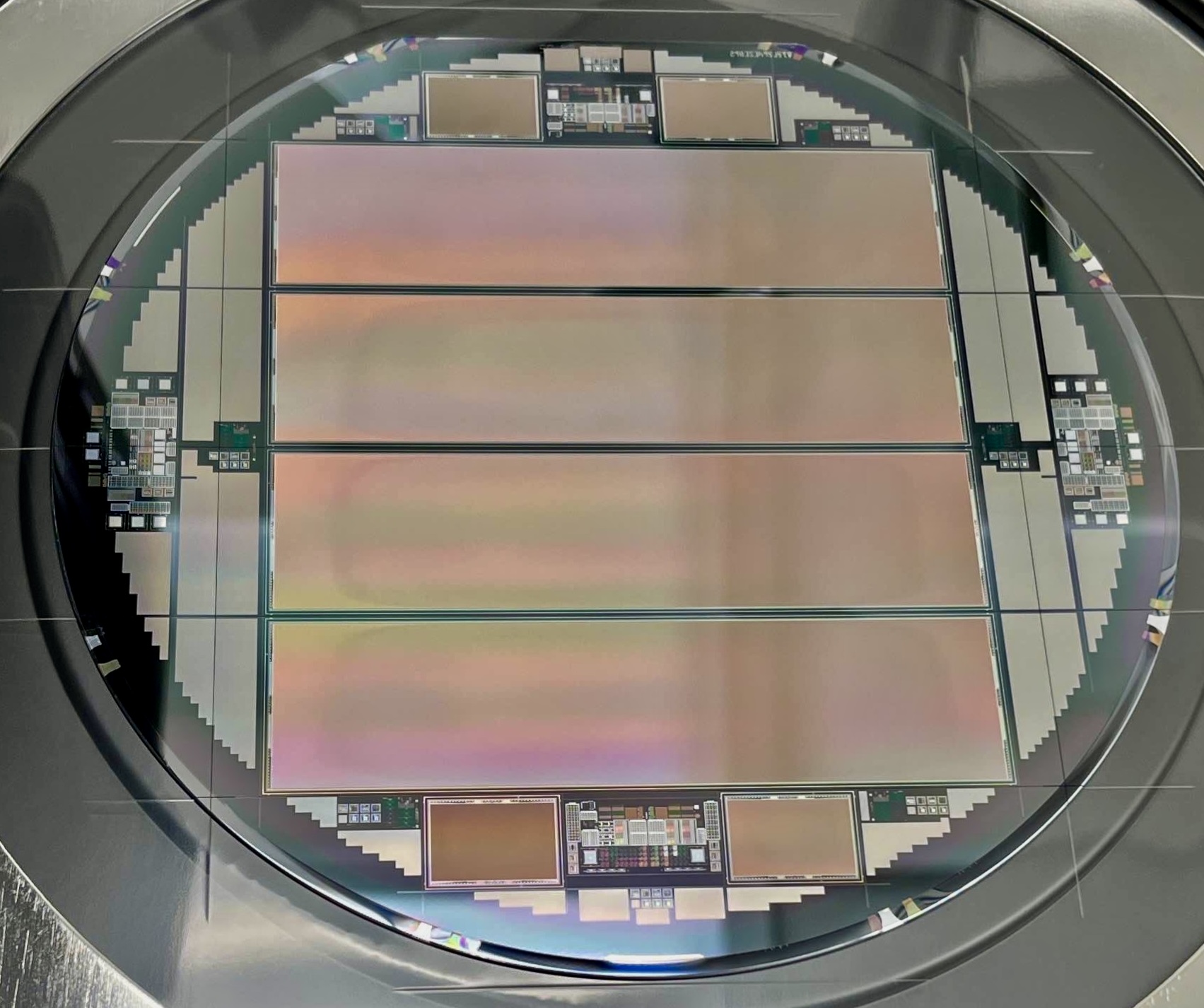}\hfill
\includegraphics[height=4.74cm]{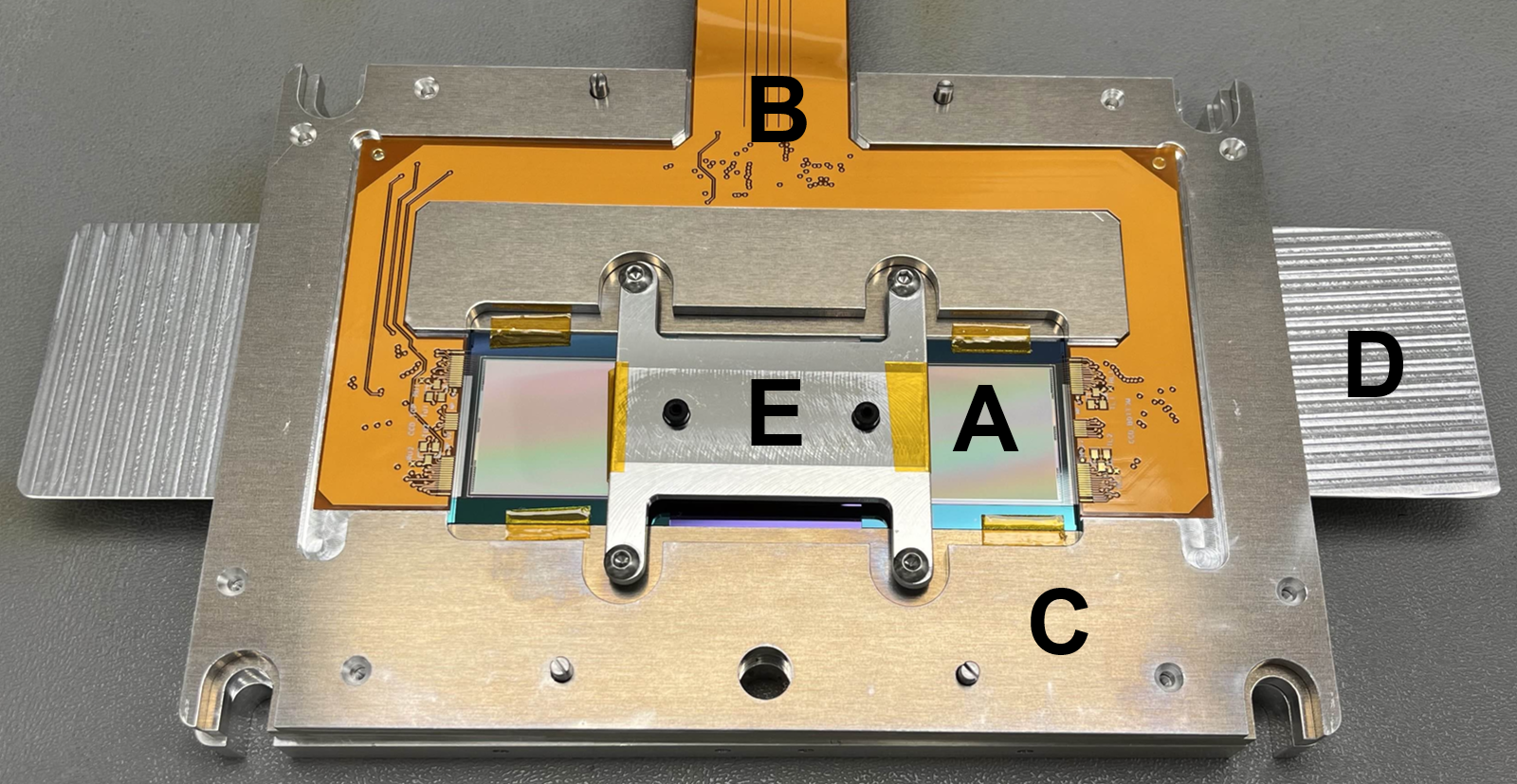}

\caption{Left: a diced wafer affixed to film containing 4 CCDs. Note the extended inactive area from the end of the device to the dicing street on each side. Right: components of the die test box: the CCD (A) sitting in a silicon pocket (purple) installed on the bottom layer, the flex cable (B) affixed to the middle layer (C), the lever plate (D) used to disassemble the box, and the presser plate (E) installed only during wire bonding and CCD retrieval. }
\label{fig:die_test_box}  
\end{figure}

The CCD die is a rectangular piece of silicon 25.0~mm wide and 106.5~mm long, which includes an extended inactive area of 5.9~mm past the end of the device on each side (Fig.~\ref{fig:die_test_box}, left).
A custom aluminum container, referred to as the die-test box, was developed as a temporary package to install individual dice in the test chamber.
Fig.~\ref{fig:die_test_box}, right shows the main components of the die-test box, which consists of three parts: the bottom layer that holds the die, the middle layer to which the flex cable is affixed, and a top lid. The die is placed within an insulating silicon pocket mounted on the bottom layer to provide electrical insulation. The middle layer is then placed on top, maintaining a clearance of 0.2~mm. The inactive area of the die overlaps with the middle layer by approximately 50~mm on each side so that the CCD pads are positioned close to the flex pads. This overlap, together with the silicon pocket, constrains the die in position.

To begin, a CCD is extracted from the diced wafer with a Techni-Pro vacuum tweezer and positioned within the silicon pocket. The middle layer and a presser plate are subsequently installed. 
The presser plate holds the die steady by applying gentle pressure via two antistatic rubber cups contacting 2~cm$^2$ of the CCD surface. 

\begin{figure}[htbp]
\includegraphics[height=6.3cm]{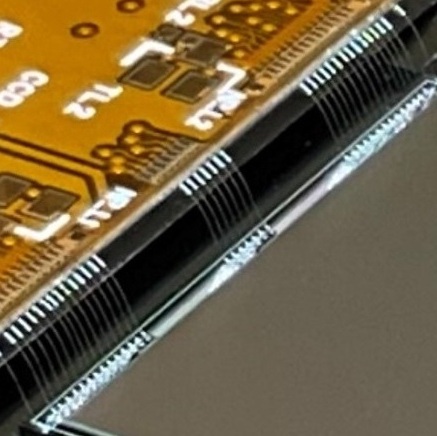}\hfill
\includegraphics[height=6.3cm]{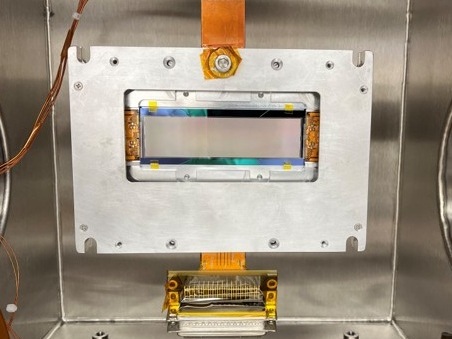}

\caption{Left: wire bonds from the CCD pads to the flex-cable pads. Right: CCD die mounted for testing. In this staged photo, a windowed lid is used to show the suspended die in a prototype package with a large 50-pin connector.
During die testing, a solid lid and a small-format Samtec 60-pin connector are used instead.
}
\label{fig:box_packaging}  %
\end{figure}

The wire connections between the 74 pairs of pads in the CCD and the flex are made by a programmable wire bonder with 25-$\upmu$m-thick aluminum (99\% Al, 1\% Si alloy) wires with a loop height of 2.5~mm (Fig.~\ref{fig:box_packaging}, left). The presser plate is removed after wire bonding and the lid is installed to ensure light tightness, completing the packaging of a single die. When the box is installed in the chamber for testing, the aluminum wires, which can hold a load of up to 0.15~N each, provide mechanical support to the die, as illustrated in Fig.~\ref{fig:box_packaging} (right). With the die suspended this way, there is minimal stress on the silicon, and heat transfer during the temperature cycle is primarily radiative.

Following testing, die retrieval is performed in reverse order of the packaging procedure. The presser plate is reinstalled before separating the middle flex layer and the bottom layer. Lever plates (Fig.~\ref{fig:die_test_box} right, D) are inserted between the layers to lift the middle layer, which is guided vertically by alignment dowel pins. The wires feature smaller joint angles on the flex end than on the die end, making the former joints stronger. This guarantees (with probability ${>}99$\%) that wire detachment occurs at the die end, which significantly simplifies the wire cleanup and consequently reduces the ESD risk to the device. The tested dice are retrieved from the bottom layer using vacuum tweezers, transferred to Gel-Pak VRLF-150C wafer carriers, and stored. The die-test boxes were reused for up to 12 dice.

\subsection{Pitch Adapter Fabrication}
\label{sec:pitchadapt}

The pitch adapter traces were laid out so that the four-CCD module can be controlled and read out with the same electronics as a single CCD.
The bias and clock signals are distributed to the four CCDs, and one of the two amplifiers closest to the flex of each CCD can be connected for readout.
The traces are low resistance (${<}300~\Omega$) and capacitance (${<}1$~nF) to minimize the distortion of the CCD signals and cross talk.
We chose to have a large window on the pitch adapter so that all CCD surfaces are exposed, minimizing the non-active volume of the DAMIC-M CCD array.
Furthermore, not having CCD surfaces covered by the pitch-adapter silicon significantly relaxes the requirement on the radio-purity of the pitch adapters.
On the other hand, the window makes the pitch adapter more fragile and requires that traces be routed through the narrow frame, which demands greater accuracy in the alignment of the trace pattern on the wafer.

\begin{figure}[t]
\centering
\includegraphics[height=6.8cm]{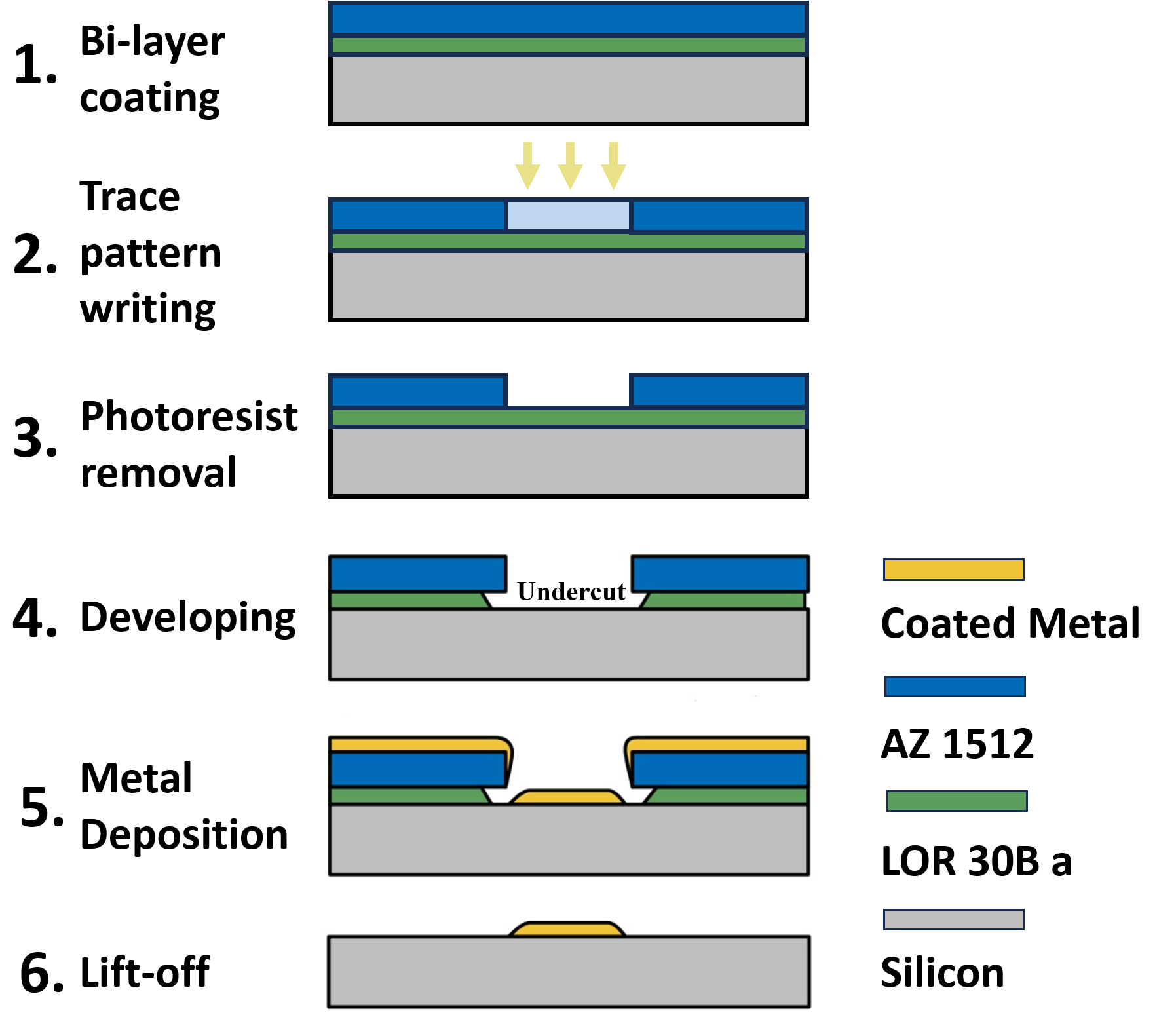}
\includegraphics[height=6.8cm]{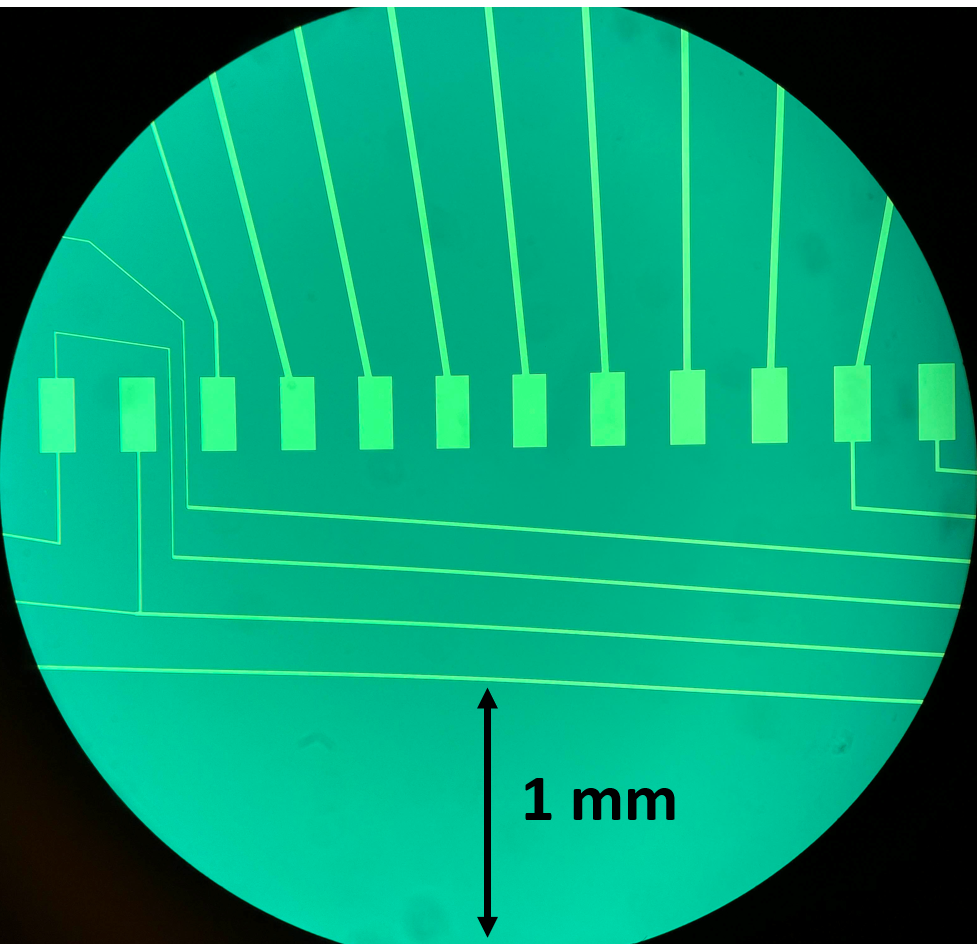}

\caption{Left: simplified schematic of the bi-layer lift-off process. (1) Two layers of photoresist are coated on the silicon surface. (2) Laser writes the trace pattern and degrades the top layer photoresist, which is subsequently dissolved in the AD-10 developer (3). The bottom layer is not photosensitive but is soluble in the developer. (4) As development proceeds, the bottom layer is gradually removed, forming an undercut profile. This undercut creates a gap between the metal trace and the photoresist after metal deposition (5), facilitating effective removal (6) of both resist layers during lift-off using acetone AZ 1:1 developer. Right: trace pattern inspected with an optical microscope.}
\label{fig:tracepa}  
\end{figure}

Prototyping and fabrication of the pitch adapter were conducted at WNF.
The process began with high resistivity 200-mm diameter silicon wafers with a thickness of 675~$\upmu$m.
A 1.6-$\upmu$m-thick layer of silicon dioxide was thermally grown via wet oxidation at temperatures exceeding 1000~$^\circ$C.
To create the trace patterns, which have a minimum width of 5.5~$\upmu$m,  a bi-layer lift-off process (Fig.~\ref{fig:tracepa}, left) was performed, consisting of LOR 30B and AZ 1512 positive photoresist as the bottom and top layers, respectively.
A Heidelberg DWL66+ direct-write laser lithography system with 20~mm writing head was used to write the pattern in approximately 40~minutes, ensuring accurate alignment.
After exposure, the photoresist was developed in an AD-10 (MIF 726) developer for 90~s. The development time was tuned to ensure a proper pattern undercut length on the bottom coating layer, which is critical to ensure the high quality of the metal deposition that follows. Before deposition, all wafers were spin-and-dried (SRD) and mild-plasma-cleaned in a reactive ion etching (RIE) system to improve surface cleanliness and ensure good metal adhesion. We inspected all wafers with an optical microscope to verify pattern integrity. 

During prototyping, we demonstrated that both electron beam physical vapor deposition (EB-PVD) and sputtering of metals produced high-quality traces suitable for reliable wire bonding.
For production, sputtering was selected due to availability of equipment at WNF.
A layer of 20~nm titanium was first deposited to improve adhesion, followed by a 980-nm-thick aluminum layer.
This was the tallest metal stack that we could reliably fabricate, resulting in trace resistances between 10~$\Omega$ and 300~$\Omega$, depending on length.
The sputtering process lasted approximately 30 minutes, followed by a 24-hour lift-off step in an acetone bath, and a 10-min soak in AZ 1:1 developer to remove the two photoresist layers. A SRD cleaning concludes the deposition process, after which all the traces underwent a second optical inspection.
A picture of the trace pattern in the microscope view is shown in Fig.~\ref{fig:tracepa}, right.

\begin{figure}[htbp]
\includegraphics[height=4.76cm]{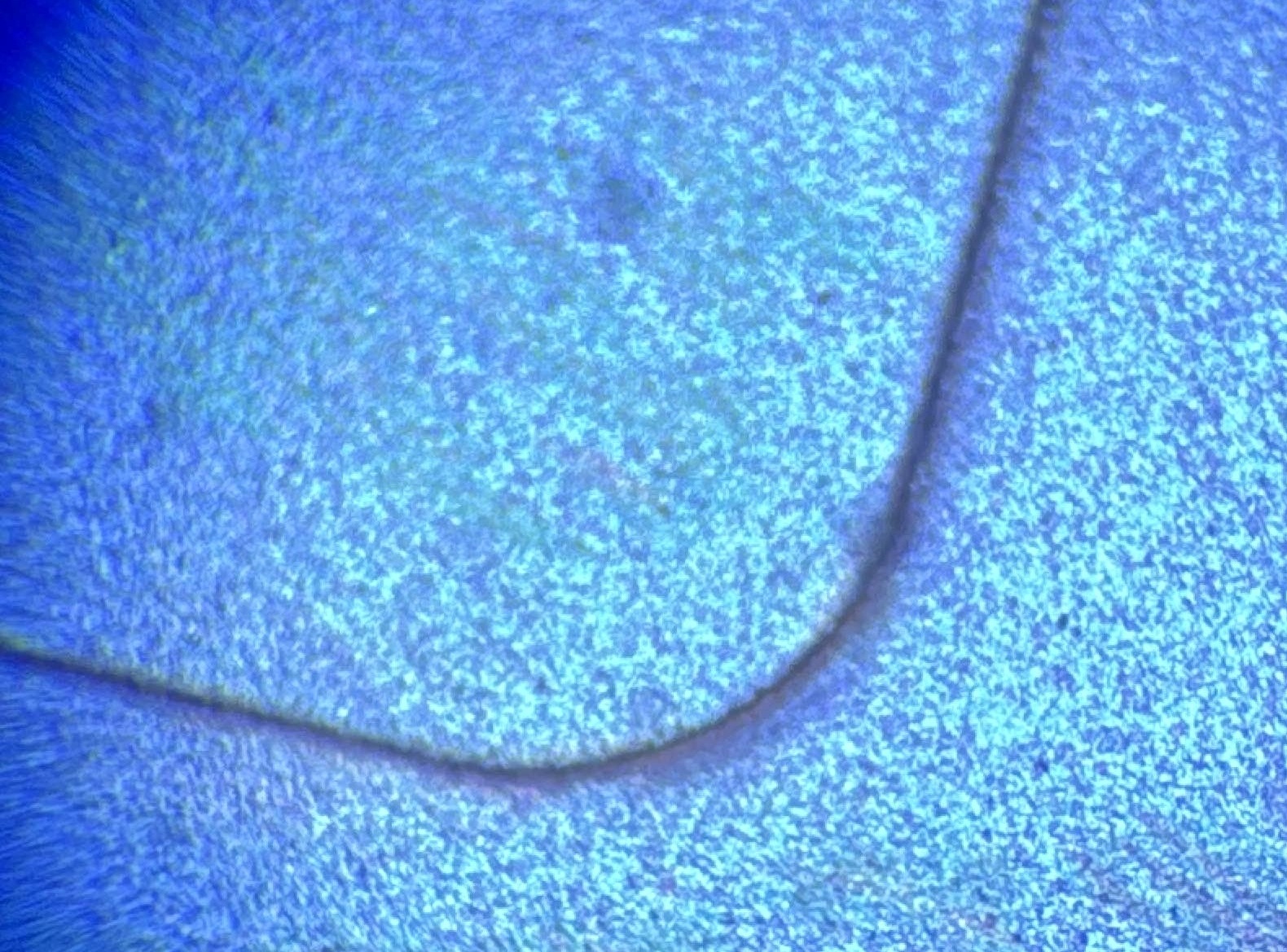}\hfill
\includegraphics[height=4.76cm]{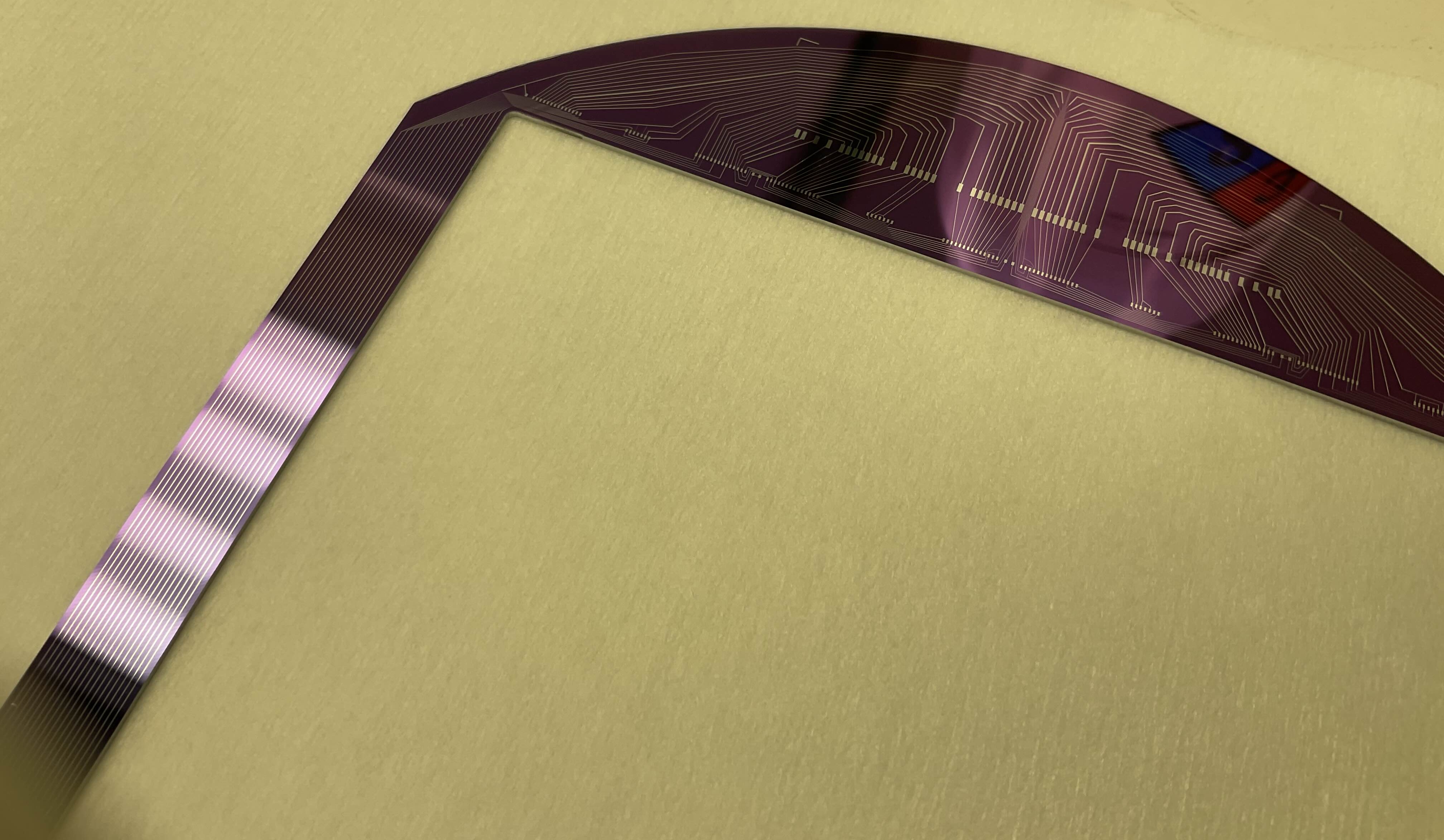}

\caption{Left: 1~mm-radius corner cut through by the laser, as viewed with a microscope. Right: window straight edges diced after laser cutting.}
\label{fig:PA_window}  %
\end{figure}

A 10~cm $\times$ 10~cm window was cut out from each pitch adapter.
Several methods were evaluated, including deep reactive ion etching (DRIE), laser cutting, and wet etching.
A hybrid approach combining laser cutting and dicing was selected to optimize costs and yield.
The cutting procedure involved two steps: laser cutting the round window corners with a radius of 1~mm, and dicing the straight edges between the corners using a precision dicing saw.
A sacrificial AZ 1512 photoresist layer was applied to protect the wafers during cutting and dicing.
Laser cutting was done at NPL using a UV laser (LPKF ProtoLaser U3) in a 5-hour routine during which the laser gradually cut through the silicon in five phases using different focal depths.
The laser power was controlled to prevent thermal damage to the silicon and ensure neat edges.
The straight cuts between the corners were performed in 1~hour with a Disco 321 dicing saw, with the wafers attached to dicing tape (Semiconductor Equipment Corp. 18074-6.50) mounted on grip rings.
This step was conducted first at WNF and, later, at EOSPACE when the WNF dicing saw was unavailable.
The last step of the fabrication was an acetone bath to remove the sacrificial photoresist layer, followed by a final visual inspection for quality control.
Figure~\ref{fig:PA_window}, right shows one window corner of the finished pitch adapter.

\subsection{Flex Cable Preparation}
\label{sec:qflex}

The low-radioactivity CCD-module flex cables were fabricated by Q-Flex in collaboration with Pacific Northwest National Laboratory (PNNL)~\cite{Arnquist:2023gtq,pnnl} in two separate productions of 20 and 8 flexes, respectively.
The flex is 1.5~m long to reach and connect to the feedthrough flange foreseen outside the DAMIC-M shield by means of a small-format Samtec 60-pin connector.

The flexes were shipped from the Q-Flex facility in Southern California to Seattle by truck to minimize cosmogenic activation of the two copper layers.
To decrease the impedance of the four output channels, there are connection pads 22~cm from the CCD side of the flex to install source followers, which consist of a 20-k$\Omega$ resistor and a junction field-effect transistor (JFET).
Both the JFETs and resistors were affixed to the flex using Epotek 301-2 epoxy and wire bonded for electrical connection. Following wire bonding, the components were encapsulated using epoxy LOCTITE EA0151, providing mechanical protection and environmental isolation. During the curing process, most of the flex surface (except for a $2\times5$\,cm$^2$ area around the source-follower installation pads) was covered with plastic sheets to minimize exposure to environmental radon. The Epotek 301-2 and LOCTITE EA0151 epoxies were cured by heating to 90~$^\circ$C for 4 and 12 hours, respectively. Once encapsulated, each flex cable was connected to a shorting connector to ground all channels for ESD safety and stored in the nitrogen-flushed cabinet until module packaging.

\subsection{CCD Module Packaging}
\label{sec:modulepack}

\begin{figure}[t]

\includegraphics[height=5.3cm]{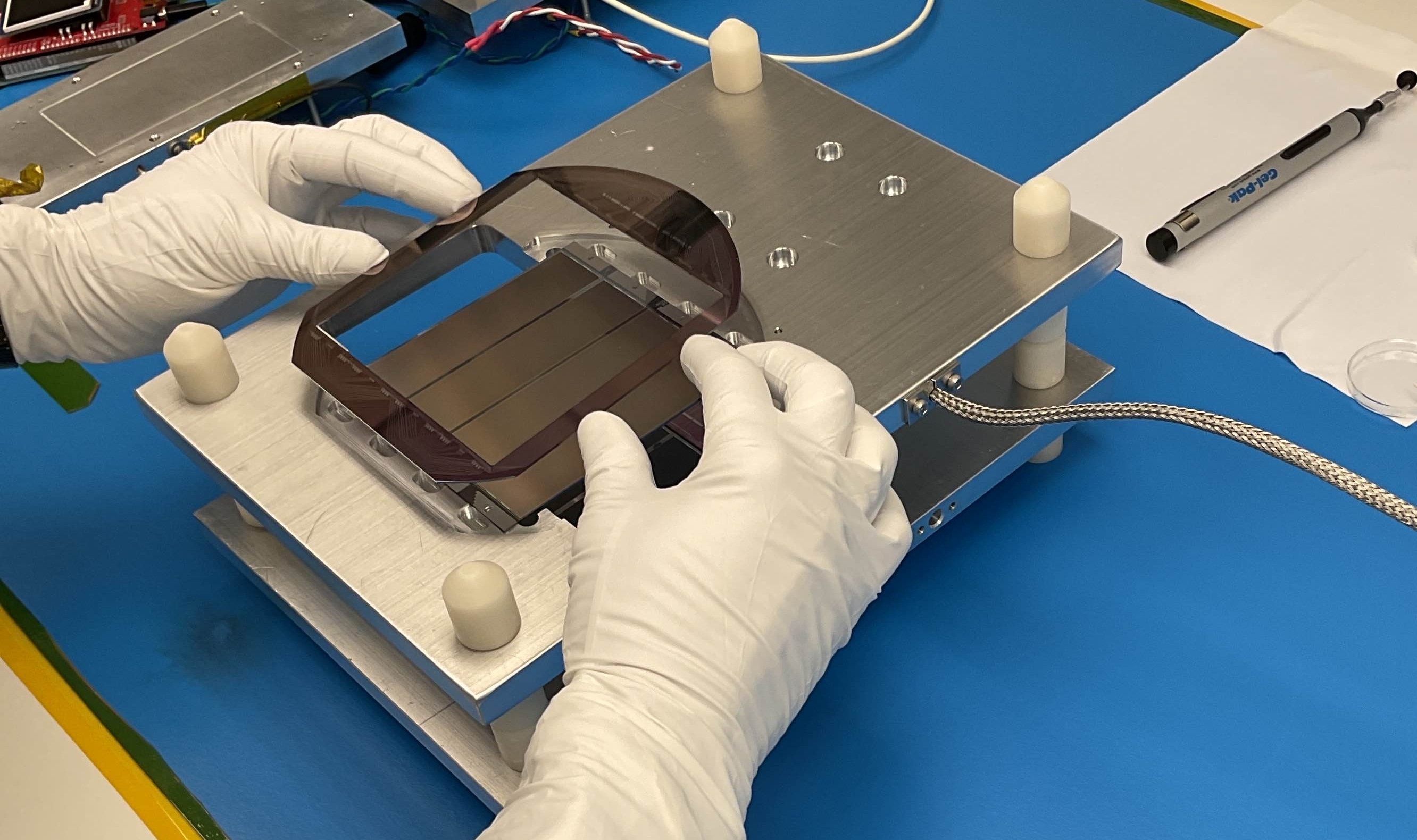}\hfill
\includegraphics[height=5.3cm]{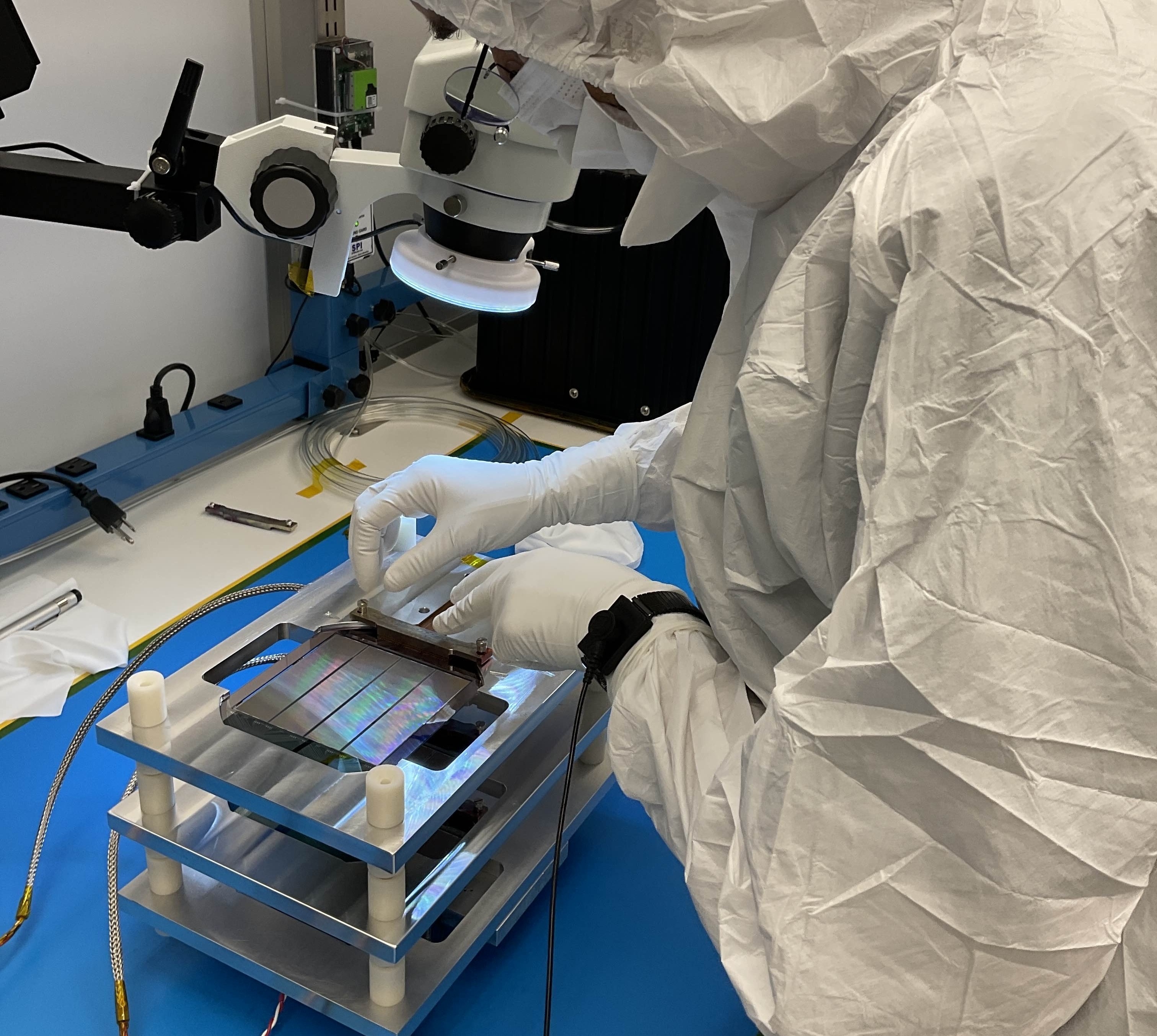}

\caption{Left: gluing of a pitch adapter onto the CCDs, which have been aligned in the jig with a vacuum tweezer. Right: installation of the copper weight to secure the flex cable during the second gluing step.}
\label{fig:gluing}  
\end{figure}

Each CCD module consists of four science-grade CCDs, one pitch adapter, and one low-radioactivity flex cable.
Assembly is performed using custom jigs and Epotek 301-2 epoxy in two stages.
In the first step, the dice are aligned in the jig, with the dice orientations determined by which amplifier was selected for connection following die testing (Sec.~\ref{sec:dietestselect}).
A small amount of epoxy is distributed onto the die's inactive areas, and the pitch adapter is laid on top (Fig.~\ref{fig:gluing}, left).
Bare-die resistors of 20~k$\Omega$ to isolate the CCD and pitch adapter grounds are also glued on the pitch adapter.
The jig is heated up to 90~$^\circ$C using a cartridge heater to cure the epoxy in 4 hours.
Three CCD modules are fabricated in each run. During curing, the jigs are stacked and held inside an ESD-safe, nitrogen-flushed enclosure (N$_2$ concentration ${>}$99.5\%).

In the second step, the flex cable is glued onto the pitch adapter.
The jig is flipped and the pitch adapter, already glued to the CCDs, is reset into the outline on the other side.
The flex cable is pressed onto the pitch adapter, with epoxy spread by a copper weight to ensure uniform pressure (Fig.~\ref{fig:gluing}, right).
Only the flex cable and pitch adapter are in contact with the jig during this step, which minimizes mechanical stress on the module.
Curing lasts 4 hours and takes place inside the nitrogen-flushed enclosure.

\begin{figure}[tbp]
\centering
\includegraphics[width=0.99\textwidth]{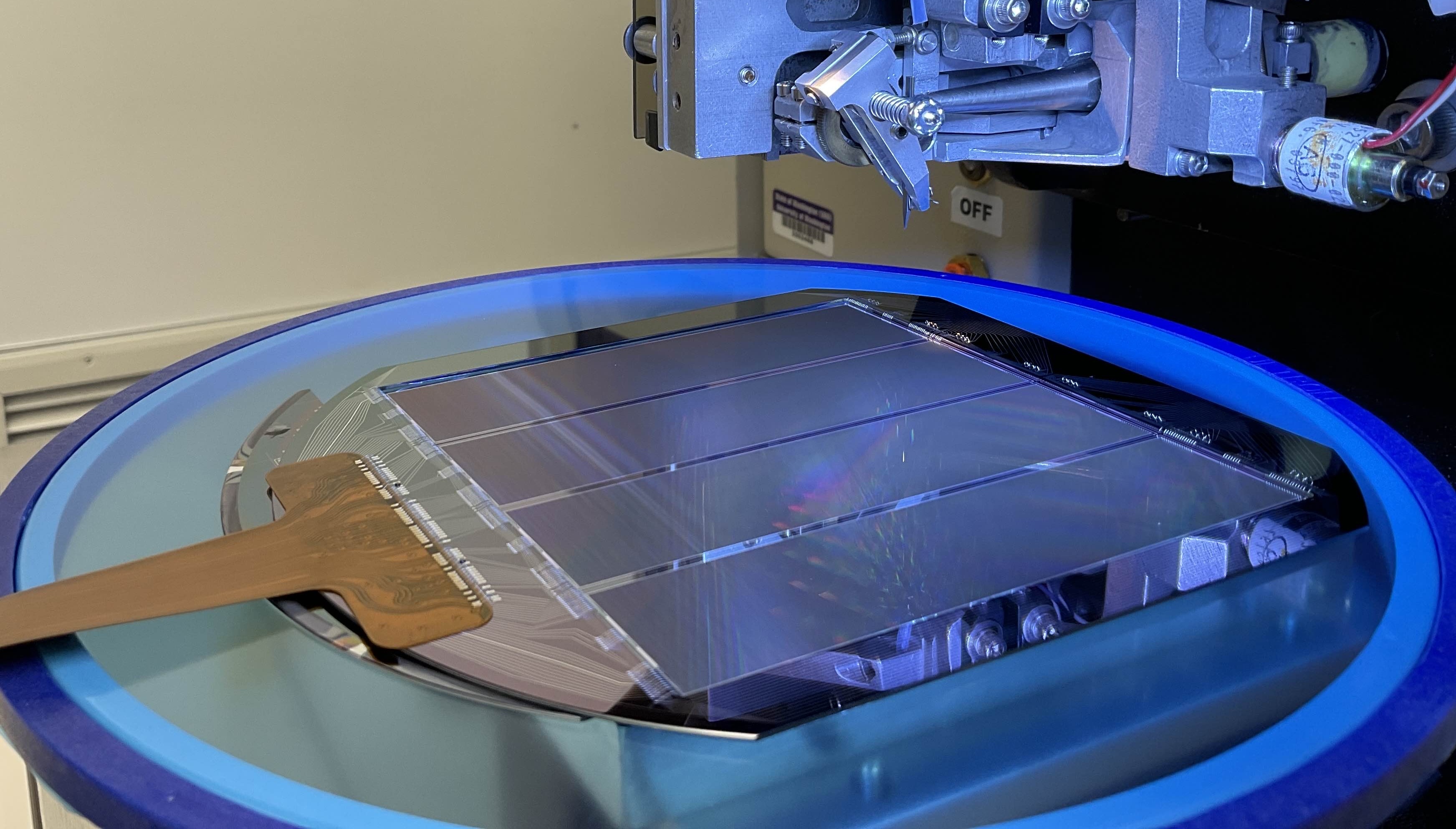}

\caption{The CCD module is affixed to dicing tape, which is mounted on the vacuum chuck of the wire bonder. The tape protects the backside of the CCDs.}
\label{fig:module_bonding}  
\end{figure}

\begin{figure}[htbp]
\centering
\includegraphics[width=0.99\textwidth]{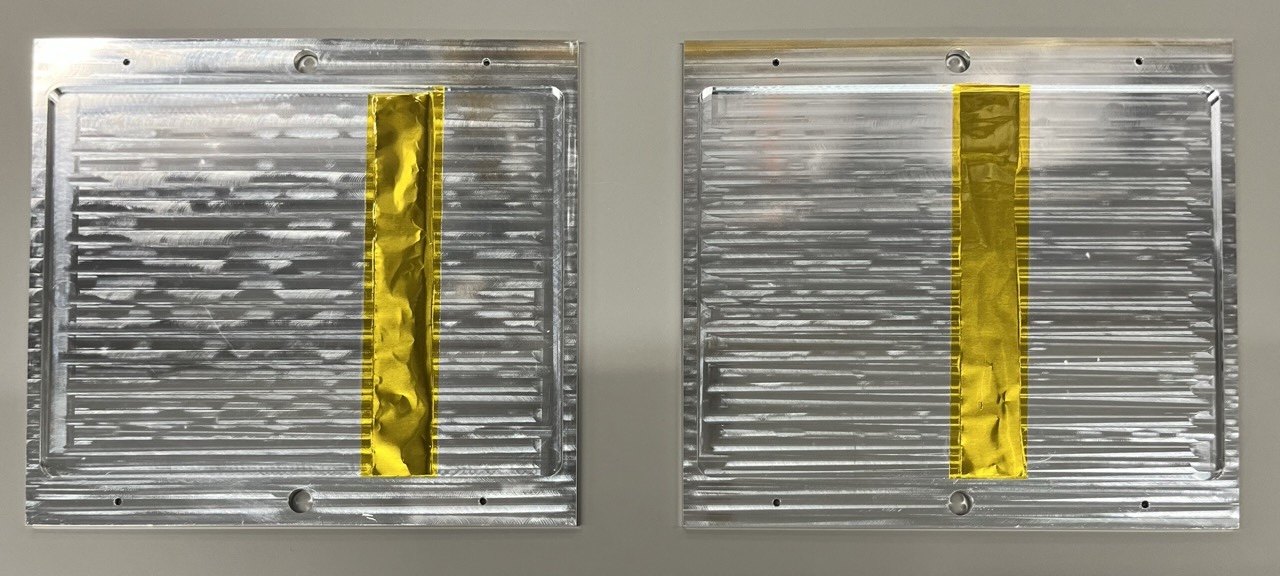}
\caption{The activated iron foils are affixed to the undersides of the top (left) and bottom (right) lid using Kapton tape (yellow). When installed, the CCDs lie horizontally, with the flex cable on the left.}
\label{fig:foilphoto}
\end{figure}

Lastly, aluminum wires are bonded from the flex cable to the pitch adapter and from the pitch adapter to the CCDs to establish signal connections, including the connections between the pitch adapter and CCD grounds through the bare-die resistor. 
The module is first affixed to dicing tape to protect the backside of the CCD (Fig.~\ref{fig:module_bonding}), and then placed on the vacuum chuck of the wire bonder.
The bonded module is removed from the tape using a vacuum die ejector, and installed in an aluminum storage box for testing.
Two iron (99.5\% Fe) foil strips, each measuring $13\times2$\,cm$^2$ in area and 25\,$\upmu$m thick, were affixed with Kapton tape to the undersides of the top and bottom lids of the box, respectively.
The strip orientation is perpendicular to the CCDs, with the front-side strip above the region of the CCDs that is furthest away from the flex, and the back-side strip offset by 2.2\,cm toward the flex (Fig.~\ref{fig:foilphoto}).
The foils were activated with thermal neutrons at the Maryland University Training Reactor to a starting $^{55}$Fe activity of 5.5\,Bq/cm$^2$.
X-ray emissions from $^{55}$Fe decay are used to characterize the response of the CCDs to ionization events, as discussed in Sec.~\ref{sec:moduletesting}.

After testing, the CCD module, secured in its storage box with Kapton tape (Fig.~\ref{fig:modser}, left), was carefully packed in double-layer antistatic bags flushed with nitrogen. The flex cable with shorting connector installed was coiled around two removable cable guides secured to the storage box to provide mechanical protection during shipment. The packaged CCD module was then transported back to NPL and stored, ready for shipment.

\section{Mitigation of Radioactive Contaminants}
\label{sec:mitigation}

This section describes the background mitigation measures we implemented during packaging and testing activities to suppress contamination from cosmogenic activation, plate-out of radon progeny, and natural uranium, thorium and potassium (U/Th/K).

\subsection{Cosmogenic Activation}
\label{sec:cosmoact}
Cosmogenic particles, in particular high-energy neutrons,\footnote{While tritium production in CCDs by muons (protons) is possible \cite{saldanha2020cosmogenic}, it is several (one) orders of magnitude lower than high-energy neutrons. The proton production rate becomes negligible in the presence of shielding.} interact with the CCDs to produce tritium by spallation of Si nuclei.
Tritium ($^3$H) emits $\beta$ electrons with energies up to $Q=18.6$~keV, and has a relatively long half-life of $12.3$~years, which makes it a persistent source of background in CCD-based dark matter experiments~\cite{DAMIC:2021crr}. 
The neutron energy threshold for tritium production is ${\sim}16$~MeV \cite{TENDL2021_Si28}. On the surface, the production rate in silicon from high-energy neutrons is estimated to be $R_{\text{trit}}\approx 112~\text{atoms per kg-day}$~\cite{saldanha2020cosmogenic}. The surface exposure budget for DAMIC-M CCDs is 90~days, after which tritium becomes the dominant background \cite{traina:tel-04063059,deDominicis:2022nls}. 

\begin{figure}[htbp]
\includegraphics[width=0.999\textwidth]{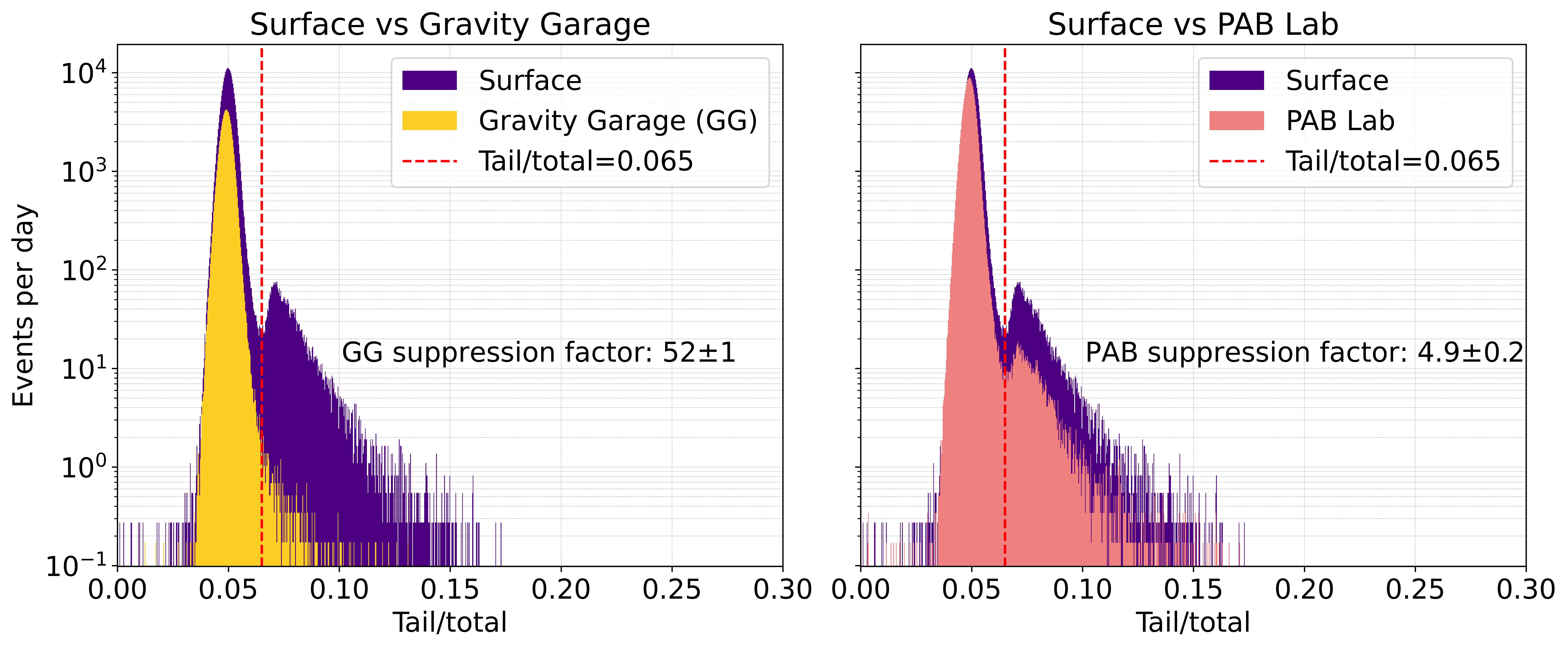}
\caption{Distribution of tail/total for scintillation signals above $16~\text{MeV}_\text{nr}$ in the Gravity Garage (left) and PAB Lab (right) compared to surface. The peak centered at tail/total $\mu = 0.05$ corresponds to cosmic muons, while the second peak to the right are the neutron signals. The neutron suppression factors were computed as the ratio of the rates of events with tail/total greater than 0.065 (dashed line), corresponding to $\mu+5\sigma$ of the surface muon peak. The uncertainty reported for the suppression factors is statistical only.}
\label{fig:Neutrons}
\end{figure}

To mitigate cosmogenic activation at UW, CCDs were stored in the Gravity Garage. This decision followed dedicated measurements of the neutron flux on-surface, in the Gravity Garage, and in the PAB lab. The measurements were conducted with a BC501A liquid scintillator cell ($5~\text{cm}\times5~\text{cm}$, cylindrical) coupled to a photomultiplier tube (PMT).
We employed Pulse Shape Discrimination (PSD) to distinguish between neutron-, muon-, and $\gamma$-induced signals.
Neutrons interact primarily via elastic scattering with protons; due to their large mass and low speed relative to electrons that generate the same number of scintillation photons, recoiling protons produce higher ionization density, which results in an increased excitation of molecular states with longer scintillation decay times~\cite{COMRIE201543}.
This leads to an elevated tail of neutron-induced scintillation pulses compared to $\gamma$ rays and muons, the latter two of which primarily interact with electrons.
For PSD, the tail-to-total ratio of the waveform is used, where the the tail integral is computed from 30 ns to 98 ns after the peak, and the total integral is computed from 14 ns before to 98 ns after the peak, capturing the tail and full waveform, respectively.
The electron-equivalent energy scale (in MeV$_{\rm ee}$) and PSD were calibrated with an $^{241}$Am$^9$Be source, which emits both high-energy neutrons and $\gamma$ rays.
We found a PMT high voltage of $-850$~V to provide excellent PSD, and large dynamic range up to 65~MeV$_{\rm ee}$ for pulses acquired with a CAEN DT-5730 14-bit digitizer.
The proton-recoil energy calibration $16~\text{MeV}_\text{nr} \approx 10~\text{MeV}_{\rm ee}$ was taken from Ref. \cite{nakao1995measurements}.
Figure~\ref{fig:Neutrons} shows a comparison of the rate of events above $16~\text{MeV}_\text{nr}$ for data collected on surface, in the Gravity Garage and in the PAB lab.
From the relative rates of neutron-like signals, we estimate the high-energy neutron flux above the tritium production threshold to be suppressed by a factor of 52~(5) in the Gravity Garage (PAB lab), compared to surface.
The muon rate is suppressed by a factor of 2.3~(1.4) in the Gravity Garage (PAB lab) with respect to surface.

\subsection{Radon Plate-Out}
\label{sec:radplate}
Part of the \textsuperscript{238}U decay chain, radon (\textsuperscript{222}Rn) is a noble gas with a half-life of 3.8 days that diffuses into air and seeps through clean room filters, where it decays sequentially into short-lived progeny \textsuperscript{218}Po, \textsuperscript{214}Pb, \textsuperscript{214}Bi, and \textsuperscript{210}Pb.
These isotopes plate out onto surfaces, including the CCD front and back surfaces and other package materials.
After several hours, all radon plate-out has decayed into  \textsuperscript{210}Pb, which, with a half-life of 22.3 years, leads to persistent surface contamination.
The $\beta$ decay of \textsuperscript{210}Pb (${Q}=63.5$~keV), followed by the $\beta$ decay of its daughter \textsuperscript{210}Bi (${Q}=1.16$~MeV) with a half-life of 5.0 days, were identified as a dominant background in DAMIC at SNOLAB~\cite{DAMIC:2021crr}.
Their decays were particularly problematic since partial charge collection in the few $\upmu$m-thick transition layer on the backside of the CCD resulted in events with degraded energy, increasing the background in the ROI.

\begin{figure}[t]
    \centering
    \includegraphics[width=0.8\textwidth]{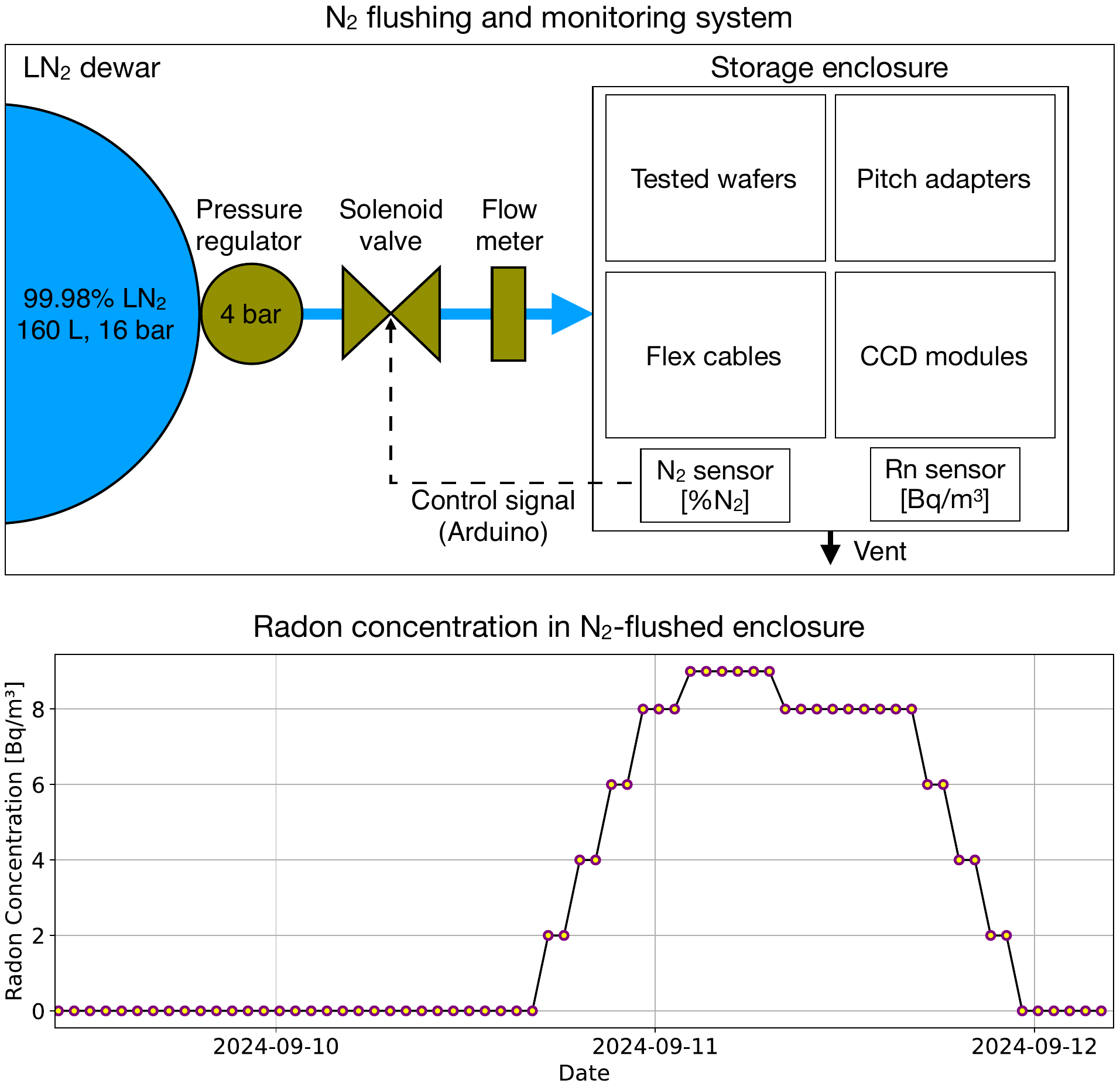}
    \caption{Top: diagram of the nitrogen flush system with salient components, labeled. Bottom: radon concentration versus time in a N$_2$-flushed enclosure. Radon concentration increases as LN$_2$ dewar empties. The concentration returns to zero as soon as dewar is refilled.}
    \label{fig:radospike}
\end{figure}

DAMIC-M CCDs were exposed to environmental air only when strictly necessary: to package and wire bond the CCDs, and during (un)mounting in the cryostats.
No exposure occurred during testing as devices were deployed in high-vacuum cryostats. During transfer between locations, all low-background items were stored in nitrogen-flushed antistatic LDPE double-bags.
To mitigate plate-out during periods of inactivity, CCDs and packaging materials were stored in enclosures flushed with liquid nitrogen boil-off.
Antistatic acrylic cabinets with stainless steel inserts and BPA-free polypropylene boxes with antistatic inserts were used in PAB and NPL, respectively.
Nitrogen was supplied by two 160-L dewars pressurized to 16~bar, connected to a flow control system designed to automatically switch from the empty to the full dewar.
A pressure regulator reduced the outlet pressure to 4~bar, followed by a solenoid valve controlled by an Arduino-based logic system.
Downstream of the valve, a flow meter ensured a stable nitrogen flow of 10~L/min into the $0.25\text{-m}^3$ storage enclosure.
Nitrogen concentration in the enclosures was actively monitored to open (close) the valve when it dropped below 98\% (exceeded 99.5\%).
We investigated potential radon contamination from the nitrogen supply by monitoring radon levels over the course of a dewar’s depletion. 
This measurement motivated the operational choice to switch dewars before they reached 25\% fill level.
Figure \ref{fig:radospike} shows a diagram of the nitrogen-flush system (top), and the radon spike measured upon dewar emptying (bottom).

\begin{figure}[t!]
    \centering
    \includegraphics[width=0.85\textwidth]{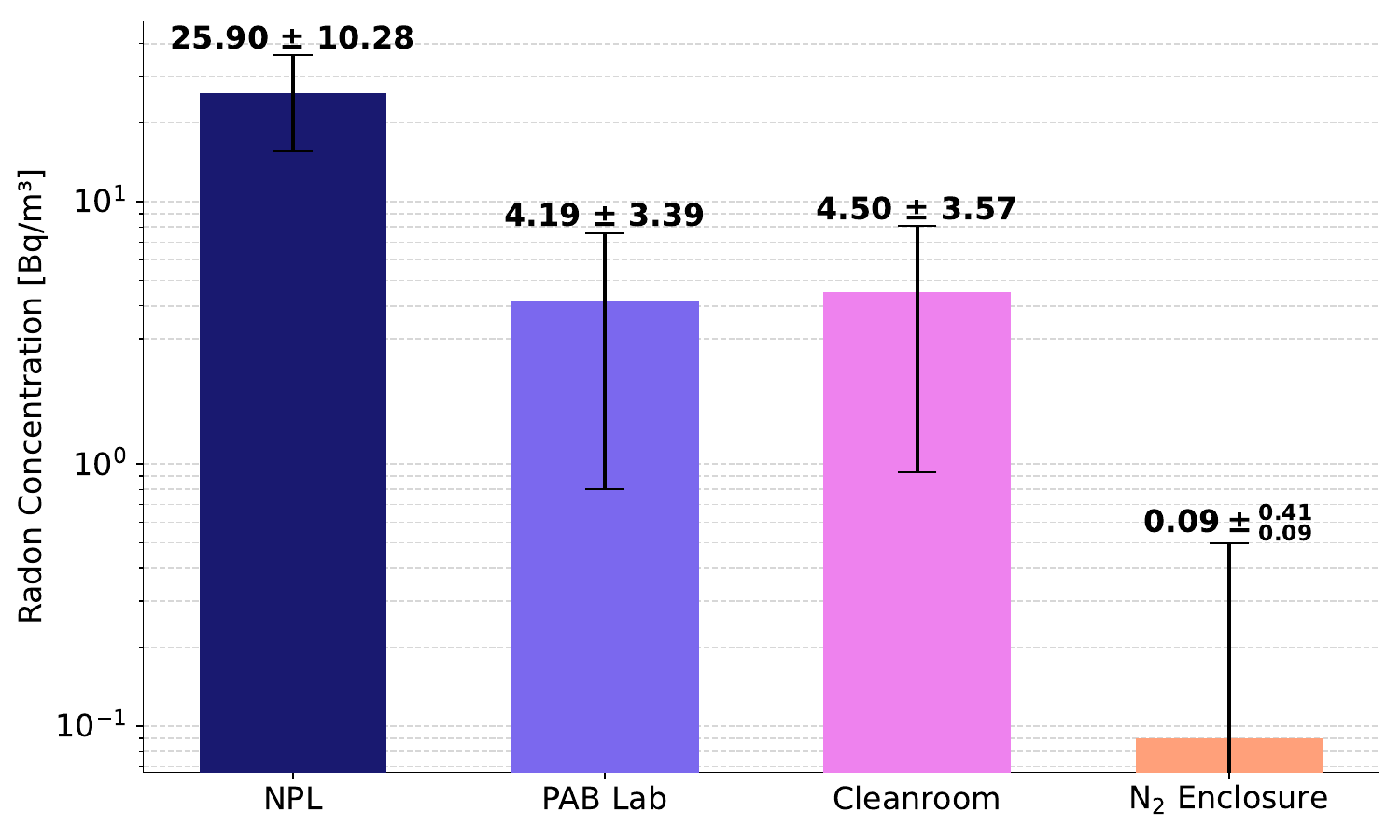}
    \caption{Radon levels in Bq/m\textsuperscript{3} at the locations of packaging and testing activities. The PAB Lab measurement was performed outside the cleanroom. Uncertainties include the daily variation in the environment.}
    \label{fig:radon_levels_logscale}
\end{figure}

We measured the radon concentration in all work locations using Airthings Wave radon sensors, acquiring at least 1000 samples per location. The highest concentration was measured in NPL, caused by poor ventilation of the building. No significant difference was observed outside and inside of the PAB cleanroom.
The results of these measurements are summarized in Fig. \ref{fig:radon_levels_logscale}.

\subsection{Uranium, Thorium and Potassium}
\label{sec:uthk}
Naturally occurring radionuclides \textsuperscript{238}U, \textsuperscript{232}Th, and \textsuperscript{40}K are commonly found as trace impurities in metals, ceramics, polymers, concrete, and most other common materials.
Their decay chains release $\alpha$, $\beta$, and $\gamma$ radiation that can result in low-energy depositions in the CCDs.
Although the main strategy against U/Th/K was the selection of CCD module components of low radioactivity, additional contamination can be introduced as dust and residues on surfaces.
Further, trace levels of $^{40}$K are found in organic residues, such as skin flakes and fingerprints. Therefore, during packaging and testing, mitigation of U/Th/K contamination focused on preventing the accumulation and transfer of dust, debris, and organic matter.

The clean room environment was the primary defense against airborne particulates, supported by a strict gowning protocol.
Particle count measurements were performed in the PAB clean room and gowning room using a Met One HHPC 2+ particle counter to validate compliance with ISO 5 standards. Measurements targeted 0.5~$\upmu$m and 5~$\upmu$m particles, with fourteen 50-L samples taken under both static and working conditions. Across all sampled locations\textemdash including on work surfaces and the clean room floor\textemdash particle counts remained well below ISO 5 limits (${<}176$ and ${<}1.45$ particles per 50 L for 0.5~$\upmu$m and 5~$\upmu$m, respectively). Transient increases were observed only during movement near the sensor, particularly in the gowning room. These results confirm that the clean room, equipped with HEPA filtration and operating at ${\sim}$300 air changes per hour, fulfills the ISO 5 standard. 

A daily cleaning protocol was implemented to maintain surface cleanliness and prevent particulate accumulation. All hand tools and work surfaces were regularly cleaned with ethanol and ultra-pure water and, when appropriate, with a 1\% Simple Green ultra-pure water solution applied either by wiping or in an ultrasonic bath to remove skin oils and grease. Extra care was taken for CCD handling tools, which were decontaminated before every use. When necessary, an ionizing air gun supplied with LN$_2$ boil-off was used to clear debris off the CCD package. ESD-safe procedures were enforced not only for device protection but also as an additional cleanliness measure, since charged surfaces attract and hold dust.

The rapid movement of the wire-bonder head over the CCDs during wire bonding was found to be the dominant source of particulates.
To estimate contamination, silicon substrates were placed on the wire bonding station and a set number of wire bonds were made.
Process blanks away from the wire bonder served as controls.
The particulates from both samples were transferred to carbon tape, suitable for imaging with a scanning electron microscope (SEM).
The number of $\upmu$m-size metallic particulates were counted in images covering a given area of the carbon tape.
From the difference between the wire bonded samples and the blanks, we measure a particulate generation rate of $<$1.6/cm$^2$ per wire bond.
The estimated added radioactivity from this source is reported in Sec.~\ref{sec:results}.

\section{CCD Die and Module Testing}
\label{sec:ccdtesting}

This section details the CCD testing protocols adopted to characterize and select individual CCD dice and, subsequently, to validate the performance of the DAMIC-M CCD modules for installation in the DAMIC-M experiment.

\subsection{CCD Testing Setups}\label{sec:testingsetup}
\begin{figure}
    \centering
    \includegraphics[width=1.0\linewidth]{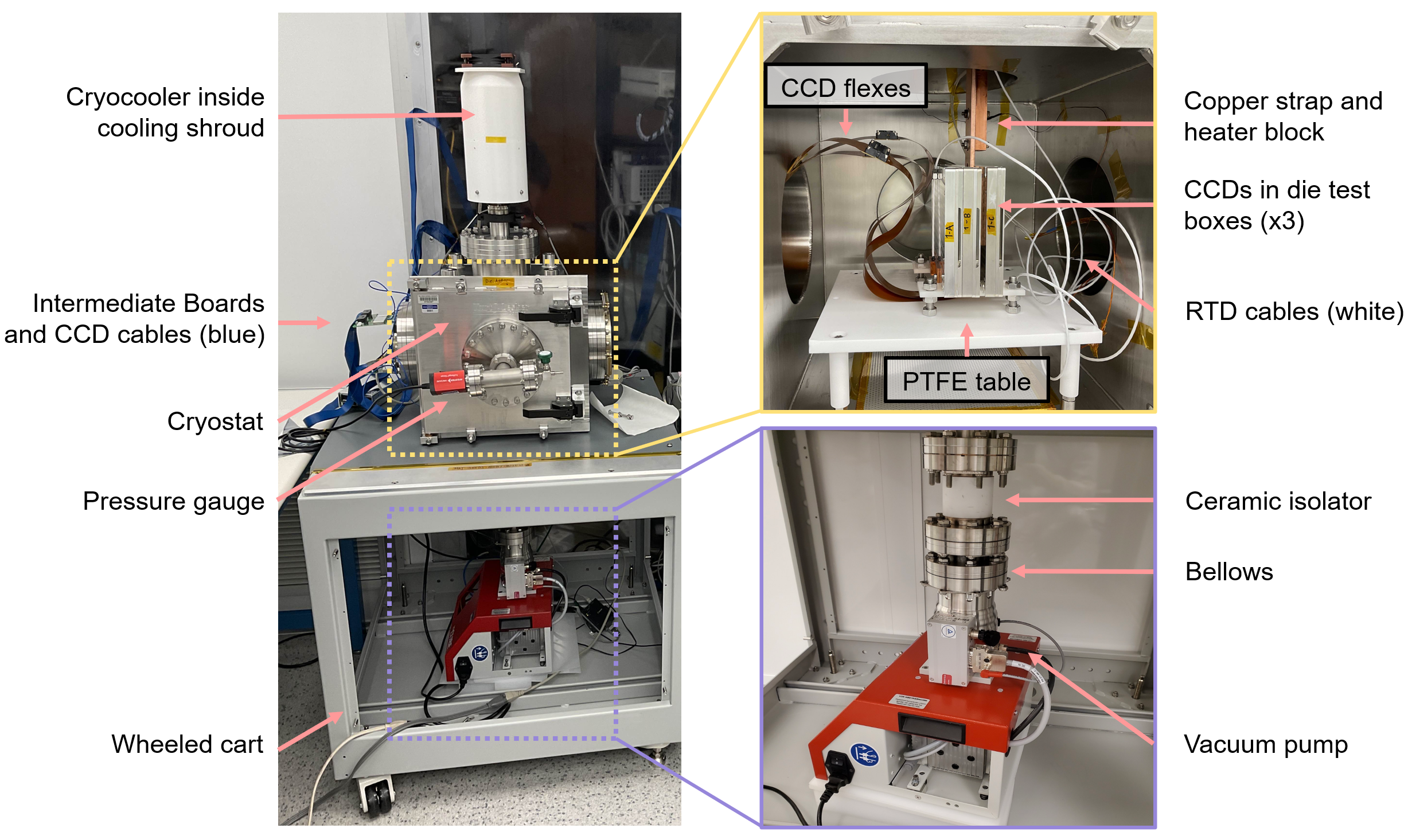}
    \caption{Setup of the CCD test chambers inside the cleanroom as described in Sec.~\ref{sec:testingsetup}, with salient components labeled.}
    \label{fig:JHUChamberExterior}
\end{figure}

The CCD test setups are shown in Fig.~\ref{fig:JHUChamberExterior}. Each setup is designed to accommodate up to four CCD dice or modules but only three were tested at a time due to the availability of electronics. The setups consist of the following components:
\begin{itemize}
    \item Cryostat and vacuum chain: each test system is based around a stainless-steel vacuum cryostat mounted to the top of a wheeled cart for easy transport. Below each cryostat is a turbomolecular pump, which is electrically isolated from the test chamber by a ceramic vacuum fitting. The residual pressure inside each cryostat is monitored by a dual Pirani/cold cathode\footnote{Since the cold cathode operates via high-voltage glow discharge, which can generate detectable photons and electromagnetic interference, we power off the gauge during CCD testing.} gauge, and reaches ${\lesssim}10^{-4}\text{\,mbar}$ at room temperature. The pumps and pressure gauges have external Ethernet connections for monitoring and control.
    
    \item Thermal chain: each cryostat features a Cryotel GT cryocooler with active vibration cancellation. The air-side cooling fins are enclosed by 3D-printed shrouds with a fan to prevent overheating. Inside the cryostats, the aluminum CCD boxes are thermally coupled to the cryocooler cold end via a copper strap. We use a layer of Kapton tape to electrically isolate the strap from the cryocooler. Resistance temperature detectors (RTDs) are positioned on the cryocooler cold end, on the thermal strap, and on the outer CCD boxes (two RTDs). The RTD and a resistive heater mounted to the copper strap are connected on a feedback loop with a Lakeshore temperature controller. A PTFE table provides mechanical support to the CCD boxes as well as thermal and electrical isolation from the chamber walls.
    
    \item Detectors and electronics chain: each CCD die or module is connected to a vacuum feedthrough via the Kapton flex, which carries the clocks, biases and amplifier output signals. Outside the chamber, three front-end boards (one per CCD die or module) plugged into the vacuum feedthrough perform the initial signal amplification. High-speed twinax cables (Samtec ERCD-030-120.0-TBR-TTL-1-C) connect the front-end boards to the main electronics racks outside the clean room.
\end{itemize}
As many of the electronics components as possible are placed in racks outside to prevent introducing dust into the cleanroom. These can be broadly grouped into two categories:
\begin{itemize}
    \item Slow control: this system monitors and controls the cryostat service equipment \cite{DAMIC-M:2024ooa}. A dedicated Linux server running custom software interfaces with the vacuum pump, pressure gauge, temperature and cryocooler controllers, enabling full remote operations. The system status is logged in a MySQL database, and alarms are raised if parameters stray outside predefined bounds.
    \item Data acquisition: this system is responsible for CCD control and data taking. There is a VME crate for each test chamber with three Acquisition and Control Modules (ACM), each connected to a front-end board on the cryostat. The ACMs generate bias and clock voltages for the CCDs, and process their output signals through four channels, one per amplifier (CCD) in die (module) testing. A dedicated clock synchronization board enables synchronous operations of the ACMs (CCD readout). A suite of Acopian linear power supplies provides low-noise DC voltage to the VME crates. An additional Linux server sets the CCD operating parameters, executes readout, and serves as local data storage. 
\end{itemize}

\subsection{CCD-Die Test and CCD Selection}
\label{sec:dietestselect}

The cryostat is first pumped down to a pressure ${\lesssim}5\times 10^{-4}\,\mathrm{mbar}$ at room temperature to prevent condensation of volatile compounds when cold. The cooldown process then begins, maintaining a cooling rate of ${<}0.5\mathrm{\,K/min}$ to avoid thermal stress on the silicon. Approximately five hours are required to reach the set temperature of 185~K. To ensure the CCDs have sufficient time to thermally equilibrate, we require that the temperature change of the CCD boxes be ${<}0.01\mathrm{\,K}$ over a five-minute interval.
We then proceed with the single-die tests, to identify science-grade amplifiers through which the entire CCD pixel array can be read out with low noise and high fidelity.

\subsubsection{CCD Image Characteristics}
\begin{figure}[t!]
    \centering
    \includegraphics[width=0.999\linewidth]{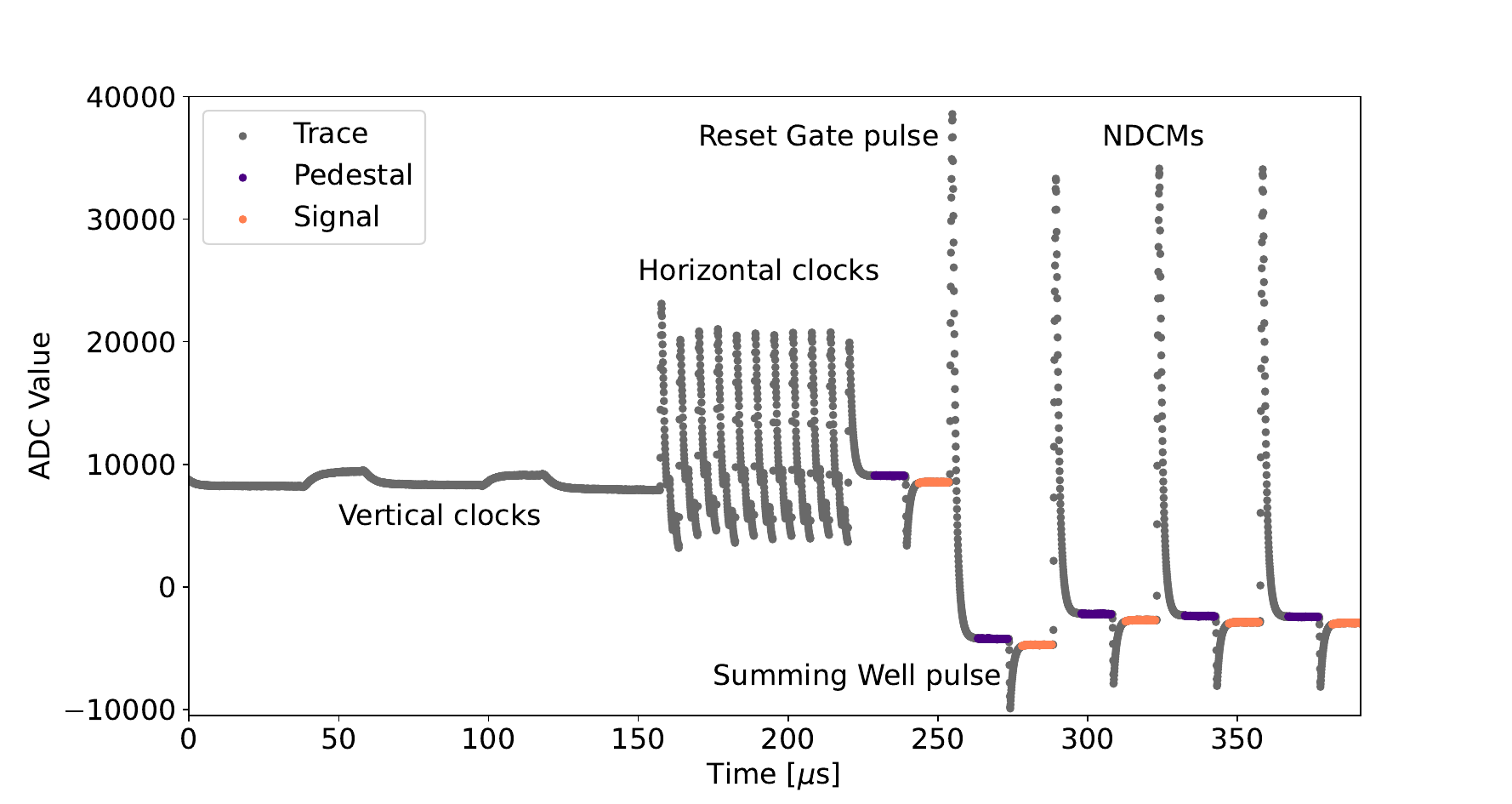}
    \caption{Output signal trace acquired with a science-grade CCD. The different features in the trace are labeled, including the response to changes in clock voltages.
    The Reset Gate and Summing Well are two of the output-stage clocks.
    CDS is performed by taking the difference of the integrated CCD output between the pedestal (purple) and signal (orange) regions, before and after the charge is transferred into the floating gate.}
    \label{fig:trace}
\end{figure}

A CCD image is acquired in two sequential stages: exposure and readout. During exposure, the CCD clocks remain at a constant value for a specified duration (exposure time), with the pixels collecting charge. Incident particles depositing energy in the silicon generate charge carriers, which are drifted to the pixels by the electric field generated by the substrate voltage, $V_\text{sub}$. During readout, the accumulated charges are transferred for measurement by clocking the 3-phase pixel gates. One row of pixels is first shifted into the serial register (SR) by the parallel clocks. The serial clocks then move charge toward the output stage, one pixel at a time.
The measurement of the charge is performed by correlated double sampling (CDS) of the floating gate, which is capacitively coupled to the readout MOSFET amplifier~\cite{Wen1974DesignAO}.
Output-stage clocks control the movement of charge between several gates in the output stage, which enable the summing of charge from multiple pixels (pixel binning) or to repetitively perform CDS of the charge in the floating gate over a number of NDCMs.
Additional clocks and biases clear the charge and reset the value of the floating gate~\cite{Tiffenberg:2017aac}.
The CCD signal trace shown in Fig. \ref{fig:trace} shows the output of the amplifier during the readout sequence.

\begin{figure}
    \centering
    \includegraphics[width=1\linewidth]{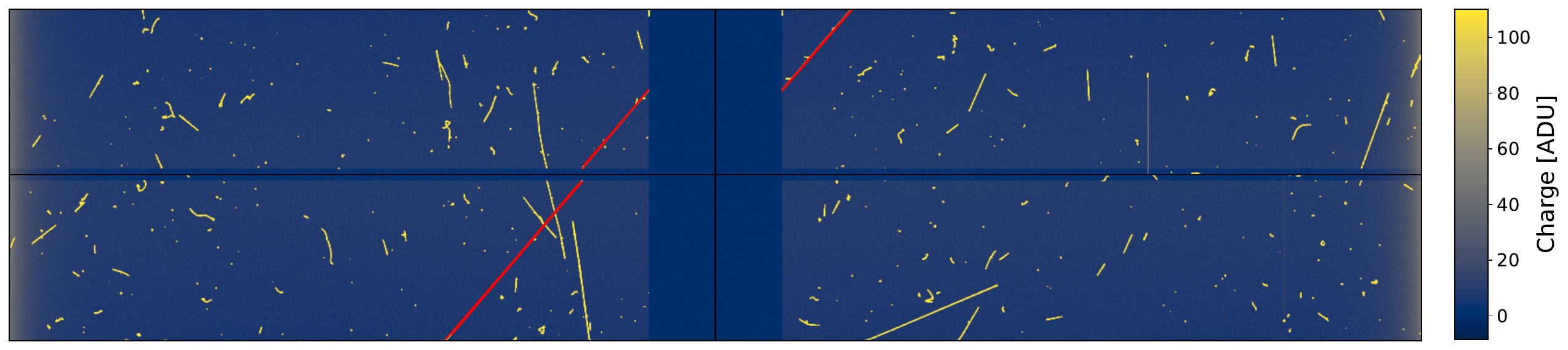}
    \caption[CCD Image Characteristics]{An image acquired using all four amplifiers of a CCD (from the upper left, clockwise: L2, U2, U1, L1), each reading a separate quadrant and collectively covering the full CCD. Key regions\textemdash active area, $x$- and $y$-overscans, and prescans\textemdash are visible in all quadrants. Transient noise from the parallel clock routine appears in the first few columns of all four images. A muon track crossing three quadrants is highlighted in red.}
    \label{fig:die_image}
\end{figure}

Once read out, the pixel values are reconstructed into a FITS image \cite{wells1979fits}. One CCD contains two serial registers (SR1 and SR2), capable of simultaneous readout through two amplifiers located on opposite sides (L and U). Each CCD thus has four amplifiers (L1, U1 for SR1, and L2, U2 for SR2), located at its corners. This setup enables synchronous readout of four separate quadrants, each covering $3072 \times 768$ pixels, collectively spanning the full array of $6144 \times 1536$ pixels.
Figure~\ref{fig:die_image} illustrates the anatomy of a CCD image. 
Each SR is 6160 pixels long, the length of one row of the pixel array plus eight additional ``prescan'' pixels between the pixel array and the amplifiers, which have no corresponding columns.
Image dimensions can be extended by continuing to read pixels beyond the physical array to create the ``overscan'' region.
The $x$-overscan consists of pixels beyond the physical dimension of the serial register. These pixels generally contain negligible charge accumulated during the $\mathcal O(100~ \text{ms})$ readout of the SR, and appear devoid of particle tracks (Fig.~\ref{fig:die_image}). Therefore, the $x$-overscan provides a measurement of the row baseline that can be later subtracted.
The $y$-overscan corresponds to rows clocked beyond the pixel array. These rows can accumulate significant exposure (up to the total readout time) as they are clocked across the CCD pixel array, and may capture particle tracks.

\subsubsection{Sources of Noise in CCD Data}

The performance of the CCDs is limited by several sources of noise intrinsic to the devices, from the electronics, and from the environment. These include electronic fluctuations, thermally generated charge, and charge introduced during transfer and readout:

\begin{itemize}
    \item {\bf Readout noise} originates from instabilities in the output signal during charge measurement.
    Reset and thermal noise \cite{johnsonnoise,PhysRev.32.110} are reduced by CDS and increasing the signal integration time.
    Flicker noise (${\sim}1/f$), e.g., by carrier trapping at interfaces, is mitigated by employing skipper readout~\cite{skipperjanesick}. Pickup noise is addressed through improved electrical grounding and shielding from sources of electromagnetic radiation.

\item {\bf Dark current (DC)} from thermal charge generation in the bulk and surface of the silicon lattice increases exponentially with temperature, by a factor of 10 for a 7\,K increase around 140\,K~\cite{Janesick:2001,2002SPIE.4669..193W}. Defects in the silicon lattice or at interfaces can lead to localized regions of increased DC, observed as persistent hot pixels or columns.

\item {\bf Clock-Induced Charge (CIC)} arises from the rapidly changing electric field caused by voltage swings of the gates during charge transfer, and depends on clock amplitude and rise time~\cite{Oscura:2023qik}. Serial clocks are the primary contributors to CIC, since they have higher frequency than parallel clocks.

\item {\bf Charge Transfer Inefficiency (CTI)} results from the incomplete transfer of charge from one pixel to the next, creating trails of charge in the images. Serial CTI is mitigated with higher clock amplitudes and time widths, while parallel CTI is more affected by long-lived traps in the pixel array and is harder to reduce by tuning the clocks. CTI worsens at lower temperatures due to reduced carrier mobility and increased emission time constants for charge traps, e.g., divacancy and C/O interstitial~\cite{shockleyread1952, 2014JInst...9C2004H}.

\item {\bf Glowing} is caused by infrared photons emitted by the readout MOSFET, and appears as a localized excess of charge in the corners of the CCD near the amplifier. It can be reduced by decreasing the drain voltage of the MOSFET, which is positively correlated to power dissipation by the amplifier.
Large defects in the pixel array or SR which draw significant current can also locally heat up the silicon and lead to glowing.

\end{itemize}

\subsubsection{CCD Die Testing Protocols}
\label{sec:dietest}

Die testing completed with 188 CCDs tested in 35 days.
Single-die tests were conducted at 185\,K to enable efficient identification of defects, and to reduce the duration of the testing routine since thermal cycling was the main bottleneck. The goal of the campaign was to characterize the performance of the pixel array, serial registers, and all four amplifiers of every CCD, and, subsequently, to identify CCDs with science-grade amplifiers for module fabrication.

Below, we present the eight tests conducted to evaluate CCD performance. Most images were acquired with synchronous readout of the four amplifiers (each reading one quadrant of the pixel array). Unless otherwise noted, acquisition was performed with 1 NDCM per pixel.
We started with the CCD biases, clock voltages, timings, and readout sequence optimized for the prototype CCD modules operated in the DAMIC-M LBC at LSM~\cite{DAMIC-M:2024ooa, DAMIC-M:2025luv}, with $V_\text{sub}$ increased to 60~V.
Fine-tuning of the clock parameters was needed, since the pitch adapter introduces additional resistance and capacitance ($RC$) to the clock lines in the module. The lower $RC$ of the clock lines in the die test allowed us to significantly increase the frequency of the parallel and the serial clocks for a shorter readout time and faster testing.

\subsubsection*{Trace}
As the first step of the testing routine, we examined the output trace from all amplifiers. We verified that the trace displays all expected features labeled in Fig.~\ref{fig:trace}, confirming proper CCD operation, and that the trace is fully captured within the ADC range of the digitizer\footnote{Saturation, which leads to complete signal loss, occurs when the analog-to-digital converter (ADC) exceeds its maximum count, determined by its bit depth. For the 17-bit system in the ACM, this limit is $2^{17} - 1 = 131071$ ADU.}. We also checked for abnormally low trace signals, which hint at amplifier malfunction.

\subsubsection*{Readout Noise}
We measure the readout noise for a single charge measurement by taking the standard deviation of the $x$-overscan pixels in images with 1 NDCM.
Science-grade amplifiers have a readout noise ${\lesssim}6$~ADU (Analog-to-Digital Units), corresponding to ${\lesssim}6~e^-$.

\subsubsection*{Composite Image}

To identify defects and glowing, we acquired 10 images each with $300$-s exposure, ensuring signal build-up while limiting the density of particle tracks. The readout time ($150$~s) was kept below the exposure time to preserve accurate spatial mapping. Baseline subtraction was performed row-by-row using the median of the $x$-overscan, and a composite image was constructed by taking the median value of each pixel across the image stack. This procedure filters out particle tracks and suppresses pixel noise, thus enhancing the visibility of defects and glowing which persist over all images.

Column defects were identified by analyzing the median value of each column and comparing it to the local running median and running median absolute deviation (MAD) computed over a 40-column window. Columns whose median deviated by more (less) than 6 times the local MAD from the local running median were flagged as hot (cold). For pixel defects, the row-wise median was first subtracted from each pixel to account for background charge. The resulting values were then used to compute a global median and a global MAD. Hot (cold) pixels were identified as those whose values deviated by more (less) than 10 times the MAD from the global median. 

\begin{figure}[t]
    \centering
    \includegraphics[width=1\linewidth]{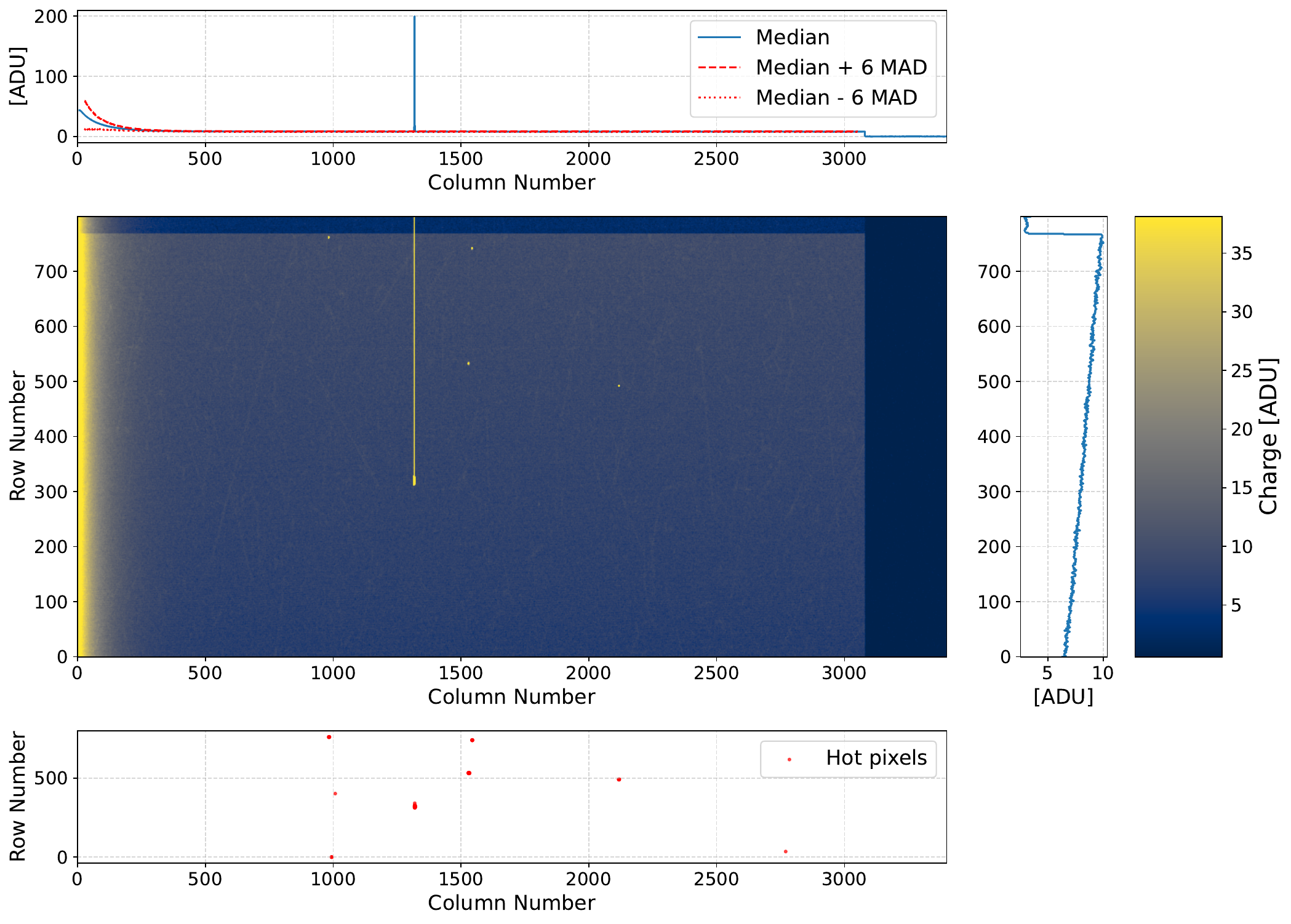}
    \caption[Composite image analysis]{A composite image of ten images from a CCD quadrant read out by the U2 amplifier. Tracks are filtered out, revealing hot columns (top panel) and hot pixels (bottom panel). The $x$ and $y$ overscan steps are visible.
    }
    \label{fig:composite_image}
\end{figure}

The result of this analysis is a defect map detailing the location, type, and count of flagged columns and pixels across the CCD.
Figure~\ref{fig:composite_image} shows a composite image of a quadrant of a CCD, highlighting the identified column and pixel defects.
Any CCD (all four amplifiers) with a large number ($\gtrsim20$) defects or which had a defect that was judged too prominent based on its spatial size and intensity (e.g., if charge spilled into neighboring pixels) was disqualified.

\subsubsection*{CTI Analysis}
The background pixel-array values in the composite image are significantly higher than in the overscan because of charge accumulated from DC, causing the steps noted in Fig.~\ref{fig:composite_image}.
Although large CTI effects create visible trails, small losses may be too subtle to detect visually.
However, since CTI causes charge to spill into subsequent pixels, it introduces charge from the pixel array into the first pixel of the overscan, which results in a value above the baseline.
We quantify CTI using the following figure of merit\footnote{Up to a factor of the number of serial or parallel transfers, this definition is essentially equivalent to the extended pixel edge response (EPER) method of determining CTI \cite{Janesick:2001}.} computed from the composite image:
\begin{equation}
    \mathrm{CTI} = \frac{S_\mathrm{OS} - S_\mathrm{baseline}}{S_\mathrm{active} - S_\mathrm{baseline}} \times 100\%.
\end{equation}
Here, $S_\mathrm{baseline}$ is the median pixel value of the last 100 overscan columns (rows) for serial (parallel) CTI; $S_\mathrm{OS}$ is the median pixel value in the first column (row) of the $x$ ($y$) overscan; and $S_\mathrm{active}$ is the median pixel value in the last 100 columns (rows) of the pixel array. Amplifiers with serial CTI$x$ or parallel CTI$y$ $\gtrsim 5\%$ were disqualified.

\subsubsection*{Stress Test}
\begin{figure}
    \centering
    \includegraphics[width=1\linewidth]{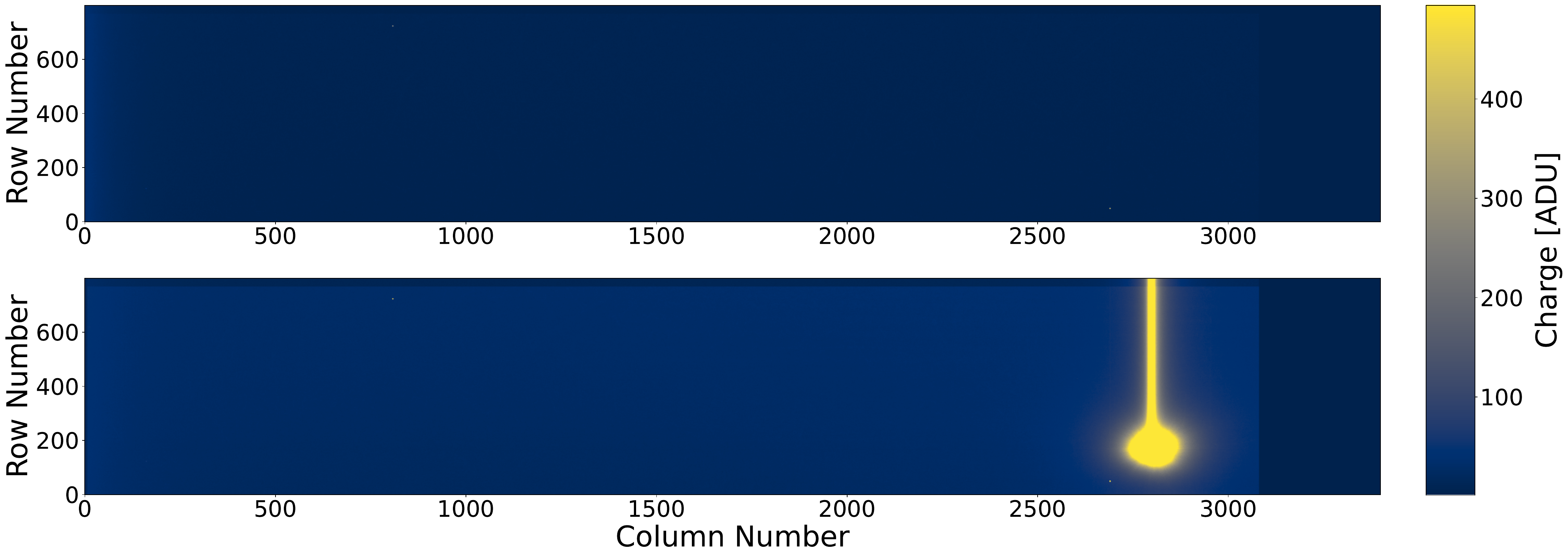}
    \caption[CCD Stress test]{Comparison of composite images acquired under nominal and stress conditions with higher $V_\text{sub}$ and $\Delta V$. A significant defect is clearly uncovered under stress conditions. The increased background charge level in the pixel array due to leakage current and CIC is also visible. This CCD (all four amplifiers) was not classified as science grade.
 
    }
    \label{fig:stress_image}
\end{figure}

Through extensive testing of pre-production CCDs, we observed that certain CCD defects, mostly from the CCD backside, are highly sensitive to V$_\text{sub}$.
To reveal latent CCD defects, we performed stress tests by operating the devices beyond nominal conditions. The voltage $V_\text{sub}$  was raised from 60~V to 80~V.
We also increased the vertical clock amplitude from $\Delta V = 3$~V to 6~V, which stimulates the release of trapped charge at silicon-oxide interfaces, thereby exposing latent front-side defects~\cite{Janesick:2001}.
Dark current was also monitored under stress conditions, as significant increases point to underlying defects.
The analysis followed the methodology developed for the composite image tests, acquiring 10 images under stress conditions and assessing defect counts, background charge levels, and CTI performance.
Figure~\ref{fig:stress_image} demonstrates how the stress test can uncover new defects and increase background DC.

\subsubsection*{Serial Register Image}
\begin{figure}
    \centering
    \includegraphics[width=1\linewidth]{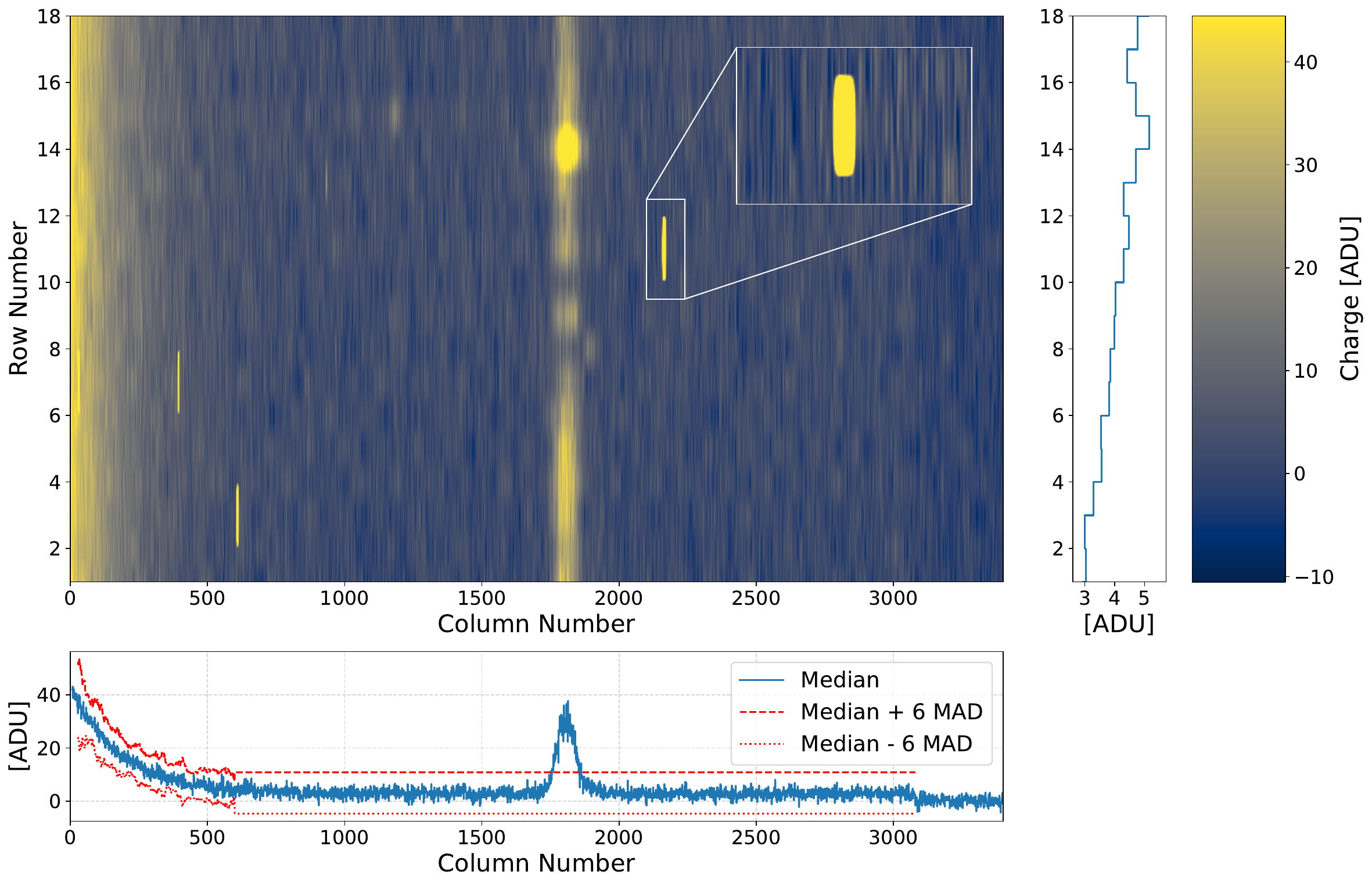}
    \caption[Serial register image analysis]{A serial register image with a prominent defect. Also shown is a zoomed-in view of one of the particle tracks in the SR (note aspect ratio). Amplifiers on this SR were not classified as science grade. 
    }
    \label{fig:sr_defect}
\end{figure}

In a serial register image, charge from the pixel array is not transferred into the SR.
Instead, the SR is exposed for 2 seconds, to allow for charge to accumulate, before being clocked and read out.
The procedure is repeated 20 times, resulting in an image where each row corresponds to a successive readout of the SR.
Serial register images do not contain two-dimensional particle tracks; instead, particle interactions that deposit charge in the SR appear as horizontal streaks (see Fig.~\ref{fig:sr_defect}).
A SR defect appears as a vertical column of excess charge in the image.
Defects were identified in both SRs of every CCD using the same MAD-based algorithm to detect column defects in the composite images.
Only amplifiers on defect-free SRs were considered for modules.

\subsubsection*{Skipper Performance}
\begin{figure}[t!]
\centering
\includegraphics[width=1\textwidth]{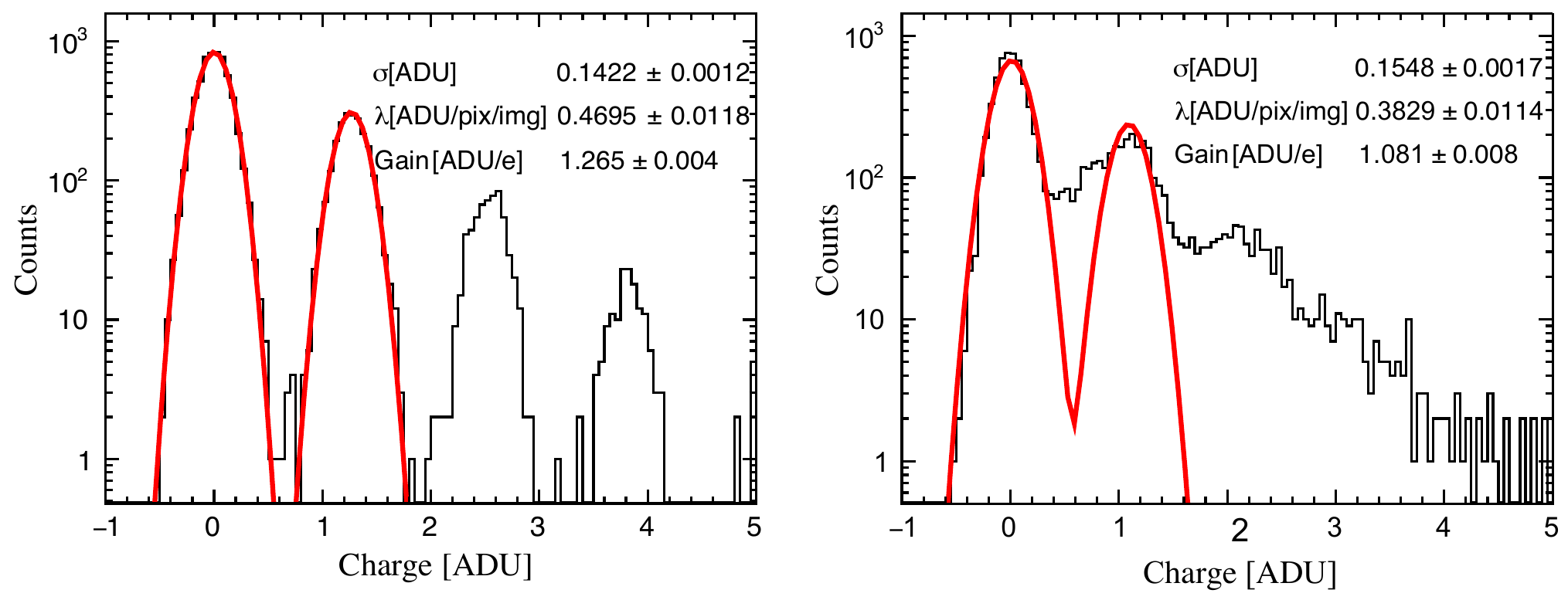}
\caption{A Poisson-convolved-Gaussian fit to the pixel-value distribution of 630 $x$-overscan pixels. Left: A science-grade CCD showing clear single-electron resolution with standard values of gain, resolution ($\sigma$), and average charge per pixel ($\lambda$). Right: A CCD with a defective amplifier, with low gain and charge between single-electron peaks.
}
\label{fig:skipper_fit}
\end{figure}

DAMIC-M CCDs are operated in skipper mode to achieve single-electron resolution.
We verified nominal skipper performance by reading out 630 $x$-overscan binned pixels (10 pixels per bin) with 1000 NDCMs.
When operating the test system at 185\,K, this provides enough charge to clearly identify the first few electron peaks.
The distribution of pixel values averaged over the NDCMs were fit with a Poisson distribution convolved with a Gaussian.
The distance between consecutive peaks provides an estimate of the amplifier gain (in ADU/$e^-$), while the width (measured as the standard deviation) of the peaks provides an estimate of the readout noise.
The Poisson rate parameter $\lambda$ provides a value for the average charge collected per pixel, which is proportional to serial register DC plus CIC.
We reject amplifiers that do not show clearly resolved charge peaks, with gain outside the range 1.2 to 1.6 ADU/$e^-$ or readout noise ${>}0.2$~ADU (${>}0.17~e^-$).
Figure~\ref{fig:skipper_fit} compares a CCD with single-electron resolution with one with poor skipper performance.

\subsubsection*{Full-Array Image}
\begin{figure}[t!]
    \centering
    \includegraphics[width=1\linewidth]{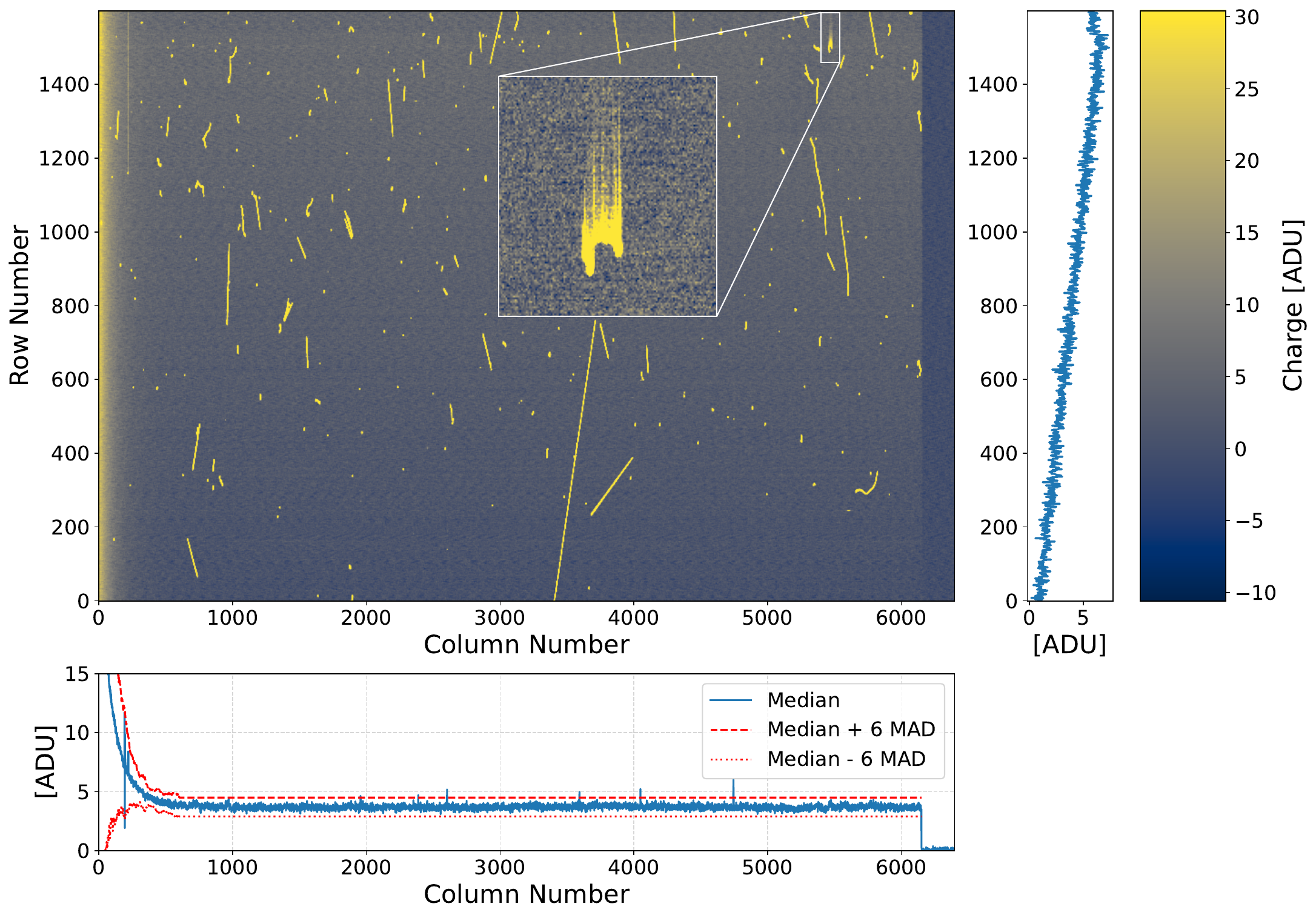}
    \caption[Full-sized CCD image analysis]{An example of full-sized image taken with amplifier L2, showing CTI$y$ in the top-right corner, far away from the amplifier, which is not possible to identify in quadrant-based testing. 
     }
    \label{fig:fullsized_image}
\end{figure}

In the tests described so far, each amplifier reads out only a single quadrant of the CCD’s active area.
While these tests allow us to discard amplifiers, quadrant-based readout does not fully characterize performance under final module conditions, where a single amplifier reads out the entire CCD.
For example, the fact that amplifier L of a SR exhibits CTI$x$, while the U amplifier does not, does not imply that the U amplifier is good, since a charge transfer problem on the L side of the SR will also lead to CTI$x$ when the full length of the SR is read out through amplifier U.
Likewise, quadrant-based readout cannot assess whether CTIy observed, for example, in the SR1 amplifiers, will also be present when the entire CCD is read out through the SR2 amplifiers.
Therefore, our final test reads out the entire CCD active area through each one of the four amplifiers.
We visually inspect these images for trailing charge after particle tracks, and calculate CTI for the composite image.
Any evidence for CTI disqualifies the amplifier. Figure~\ref{fig:fullsized_image} shows an amplifier that only exhibited CTIy issues in full-image readout, leading to its rejection.

\subsubsection{Quality Assessment}
\label{sec:quality_assessment_dice}

Following the criteria presented above, we select science-grade amplifiers with adequate gain and noise, and which read out particle tracks without distortion across the entire CCD, with no indication of CTI, with few defects and no glowing.
All tested CCDs were then classified based on the number of science-grade amplifiers as follows:

\begin{itemize}
    \item Tier-1: The highest quality CCD featuring four science-grade amplifiers. This configuration provides maximum flexibility in module integration since the CCD can be glued in either orientation and any amplifier can be connected.
    \item Tier-2: The CCD has two science-grade amplifiers on the U or L side, allowing either amplifier to be connected once the CCD is glued in the correct orientation.
    \item Tier-3: The CCD features at least one science-grade amplifier but no two of them on either the U or L side, requiring a specific amplifier to be connected once the CCD is glued.
    \item Tier-4: The CCD features no science-grade amplifiers, so it is not suitable for module fabrication.
\end{itemize}

\subsection{CCD Module Test}
\label{sec:moduletesting}

Once the selected CCDs were integrated into modules, the modules were tested to confirm successful fabrication, and performance consistent with the single-die test.
The CCD biases, clock voltages, timings and readout sequence were those optimized for the prototype modules operated in the LBC at LSM~\cite{DAMIC-M:2024ooa, DAMIC-M:2025luv}, except for $V_\text{sub}=60$~V, as in the die test.
The modules were also tested at lower temperatures, closer to the DAMIC-M operating temperature, to confirm the resilience of the package and the expected performance.
Evaluation at UW was devised to identify any issues to address during production since the final, exhaustive characterization of the DAMIC-M modules will be performed underground at LSM.

\subsubsection*{High Temperature Tests}
All single-die tests except the stress test (see Sec.~\ref{sec:dietest}) were repeated on the CCD modules at 185\,K to confirm CCD performance: skipper-amplifier response, defects, CTI, etc.
The four CCDs in the module were read out simultaneously, each through the selected amplifier, so the full array of the CCD was read out instead of quadrants.
The evolution of defects was monitored, and persistent defects were used as a ``fingerprint'' to confirm that the CCDs and amplifiers were glued and connected correctly.

\subsubsection*{Low Temperature Tests}
\begin{figure}[h!]
    \centering
\includegraphics[width=1\linewidth]{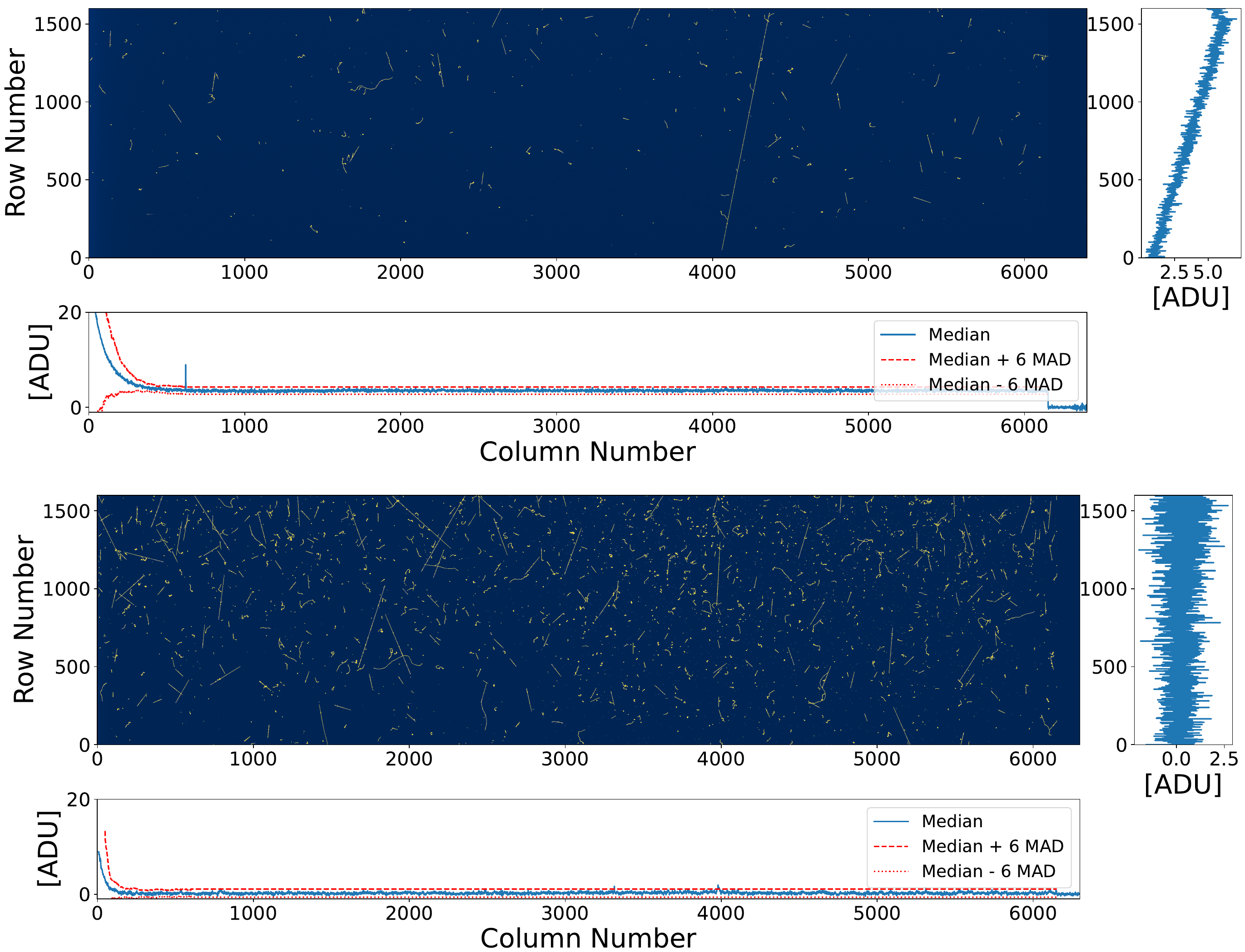}
    \caption[Module Vs. Die analysis comparison]{Comparison of full-array images of the same CCD from the die test at 185~K (top) and as part of a DAMIC-M module at 135~K (bottom). Note the much lower background charge levels at 135~K. The defect previously observed in Column 621 is no longer visible at lower temperatures. The significant increase in the number of tracks in the module is due to the longer readout time. Point-like $^{55}$Fe X-ray events from the activated iron foil installed in the module can be observed on the right-hand side of the image.
    
    }
    \label{fig:module_comparison}
\end{figure}

Our packaging schedule allowed for each module to be tested over two days (including thermal cycles).
This provided time for additional low-temperature tests at 135 K, closer to the DAMIC-M detector conditions.
One important check was to confirm single-electron resolution, which can degrade at lower temperature.
Since DC decreases with temperature, to estimate skipper performance, 10 pixels of the SR were transferred into the floating gate before measurement to ensure sufficient charge in the imaged pixels.
A composite median image, generated from three full-array images, allowed us to confirm the reduced intensity of defects.
The lower DC resulted in a decreased level of the background pixel charge, limiting our sensitivity of the CTI Analysis.
Therefore, CTI evaluation was performed by visual inspection of trails of charge following tracks in full-array images.
Figure~\ref{fig:module_comparison} shows a comparison between a full-array image acquired with the same CCD during die testing and in the module at low temperature.

\subsubsection*{Iron foil events}
X rays from the radioactive decay of $^{55}$Fe in the iron foils attached to the storage box lids (see Sec.~\ref{sec:modulepack}) produce front and backside ionization events.
These events will serve as a valuable tool for studying the energy response, CTI and lateral charge diffusion in the CCDs during characterization at LSM~\cite{Janesick:2001}.
A preliminary analysis was performed with the full-array images acquired at 135\,K.
The images were processed, with contiguous pixels above background noise grouped together into ``clusters.''
The sum of the pixel values in the cluster is proportional to the energy of the event, while the charge-weighted mean and standard deviation of the pixel coordinates in the cluster provide the location and lateral spread, respectively. 
The lateral spread $\sigma_{xy}$ is positively correlated with the interaction distance from the CCD pixel array, with front-side clusters only a few pixels in size, and backside clusters significantly more diffuse (larger $\sigma_{xy}$)~\cite{DAMIC:2016lrs}.

\begin{figure}[h!]
\centering
\includegraphics[width=0.9\textwidth]{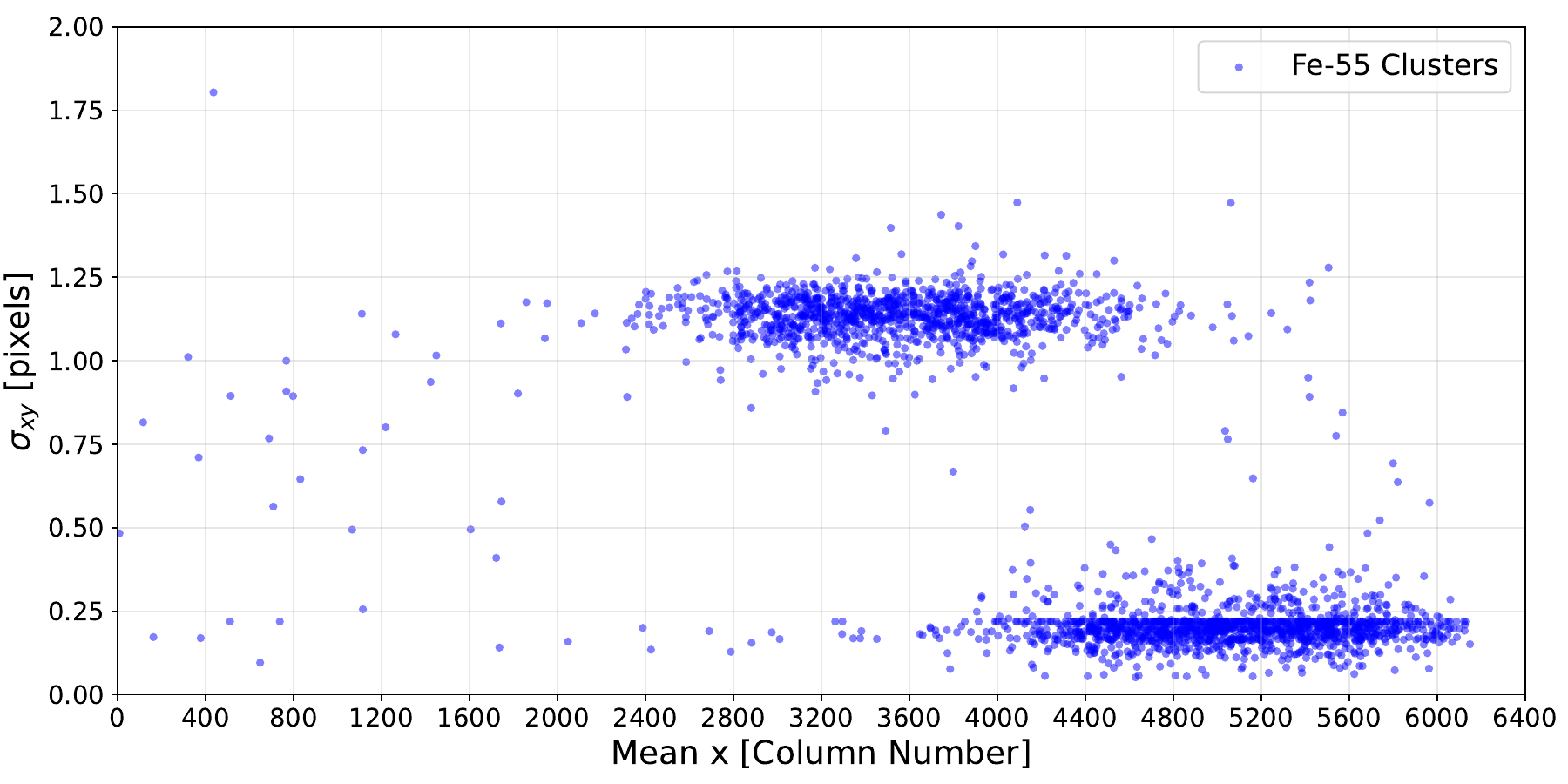}
\includegraphics[width=0.8\textwidth]{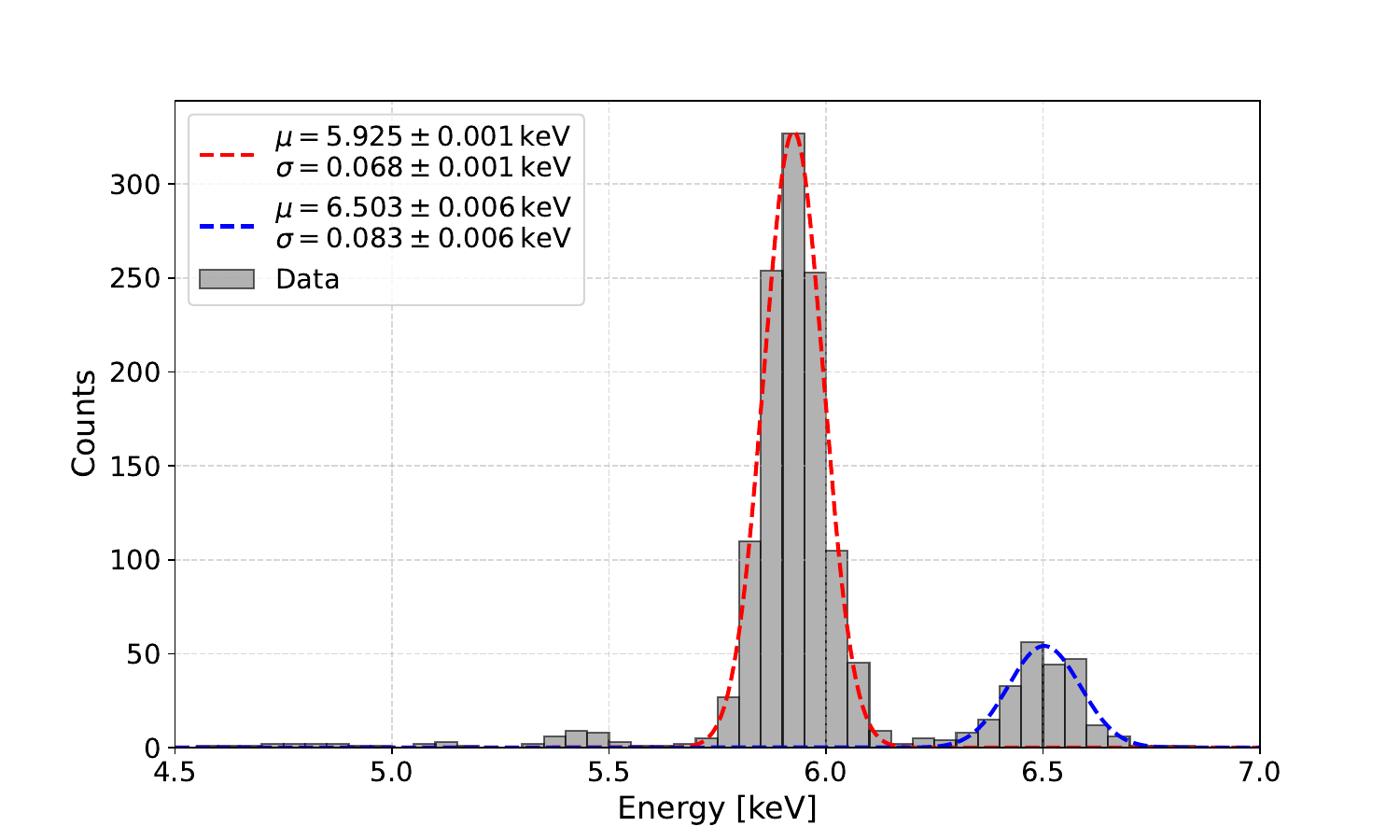}
\caption{Top: Clusters with energies between 4.5 and 7.0\,keV, mostly consisting of $^{55}$Fe events. Front and backside clusters can be clearly distinguished based on their $x$-location and lateral spread $\sigma_{xy}$. Bottom: Energy spectrum of front-side $^{55}$Fe events. The two characteristic Mn $K_\alpha$ and $K_\beta$ X-ray lines were fit to Gaussian functions, with best-fit parameters in the legend.
}
 
\label{fig:ironfoil}
\end{figure}

The top panel of Fig.~\ref{fig:ironfoil} shows a scatter plot of $\sigma_{xy}$ versus $x$ coordinate of clusters with energies between 4.5 and 7.0\,keV, calibrated using the gain of the amplifier, which consist mostly of the $^{55}$Fe events.
Two populations of clusters with different $\sigma_{xy}$, corresponding to the front and backside events, can be clearly identified.
These populations are also displaced along the $x$ axis because of the physical displacement of the foils.
The bottom panel presents the energy spectrum of the front-side clusters, selected as those with  $\sigma_{xy}<0.7$\, pixels, which feature higher signal-to-noise ratio because the charge is collected in fewer pixels.
The Mn $K_\alpha$ (5.9\,keV) and $K_\beta$ (6.5\,keV) emission lines are clearly visible with the expected energies and relative intensities~\cite{2008NDS...109..787J}. We reproduced these plots for all modules to confirm the response of CCDs to the $^{55}$Fe calibration source.

\section{Results}
\label{sec:results}

This section presents the results of the production of the first set of DAMIC-M CCD modules at the University of Washington, both in terms of their radioactivity and performance.

\subsection{Radioactive Contamination}
\label{sec:radioresults}

We present our results on radioactive contamination with reference to DAMIC at SNOLAB, the pioneering low-background CCD detector, for which an extensive study on radioactive backgrounds was performed~\cite{DAMIC:2021crr}.
We begin by summarizing in Table~\ref{tab:assay} the bulk U/Th/K radioactivity of all components in the DAMIC-M CCD modules.
The quality of the silicon is expected to remain the same as DAMIC, while there is a significant reduction in the background from the flex cable, which from simulations is expected to only contribute $0.03~\text{events}\cdot \text{keV}^{-1} \text{kg}^{-1} \text{day}^{-1}$ to the background spectrum in the ROI.
The dominant contribution to the background rate may now come from the wire bonds, with an upper limit $<0.08~\text{events}~\cdot~\text{keV}^{-1} \text{kg}^{-1} \text{day}^{-1}$.
The contribution from other components, namely the epoxies and bare-die components, is negligible.

\begin{table}[t]
\centering
\small
\begin{tabular}{lccccccc}
\toprule
\textbf{Component} & \textbf{Mass [g]} & \multicolumn{3}{c}{{\bf DAMIC-M Activity} \textbf{[$\bm{\upmu}$Bq/kg]}} & \multicolumn{3}{c}{{\bf SNOLAB Activity} \textbf{[$\bm{\upmu}$Bq/kg]}} \\
             &      & \textbf{$^{238}$U} & \textbf{$^{232}$Th} & \textbf{$^{40}$K} & \textbf{$^{238}$U} & \textbf{$^{232}$Th} & \textbf{$^{40}$K} \\
\midrule
CCDs         &  16.6 & {<}9.86 & {<}6.56 & {<}0.42 & {<}9.86 & {<}6.56 & {<}0.42 \\
Pitch adapter  & 8.65 & {<}15 & {<}7 & {<}2 & --- & --- & --- \\
Wire-bonds   & $1.5\times10^{-3}$ & $2.4\times10^5$ & $2.1\times10^5$ & <$10^7$ & --- & --- & --- \\
Flex cable     & 18.80 & $1.2\times 10^3$ & $3.3\times 10^2$ & $1.1\times 10^4$ & $5.8\times 10^4$ & $3.2\times 10^3$ & $2.9\times 10^4$ \\
Source follower  & 0.610 & {<}85 & {<}39 & {<}119 & --- & --- & --- \\
Epotek 301-2   & 0.140 & {<}41 & {<}17 & 90 & {<}220 & {<}45 & {<}130 \\
\bottomrule
\end{tabular}
\caption{Estimated U/Th/K radioactivities of the CCD module components compared to DAMIC at SNOLAB results~\cite{DAMIC:2021crr}. The estimates for the CCD and pitch adapter silicon are assumed to be the same as in DAMIC at SNOLAB. The pitch adapter activities include the aluminum traces and two resistors (0.02\% and 0.4\% of the total mass, respectively). Samples of the aluminum of the traces, wire-bond alloy, flex cables, Epotek 301-2 and bare-die components used for module production were assayed. The flex entries are averages weighted by the batch size of the two ICP-MS assays (20 and 8 cables, respectively). The source-follower entries were computed as the mass-weighted average of the assay results for each of its components (4 JFETs, 4 resistors and LOCTITE epoxy). Results of the assays of individual components are available on the \href{https://www.radiopurity.org}{radiopurity.org} data base.}
\label{tab:assay}
\end{table}

To assess the radioactivity introduced during packaging and testing, we considered the sources of contamination discussed in Sec.~\ref{sec:mitigation}: cosmogenic tritium production, plate-out of radon progeny, particulates and dust.

For exposures that are short compared to the tritium half-life, the introduced activity of cosmogenic tritium in the CCD bulk silicon is given by
\begin{equation}
A_{\text{trit}} = \frac{R_{\text{trit}}}{\tau_{\text{trit}}} \cdot \left( t_{\text{surf}} + \frac{t_{\text{NPL}}}{C_{\text{NPL}}} + \frac{t_{\text{PAB}}}{C_{\text{PAB}}} \right),
\label{eq:tritium_production}
\end{equation}
with the activation rate $R_{\text{trit}} = 112 \pm 24$~atoms/kg/day~\cite{saldanha2020cosmogenic}, the $^3$H lifetime $\tau_{\text{trit}}=17.5$~years, the average CCD exposure times on surface during transport, in storage at NPL and in the PAB lab: $t_{\text{surf}}~=~0.07 \pm 0.02$  ~days, $t_{\text{NPL}}~=~173.64 \pm 0.02$~days, and $t_{\text{PAB}}~=~6.85 \pm 0.01$~days, respectively, and the corresponding high-energy neutron suppression factors: $C_{\text{NPL}}~=~52\pm 1$ and $C_{\text{PAB}}~=~4.9\pm0.2$ from Sec.~\ref{sec:cosmoact}. The term in parenthesis corresponds to a surface-equivalent exposure at UW of 4.8~days, resulting in an increase in bulk tritium activity of: \begin{equation}
    A_{\text{trit}}=0.96 \pm 0.20 \,\upmu\text{Bq/kg}.
\end{equation}
This is only $\sim5\,\%$ of the DAMIC-M tritium budget, which accounts for 90 days of surface exposure during CCD fabrication and transport.

Surface contamination with $^{210}$Pb mostly occurs on the CCD front surfaces during the $120 \pm 22$ minutes that the CCDs were handled in the PAB clean room, which is the only period when the CCDs were not in a nitrogen-flushed environment.
The $^{210}$Pb deposition on the back surfaces of the CCDs is orders of magnitude smaller since the back surfaces were covered except for the few seconds that the CCDs were transferred from one container or holder to the next.
We estimate the $^{210}$Pb contamination from Ref.~\cite{Chernyak:2023izo}, which measured the deposition of $^{222}$Rn progeny on silicon photomultipliers (SiPMs) under environmental conditions similar to ours.
Specifically, our clean room is most comparable to ``fan on high, tent closed'' configuration, given the air exchange rate of 5/min (Sec.~\ref{sec:uthk}).
Considering the radon activity in our clean room (Sec.~\ref{sec:radplate}), scaling by our clean room surface-to-volume of 2.4 m$^{-1}$, and accounting for the $^{210}$Pb lifetime of $\tau_{\text{lead}} = 32.2$~years, we obtain an increase in the activity of $^{210}$Pb on the CCD front surfaces of: 
\begin{equation}
A_{\text{lead}} = 0.39 \pm 0.31~\text{nBq/cm}^2.
\end{equation}

For a CCD module, particulates accumulate from a total of 648 bonds, 148 per CCD and 56 bonds between flex and pitch adapter. This corresponds to an average of ${\sim}390$~particulates/cm$^2$ on the CCD module, based on the measurements described in Sec.~\ref{sec:uthk} and under the assumption that the wire bonder sheds particulates uniformly across the entire module surface. The average particulate radius is estimated as $r = 3.3\,\upmu$m from the microscope observations.
Although we have not confirmed the origin of the particulates, we assume that they come from the wire, with density $\rho = 2.7\,\text{g/cm}^3$.
This yields a total particulate mass density of $\sigma_{\text{part}} = 0.16\pm0.17\,\upmu$g/cm$^2$ on the CCD module.
Using the measured radioactivity of the wire provided in Table~\ref{tab:assay}, we obtain:
\[
\begin{aligned}
A_{^{232}\text{Th}} &= 38\pm41\,\text{pBq/cm}^2,~ 
A_{^{238}\text{U}} = 33\pm35 \,\text{pBq/cm}^2,~ 
A_{^{40}\text{K}}  <2 \,\text{nBq/cm}^2.
\end{aligned}
\]
For comparison, the accumulation of radioactivity from dust in a similar-class clean room to ours in a nearby geographical location can be found in Ref.~\cite{di2021direct,di2023evaluation}.
For the 2~hours that the CCDs were exposed to the clean room environment, the accumulated radioactivity is a factor of $10^3$ lower than the possible contribution from wire-bonder particulates.

Table \ref{tab:activity} summarizes the estimated additional contamination, in comparison with DAMIC-M budget and the measured backgrounds in DAMIC at SNOLAB. The DAMIC-M budget was set so that any one background source (first column) contributes $0.1~\text{events}\cdot \text{keV}^{-1} \text{kg}^{-1} \text{day}^{-1}$ to the energy spectrum in the ROI.


\begin{table}[t]
\centering
\small
\begin{tabular}{lcccc}
\toprule
\textbf{Source} & \textbf{Isotope} & \textbf{Added Activity} & \textbf{DAMIC-M Budget} & \textbf{SNOLAB Result} \\
\midrule
Tritium (cosmogenic)     & $^{3}$H       &
$0.96 \pm 0.20$\,$\upmu$Bq/kg     & 20\,$\upmu$Bq/kg             &$330 \pm 90$\,$\upmu$Bq/kg\\
\midrule
Radon (surface)          & $^{210}$Pb    & $0.39 \pm 0.31$\,nBq/cm$^2$ &{4\,nBq/cm$^2$} & $69 \pm 12$\,nBq/cm$^2$ \\
\midrule
                         & $^{238}$U    & $33\pm35$\,pBq/cm$^2$ & \multirow{2}{*}{2\,nBq/cm$^2$} & ---  \\
Particulates             & $^{232}$Th     & $38\pm41$\,pBq/cm$^2$ &                                 & --- \\
                         & $^{40}$K      & ${<}2$\,nBq/cm$^2$    & $20$\,nBq/cm$^2$                & --- \\
\bottomrule
\end{tabular}
\caption{Radioactivity estimates from tritium, radon plate-out, and wirebonder particulates, compared with DAMIC-M budget and DAMIC at SNOLAB results~\cite{DAMIC:2021crr}. The DAMIC-M budget reported for particulates corresponds to a total surface activity chosen such that
the combined $^{232}$Th and $^{238}$U contributions account for 0.05~$\text{events}\cdot \text{keV}^{-1} \text{kg}^{-1} \text{day}^{-1}$, while the $^{40}$K contribution
accounts for the remaining 0.05~$\text{events}\cdot \text{keV}^{-1} \text{kg}^{-1} \text{day}^{-1}$.
}
\label{tab:activity}
\end{table}

\subsection{CCD Dice}
\label{sec:ccdtesting_results}

Table~\ref{tab:grade_count_amp} summarizes the results from the tests of the 752 amplifiers.
An amplifier was either classified as science grade, or rejected according to the criteria listed in the table, determined following the tests described in Sec.~\ref{sec:dietest}.
Pixel-array defects were identified in the Composite Image (including Stress Test) and confirmed in the Full-Array Image.
Serial register (SR) defects were most often identified in the Serial Register Image, although when particularly intense they were also observed in the Composite Image.
Serial and parallel CTI were determined from the CTI Analysis and by visual inspection of tracks in the Full-Array Image.
A frequent problem for amplifier U1 was the presence of a large number of alternating hot/cold columns in the Composite Image, which investigations showed to be related to inefficient charge transfer from the pixel array into the SR.
Common issues with amplifier performance included large readout noise, low amplifier gain and pixels with values in between the discrete charge peaks.
Finally, a fraction of the amplifiers did not have the basic functionality to produce images with a stable baseline and particle tracks, with images sometimes missing large sections of the pixel array or exhibiting anomalous (or flat-line) output traces.

\begin{table}[t]
    \centering
    \setlength{\tabcolsep}{15pt}
    \begin{tabular}{lccccc}
        \toprule
         & \textbf{L1} & \textbf{L2} & \textbf{U1} & \textbf{U2} & \textbf{Total} \\
                 
        \midrule
        Science-grade & 70 & 79 & 33 & 83  & 265 \\
        \midrule
        Pixel-array defects & 48 & 40 & 43 & 39 & 170 \\
        SR defects & 14 & 11 & 18 & 9 & 52 \\
        CTI & 14 & 24 & 31 & 15 & 84 \\
        Parallel transfer into SR & 0 & 0 & 70 & 1 & 71 \\
        Amplifier performance & 17 & 10 & 7 & 6 & 40 \\
        Basic functionality &  27 & 26 & 28 & 26 & 107 \\
        \bottomrule
    \end{tabular}
    \caption{Tally of science-grade amplifiers (out of a total 188 for each L1, L2, U1 and U2) and the main reasons for why amplifiers were disqualified. Note that some amplifiers exhibited more than one problem so each of the L1--U2 columns may add to more than 188.}
    \label{tab:grade_count_amp}
\end{table}
\begin{table}[t]
    \centering
    \setlength{\tabcolsep}{30pt}
    \begin{tabular}{l l}
        \toprule
        \textbf{CCD tier} & \textbf{Total [\%]} \\
        \midrule
        Tier 1  & 16  [8.5\%] \\
        Tier 2  & 31 [16.5\%] \\
        Tier 3  & 66 [35.1\%] \\
        Tier 4  & 75 [39.9\%] \\
        \midrule
        Eligible for modules [Tiers 1--3] & 113 [60.1\%] \\
        \bottomrule
    \end{tabular}
    \caption{CCD tier distribution showing counts, percentages (with respect to the 188 devices), and the subset eligible for modules.}
    \label{tab:ccd_distribution}
\end{table}

In Table~\ref{tab:ccd_distribution}, we list the number of CCDs in each of Tiers 1--4 defined in Sec~\ref{sec:quality_assessment_dice}.
A total of 113 CCDs ($60$\% of the 188 tested) were eligible for module packaging.
We found that the yield of science-grade amplifiers, the observed problems and the CCD tier were approximately constant over time, indicating consistent production and testing processes.

\subsection{CCD Modules}

We report the main observations from the tests of the 28 modules (112 CCDs) described in Sec.~\ref{sec:moduletesting}. Overall, the CCDs exhibited the same excellent performance observed in the die tests, except for the cases noted below.

\subsubsection*{Defects}
A comparison between die and module tests showed that not all defects first identified in the die test were also present in the module. This is common for the first few thermal cycles of a CCD, as defects often appear due to thermal stress or disappear from annealing of the crystal.
However, a handful of localized defects always remained in every CCD, which was used to confirm that all CCDs were glued in the intended position and orientation in the module.
Two CCDs in separate modules developed very bright defects in the pixel array. In one case, the defect caused considerable glowing that affected all CCDs in the module. We attempted to salvage the module by removing the wire connections to the $V_\text{sub}$ pads of the problematic CCD. However, the physical contact between the CCD die to its neighbors was sufficient to establish electrical contact between the CCD backsides, through which $V_\text{sub}$ remained applied. Following this event, we introduced a small gap between CCD dice in the module, although such bright defects were not encountered again. Finally, small SR defects were identified in three CCDs in separate modules. Their impact on CCD performance will be studied in more detail at LSM. In the worst case, they will result in the loss of the $x$ region of the CCD past the defect.

\begin{figure}[h]
    \centering
    \includegraphics[width=0.49\linewidth]{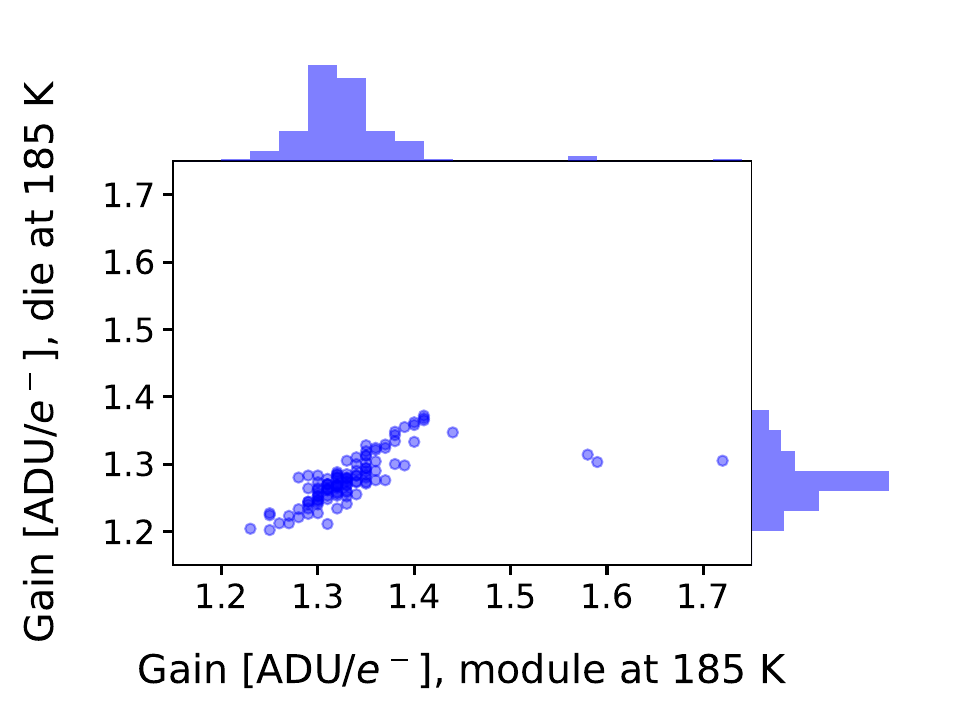}
    \includegraphics[width=0.49\linewidth]{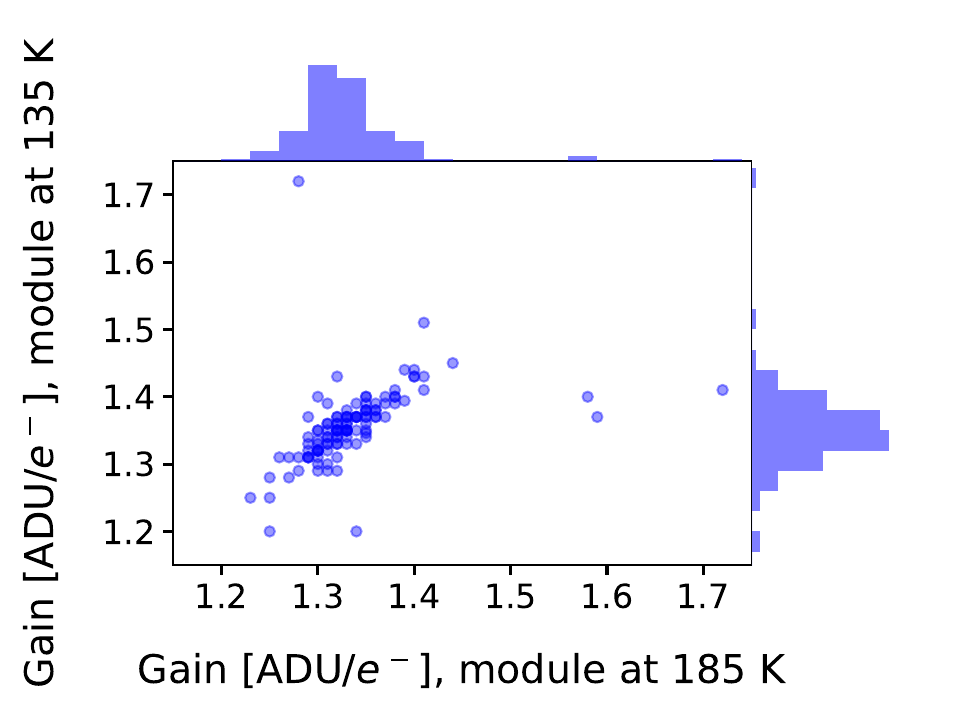}

    \includegraphics[width=0.495\linewidth]{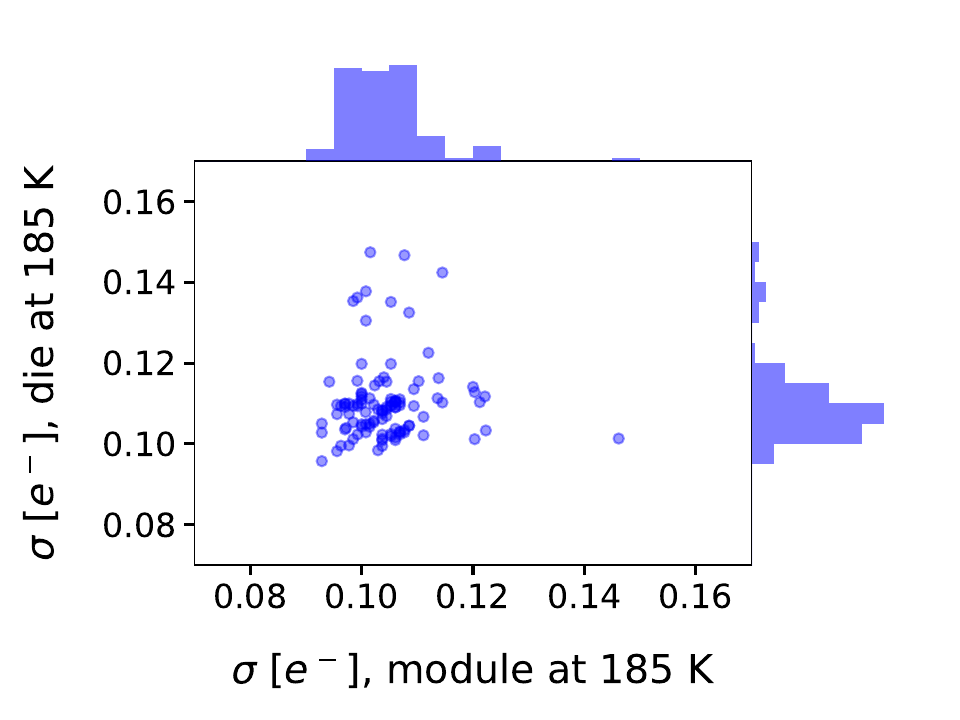}
    \includegraphics[width=0.495\linewidth]{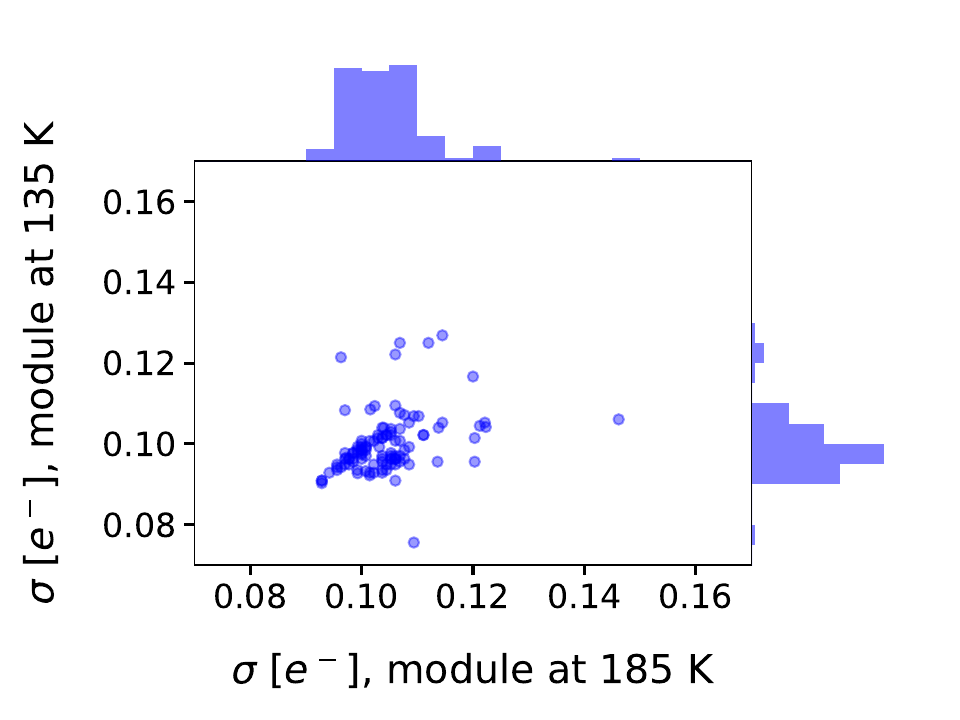}

    \caption{Top: scatter plots of the amplifier gains from the die test (left) and module test at 135~K (right) versus the module test at 185 K. In both comparisons, the data show strong linear correlation, as expected, since the gain is intrinsic to the readout amplifier. The range for science-grade amplifiers is 1.2--1.6 ADU/$e^-$. Bottom: same as top, but for the 1000-NDCM readout noise ($\sigma$). Although there is weaker correlation between the die and module tests (see text), all amplifiers in the modules satisfy the science-grade criterion of ${<}0.2$ ADU (${<}0.17\,e^-$).} 
    \label{fig:scatterplots}
\end{figure}

\subsubsection*{Charge Transfer}
The CTI Analysis at high temperature (185~K) confirmed that no CCD had CTI$x$ $\gtrsim1\%$, as expected from the die test.
However, two CCDs in separate modules exhibited localized CTI$y$, visible as charge trails on particle tracks. For one of them, the problem was already present in the die test but was missed by the visual inspection. At low temperature (135~K), one CCD developed the issue with parallel transfer into the SR described in Sec.~\ref{sec:ccdtesting_results}, while seven more started showing slight CTI$y$. To some extent, such behavior is expected since charge transfer worsens at lower temperatures, 
but it should be resolved by tuning clock parameters at LSM.

\subsubsection*{Amplifier Performance}
All amplifiers showed consistent results from the Skipper Performance tests completed during die and module testing. Since amplifiers are most sensitive to ESD, this confirms appropriate handling and ESD-safety precautions. All but one of the amplifiers exhibited excellent performance at low temperature (135~K).
This amplifier had the lowest gain at 1.2~ADU/$e^-$, the lower bound of the acceptable range, below which we commonly observed the loss of single-electron resolution at low temperature. Figure~\ref{fig:scatterplots} shows scatter plots of the gain and single-electron resolution. The gain and resolution in the die test at 185~K (left panels) and in the module test at 135~K (right panels) are plotted against the values in the module test at 185~K as reference.
The gain of the amplifiers was strongly correlated in all configurations, demonstrating that variations in the gain are intrinsic to the amplifier and that our initial die tests were indeed indicative of amplifier performance. On the other hand, the correlation between the single-electron resolution in the die and module test is weaker. Since most of the amplifiers were tested through different electronics channels between the die and module tests, it is likely that this caused the ${\sim}$20\% variations in single-electron resolution between die and module tests. This argument is supported by the tighter correlation in the resolution between the module tests at different temperatures. In any case, all CCD module amplifiers meet the science-grade resolution criterion of ${<}0.2$ ADU (${<}0.17\,e^-$).

\subsubsection*{Serial-register charge}
The average charge per pixel ($\lambda$) from the Skipper Performance test is dominated by DC in the SR, with an important contribution from environmental radiation.
The contribution from CIC should be negligible since we use the clock parameters optimized for the prototype modules in the LBC~\cite{DAMIC-M:2025luv}.
As expected, we measured similar $\lambda$ across all CCDs, with no outliers, which would hint at underlying problems in a particular CCD.
In the module tests, we found that the median $\lambda$ decreases by a factor ${\sim}10^3$ between 185~K and 135~K. Statistical mechanics predicts a decrease in DC by a factor of ${\sim}10^6$ between 185~K and 135~K ~\cite{Janesick:2001,2002SPIE.4669..193W,10.1063/1.1663501}, three orders of magnitude lower than observed.
This discrepancy has been observed before when CCDs are operated on the surface and may be attributed to environmental radiation, but it could also be caused by surface leakage currents that do not decrease as much with temperature.
Detailed studies in the much lower-background environment of LSM will help disentangle these effects.

\section{Conclusion}
\label{sec:conclusion}

The DAMIC-M Collaboration completed the first production of CCD modules at the University of Washington (UW), between July and December 2024.
A total of 188 individual CCDs were first tested in a single-die temporary package.
From these, 112 were selected to fabricate 28 CCD modules.
Each CCD module was then tested to confirm science-grade performance, consistent with the single-die results.
One module was lost due to the appearance of a very bright defect, while a second may develop a similar problem.
Of the remaining 26 modules, only a handful of CCDs exhibited some issues.
These will be revisited during final testing at LSM and we expect them to be mostly resolved by optimizing CCD parameters.
The activities were carried out in controlled environments across all stages of the campaign to suppress radioactive contamination.
The estimated added radioactivity 
is well within DAMIC-M radio-purity requirements.

The 28 CCD modules are currently stored in a nitrogen-flushed environment under 4800-m-water-equivalent overburden at LSM.
A second campaign is starting underground to perform the final, accurate characterization and optimization of all CCDs.
The deployment of 26 of these CCD modules in the DAMIC-M detector is foreseen in early 2026.

\section{Acknowledgments}

We would like to thank the Center for Experimental Nuclear Physics and Astrophysics (CENPA) at the University of Washington (UW) and its staff for support through laboratory space, logistical and technical services.
In particular, Director Prof. David Hertzog for prioritizing the project, Prof. Jens Gundlach and Dr. Michael Ross for providing access to the Gravity Garage, and Prof. Peter Kammel and Prof. Alejandro Garcia for the equipment and assistance with the cosmogenic neutron measurement.
We thank the Washington Nanofabrication Facility (WNF) and its staff for their assistance and patience through pitch adapter production.
We thank EOSPACE President Dr. Suwat Thaniyavarn, KK Wong and Kelvin Lim for providing much needed access to their dicing saw.
We thank Bert Harrop at Princeton University for introducing us to the bi-layer lift-off process.

The CCD Laboratory at the UW was set up with contributions from the Department of Physics and the College of Arts \& Sciences.
The UW DAMIC-M group was supported through Grant No. NSF PHY-2413014.
Part of this work was conducted at the WNF, a National Nanotechnology Coordinated Infrastructure (NNCI) site at the UW with partial support from the National Science Foundation (NSF) via awards NNCI-1542101 and NNCI-2025489.
The University of Chicago  was supported through Grant No. NSF PHY-2413013 and by the Kavli Institute for Cosmological Physics through an endowment from the Kavli Foundation.
The DAMIC-M group at Johns Hopkins University was supported by the Krieger School of Arts \& Sciences.
The development of the radiopure flex cables was carried out under the U.S. Department of Energy (DOE) SBIR Program (Award No. DE-SC0021547) through a collaboration between Q-Flex Inc. and  Pacific Northwest National Laboratory (PNNL). PNNL is a multi-program national laboratory operated for the DOE by Battelle Memorial Institute under contract number DE-AC05-76RL01830.

The DAMIC-M project has received funding from the European Research Council (ERC) under the European Union’s Horizon 2020 research and innovation programme Grant Agreement No. 788137, and from NSF through Grant No. NSF PHY-1812654. IFCA was supported by the project DMpheno2lab (PID2022-139494NB-I00) financed by MCIN/AEI/10.13039/501100011033/FEDER, EU.
The CCD development work at Lawrence Berkeley National Laboratory MicroSystems Lab was supported in part by the Director, Office of Science, of the U.S. Department of Energy under Contract No. DE-AC02-05CH11231. We thank Teledyne DALSA Semiconductor for CCD fabrication.

\bibliography{biblio}

\providecommand{\href}[2]{#2}\begingroup\raggedright\begin{thebibliography}{10}

\bibitem{Bertone:2016nfn}
G.~Bertone and D.~Hooper, \emph{{History of dark matter}},
  \href{https://doi.org/10.1103/RevModPhys.90.045002}{\emph{Rev. Mod. Phys.}
  {\bfseries 90} (2018) 045002}
  [\href{https://arxiv.org/abs/1605.04909}{{\ttfamily 1605.04909}}].

\bibitem{Drukier:1984vhf}
A.~Drukier and L.~Stodolsky, \emph{{Principles and applications of a neutral
  current detector for neutrino physics and astronomy}},
  \href{https://doi.org/10.1103/PhysRevD.30.2295}{\emph{Phys. Rev. D}
  {\bfseries 30} (1984) 2295}.

\bibitem{Goodman:1984dc}
M.W.~Goodman and E.~Witten, \emph{{Detectability of certain dark matter
  candidates}}, \href{https://doi.org/10.1103/PhysRevD.31.3059}{\emph{Phys.
  Rev. D} {\bfseries 31} (1985) 3059}.

\bibitem{Billard:2021uyg}
J.~Billard et~al., \emph{{Direct detection of dark matter\textemdash{}APPEC
  committee report}},
  \href{https://doi.org/10.1088/1361-6633/ac5754}{\emph{Rept. Prog. Phys.}
  {\bfseries 85} (2022) 056201}
  [\href{https://arxiv.org/abs/2104.07634}{{\ttfamily 2104.07634}}].

\bibitem{Boehm:2003ha}
C.~Boehm, P.~Fayet and J.~Silk, \emph{{Light and heavy dark matter particles}},
  \href{https://doi.org/10.1103/PhysRevD.69.101302}{\emph{Phys. Rev. D}
  {\bfseries 69} (2004) 101302}
  [\href{https://arxiv.org/abs/hep-ph/0311143}{{\ttfamily hep-ph/0311143}}].

\bibitem{Hooper:2008im}
D.~Hooper and K.M.~Zurek, \emph{{A natural supersymmetric model with MeV dark
  matter}}, \href{https://doi.org/10.1103/PhysRevD.77.087302}{\emph{Phys. Rev.
  D} {\bfseries 77} (2008) 087302}
  [\href{https://arxiv.org/abs/0801.3686}{{\ttfamily 0801.3686}}].

\bibitem{Pospelov:2007mp}
M.~Pospelov, A.~Ritz and M.B.~Voloshin, \emph{{Secluded WIMP dark matter}},
  \href{https://doi.org/10.1016/j.physletb.2008.02.052}{\emph{Phys. Lett. B}
  {\bfseries 662} (2008) 53} [\href{https://arxiv.org/abs/0711.4866}{{\ttfamily
  0711.4866}}].

\bibitem{Knapen:2017xzo}
S.~Knapen, T.~Lin and K.M.~Zurek, \emph{{Light dark matter: models and
  constraints}}, \href{https://doi.org/10.1103/PhysRevD.96.115021}{\emph{Phys.
  Rev. D} {\bfseries 96} (2017) 115021}
  [\href{https://arxiv.org/abs/1709.07882}{{\ttfamily 1709.07882}}].

\bibitem{Privitera:2024tpq}
{\scshape DAMIC-M} collaboration, \emph{{The DAMIC-M experiment: status and
  first results}}, \href{https://doi.org/10.22323/1.441.0066}{\emph{PoS}
  {\bfseries TAUP2023} (2024) 066}.

\bibitem{DAMIC-M:2024ooa}
{\scshape DAMIC-M} collaboration, \emph{{The DAMIC-M Low Background Chamber}},
  \href{https://doi.org/10.1088/1748-0221/19/11/T11010}{\emph{JINST} {\bfseries
  19} (2024) T11010} [\href{https://arxiv.org/abs/2407.17872}{{\ttfamily
  2407.17872}}].

\bibitem{DAMIC-M:2025luv}
{\scshape DAMIC-M} collaboration, \emph{{Probing Benchmark Models of
  Hidden-Sector Dark Matter with DAMIC-M}},
  \href{https://doi.org/10.1103/2tcc-bqck}{\emph{Phys. Rev. Lett.} {\bfseries
  135} (2025) 071002} [\href{https://arxiv.org/abs/2503.14617}{{\ttfamily
  2503.14617}}].

\bibitem{DAMIC:2021crr}
{\scshape DAMIC} collaboration, \emph{{Characterization of the background
  spectrum in DAMIC at SNOLAB}},
  \href{https://doi.org/10.1103/PhysRevD.105.062003}{\emph{Phys. Rev. D}
  {\bfseries 105} (2022) 062003}
  [\href{https://arxiv.org/abs/2110.13133}{{\ttfamily 2110.13133}}].

\bibitem{DAMIC:2021esz}
{\scshape DAMIC} collaboration, \emph{{Results on low-mass weakly interacting
  massive particles from a 11 kg d target exposure of DAMIC at SNOLAB}},
  \href{https://doi.org/10.22323/1.395.0539}{\emph{PoS} {\bfseries ICRC2021}
  (2021) 539} [\href{https://arxiv.org/abs/2108.05983}{{\ttfamily
  2108.05983}}].

\bibitem{topsil}
{Topsil}. \url{https://www.topsil.com/en/}.

\bibitem{sehe}
{Shin-Etsu Handotai Europe}. \url{https://www.sehe.com}.

\bibitem{dalsa}
{Teledyne-DALSA}. \url{https://www.teledynedalsa.com/en/home/}.

\bibitem{wnf}
{Washington Nanofabrication Facility (WNF)}. \url{https://www.wnf.uw.edu}.

\bibitem{Arnquist:2023gtq}
I.J.~Arnquist, M.L.~di~Vacri, N.~Rocco, R.~Saldanha, T.~Schlieder, R.~Patel
  et~al., \emph{{Ultra-low radioactivity flexible printed cables}},
  \href{https://doi.org/10.1140/epjti/s40485-023-00104-6}{\emph{EPJ Tech.
  Instrum.} {\bfseries 10} (2023) 17}
  [\href{https://arxiv.org/abs/2303.10862}{{\ttfamily 2303.10862}}].

\bibitem{pnnl}
{Pacific Northwest National Laboratory (PNNL)}. \url{https://www.pnnl.gov}.

\bibitem{saldanha2020cosmogenic}
R.~Saldanha et~al., \emph{{Cosmogenic activation of silicon}},
  \href{https://doi.org/10.1103/PhysRevD.102.102006}{\emph{Phys. Rev. D}
  {\bfseries 102} (2020) 102006}
  [\href{https://arxiv.org/abs/2007.10584}{{\ttfamily 2007.10584}}].

\bibitem{TENDL2021_Si28}
A.~Koning and D.~Rochman, ``{TENDL-2021: TALYS-based evaluated nuclear data
  library}.'' IAEA Nuclear Data Services, 2021.

\bibitem{traina:tel-04063059}
M.~Traina, \emph{{Search for light dark matter and exploration of the hidden
  sector with the DAMIC at SNOLAB and DAMIC-M charge-coupled devices}}, Ph.D.
  thesis, {Sorbonne Universit{\'e}}, Oct., 2022.

\bibitem{deDominicis:2022nls}
C.~de~Dominicis, \emph{{Search for light dark matter with DAMIC-M experiment}},
  Ph.D. thesis, SUBATECH, Nantes, 2022.

\bibitem{COMRIE201543}
A.~Comrie, A.~Buffler, F.~Smit and H.~W{\"o}rtche, \emph{{Digital neutron/gamma
  discrimination with an organic scintillator at energies between 1 MeV and 100
  MeV}},
  \href{https://doi.org/https://doi.org/10.1016/j.nima.2014.10.058}{\emph{Nucl.
  Instrum. Meth. A} {\bfseries 772} (2015) 43}.

\bibitem{nakao1995measurements}
N.~Nakao, T.~Nakamura, M.~Baba, Y.~Uwamino, N.~Nakanishi, H.~Nakashima et~al.,
  \emph{{Measurements of response function of organic liquid scintillator for
  neutron energy range up to 135 MeV}},
  \href{https://doi.org/10.1016/0168-9002(95)00193-X}{\emph{Nucl. Instrum.
  Meth. A} {\bfseries 362} (1995) 454}.

\bibitem{Wen1974DesignAO}
D.~Wen, \emph{Design and operation of a floating gate amplifier},
  \href{https://doi.org/10.1109/JSSC.1974.1050535}{\emph{IEEE J. Solid-State
  Circuits} {\bfseries 9} (1974) 410}.

\bibitem{Tiffenberg:2017aac}
{\scshape SENSEI} collaboration, \emph{{Single-electron and single-photon
  sensitivity with a silicon Skipper CCD}},
  \href{https://doi.org/10.1103/PhysRevLett.119.131802}{\emph{Phys. Rev. Lett.}
  {\bfseries 119} (2017) 131802}
  [\href{https://arxiv.org/abs/1706.00028}{{\ttfamily 1706.00028}}].

\bibitem{wells1979fits}
D.C.~{Wells}, E.W.~{Greisen} and R.H.~{Harten}, \emph{{FITS -- a Flexible Image
  Transport System}}, {\emph{Astron. Astrophys. Suppl. Ser.} {\bfseries 44}
  (1981) 363}.

\bibitem{johnsonnoise}
J.B.~Johnson, \emph{Thermal agitation of electricity in conductors},
  \href{https://doi.org/10.1103/PhysRev.32.97}{\emph{Phys. Rev.} {\bfseries 32}
  (1928) 97}.

\bibitem{PhysRev.32.110}
H.~Nyquist, \emph{Thermal agitation of electric charge in conductors},
  \href{https://doi.org/10.1103/PhysRev.32.110}{\emph{Phys. Rev.} {\bfseries
  32} (1928) 110}.

\bibitem{skipperjanesick}
J.R.~Janesick, T.S.~Elliott, A.~Dingiziam, R.A.~Bredthauer, C.E.~Chandler,
  J.A.~Westphal et~al., \emph{{New advancements in charge-coupled device
  technology: subelectron noise and 4096 $\times$ 4096 pixel CCDs}},  in
  \emph{Charge-Coupled Devices and Solid State Optical Sensors}, M.M.~Blouke,
  ed., vol.~1242, pp.~223 -- 237, International Society for Optics and
  Photonics, SPIE, 1990, \href{https://doi.org/10.1117/12.19452}{DOI}.

\bibitem{Janesick:2001}
J.R.~Janesick, \emph{Scientific charge coupled devices}, SPIE (2001),
  \href{https://doi.org/10.1117/3.374903}{10.1117/3.374903}.

\bibitem{2002SPIE.4669..193W}
R.~{Widenhorn}, M.M.~{Blouke}, A.~{Weber}, A.~{Rest} and E.~{Bodegom},
  \emph{{Temperature dependence of dark current in a CCD}},  in \emph{Sensors
  and Camera Systems for Scientific, Industrial, and Digital Photography
  Applications III}, M.M.~{Blouke}, J.~{Canosa} and N.~{Sampat}, eds.,
  vol.~4669 of \emph{Society of Photo-Optical Instrumentation Engineers (SPIE)
  Conference Series}, pp.~193--201, Apr., 2002,
  \href{https://doi.org/10.1117/12.463446}{DOI}.

\bibitem{Oscura:2023qik}
{\scshape Oscura} collaboration, \emph{{Skipper-CCD sensors for the Oscura
  experiment: requirements and preliminary tests}},
  \href{https://doi.org/10.1088/1748-0221/18/08/P08016}{\emph{JINST} {\bfseries
  18} (2023) P08016} [\href{https://arxiv.org/abs/2304.04401}{{\ttfamily
  2304.04401}}].

\bibitem{shockleyread1952}
W.~Shockley and W.~Read~Jr, \emph{Statistics of the recombinations of holes and
  electrons}, \href{https://doi.org/10.1103/PhysRev.87.835}{\emph{Phys. Rev.}
  {\bfseries 87} (1952) 835}.

\bibitem{2014JInst...9C2004H}
D.J.~{Hall}, N.J.~{Murray}, J.P.D.~{Gow}, D.~{Wood} and A.~{Holland},
  \emph{{Studying defects in the silicon lattice using CCDs}},
  \href{https://doi.org/10.1088/1748-0221/9/12/C12004}{\emph{JINST} {\bfseries
  9} (2014) C12004}.

\bibitem{DAMIC:2016lrs}
{\scshape DAMIC} collaboration, \emph{{Search for low-mass WIMPs in a 0.6 kg
  day exposure of the DAMIC experiment at SNOLAB}},
  \href{https://doi.org/10.1103/PhysRevD.94.082006}{\emph{Phys. Rev. D}
  {\bfseries 94} (2016) 082006}
  [\href{https://arxiv.org/abs/1607.07410}{{\ttfamily 1607.07410}}].

\bibitem{2008NDS...109..787J}
H.~{Junde}, \emph{{Nuclear data sheets for A = 55}},
  \href{https://doi.org/10.1016/j.nds.2008.03.001}{\emph{Nucl. Data Sheets}
  {\bfseries 109} (2008) 787}.

\bibitem{Chernyak:2023izo}
D.~Chernyak, J.~Howell, D.~Majumdar, N.~Mukherjee, O.~Nusair and A.~Piepke,
  \emph{{Comprehensive study of radon progeny attachment to surfaces}},
  \href{https://doi.org/10.1103/PhysRevC.107.065802}{\emph{Phys. Rev. C}
  {\bfseries 107} (2023) 065802}
  [\href{https://arxiv.org/abs/2301.07786}{{\ttfamily 2301.07786}}].

\bibitem{di2021direct}
M.L.~di~Vacri, I.J.~Arnquist, S.~Scorza, E.W.~Hoppe and J.~Hall, \emph{{Direct
  method for the quantitative analysis of surface contamination on ultra-low
  background materials from exposure to dust}},
  \href{https://doi.org/10.1016/j.nima.2021.165051}{\emph{Nucl. Instrum. Meth.
  A} {\bfseries 994} (2021) 165051}
  [\href{https://arxiv.org/abs/2006.12746}{{\ttfamily 2006.12746}}].

\bibitem{di2023evaluation}
M.L.~di~Vacri, S.~Scorza, A.~French, N.D.~Rocco, T.D.~Schlieder, I.J.~Arnquist
  et~al., \emph{{Evaluation of SNOLAB background mitigation procedures through
  the use of an ICP-MS based dust monitoring methodology}},
  \href{https://doi.org/10.1016/j.nima.2023.168700}{\emph{Nucl. Instrum. Meth.
  A} {\bfseries 1056} (2023) 168700}
  [\href{https://arxiv.org/abs/2308.12253}{{\ttfamily 2308.12253}}].

\bibitem{10.1063/1.1663501}
W.~Bludau, A.~Onton and W.~Heinke, \emph{Temperature dependence of the band gap
  of silicon}, \href{https://doi.org/10.1063/1.1663501}{\emph{J. Appl. Phys.}
  {\bfseries 45} (1974) 1846}.

\end{thebibliography}\endgroup
\end{document}